\documentclass[prx,aps,twocolumn,showpacs,preprintnumbers,superscriptaddress, 10pt]{revtex4-2}

\usepackage{graphicx}% Include figure files
\usepackage{amssymb}
\usepackage{amsmath}
\usepackage{times}
\usepackage{graphicx}
\usepackage{braket}
\usepackage{framed}
\usepackage{empheq}

\usepackage{color}
\newcommand{\rot}[1]{{\color{black} #1}}

\newcommand{\blau}[1]{{\color{black} #1}}

\usepackage{enumitem}
\usepackage{tabularx}
\usepackage{bbm}
\usepackage{hyperref}

\usepackage[disable]{todonotes} %disables all todonotes
\newcommand{\mat}[1]{\mathrm{#1}}

\newcommand{\atantwo}{\text{atan2}}
\newcommand{\muext}{\mu}

\newcommand{\pd}[2]{\frac{\partial#1}{\partial#2}}
\newcommand{\var}[1]{\text{var}(#1)}
\newcommand{\cov}[1]{\text{cov}(#1)}

\newcommand{\lrk}[1]{\left\langle #1 \right\rangle}
\newcommand{\lrrund}[1]{\left( #1 \right)}
\newcommand{\lreckig}[1]{\left[ #1 \right]}

\newcommand{\od}[2]{\frac{\mathrm{d}#1}{\mathrm{d}#2}}

\newcommand{\bs}[1]{\boldsymbol{#1}}

\newcommand{\PM}[1]{\begin{pmatrix}#1\end{pmatrix}}
\newcommand{\phimax}{r_\text{m}}
\newcommand{\betaeff}{\beta_\text{eff}}
\newcommand{\Mat}[1]{\operatorname{#1}}
\newcommand{\Vect}[1]{\bs{#1}}
\newcommand{\Id}{\Mat{I}}

\newcommand{\commentout}[1]{}

\newcommand{\sigmaext}{\sigma_{\text{ext}}}

\renewcommand{\Re}{\text{Re}}

\usepackage{pifont}
\newcommand{\cmark}{\ding{51}}%
\newcommand{\xmark}{\ding{55}}%
\usepackage{placeins}
\DeclareMathSymbol{\shortminus}{\mathbin}{AMSa}{"39}

\newcommand{\erf}{\mathrm{erf}}

\newcommand{\R}{\mathbb{R}}
\newcommand{\N}{\mathbb{N}}

\begin{document}

%\title{Mesoscopic dynamics of spiking neural networks with annealed random connectivity}
%\title{Mean-field dynamics of finite-size spiking neural networks with annealed random connectivity}
%\title{Annealed mean-field approximation for finite-size spiking neural networks with random connectivity}
%\title{How random connectivity shapes neural variability: Annealed mean-field theory of finite-size spiking neural networks}
%\title{How connection probability shapes fluctuations of neural population dynamics}
%\title{\rot{How random connectivity shapes fluctuations of finite-size neural populations}}
\title{\rot{How random connectivity shapes the fluctuating dynamics of finite-size neural populations}}

\author{Nils E. Greven}
\affiliation{Institute of Mathematics, Technical University Berlin, 10623 Berlin, Germany.}
\affiliation{Bernstein Center for Computational Neuroscience Berlin, 10115 Berlin, Germany.}

\author{Jonas Ranft}
\email[Contact author: ]{jonas.ranft@ens.psl.eu}
\affiliation{Institut de Biologie de l'ENS, Ecole normale supérieure, PSL University, CNRS, Paris.}

\author{Tilo Schwalger}
\email[Contact author: ]{tilo.schwalger@bccn-berlin.de}
\affiliation{Institute of Mathematics, Technical University Berlin, 10623 Berlin, Germany.}
\affiliation{Bernstein Center for Computational Neuroscience Berlin, 10115 Berlin, Germany.}

\date{\today}

\begin{abstract}
\rot{Mesoscopic models of finite-size} neuronal populations \rot{are crucial} to understand the dynamics of neural networks in the brain, \rot{especially their fluctuations} and response to stimuli. \rot{However, current theories to derive such models are based on the assumption of homogeneous all-to-all (full) connectivity, neglecting the variance in the connectivity of biologically realistic networks with} connection probabilities $p<1$ (non-full connectivity). To gain insight into the different fluctuation mechanisms underlying the neural variability of populations of spiking neurons, we derive and analyze a stochastic mean-field model for finite-size networks of Poisson neurons with \rot{random connectivity (including non-full connectivity)}, external noise and disordered mean inputs. We treat the quenched disorder of the connectivity by an annealed approximation enabling a \rot{doubly stochastic description of synaptic inputs for finite network size. A further reduction leads} to a low-dimensional closed system of coupled Langevin equations for the mean and variance of the neuronal membrane potentials as well as a variable capturing finite-size fluctuations arising specifically in the case \rot{of connectivity disorder}. Comparing to microscopic simulations, we find that the mesoscopic model describes the fluctuations and nonlinearities well and outperforms previous mesoscopic models that neglected the \rot{variance in the} connectivity. The joint effect of connectivity disorder and finite network size can be analytically understood by a softening of the effective nonlinearity and the multiplicative character of spiking noise. The mesoscopic theory shows that quenched disorder can stabilize the asynchronous state, and it correctly predicts large quantitative and non-trivial qualitative effects of connection probability on the variance of the population firing rate and its dependence on stimulus strength. Our theory thus elucidates how disordered connectivity shapes nonlinear dynamics and fluctuations of neural populations at the mesoscopic scale and showcases a useful mean-field method to treat \rot{random} connectivity in finite-size, spiking neural networks.
\end{abstract}
\maketitle

\section{Introduction}

Biological systems often exhibit significant fluctuations and variability in their dynamics, in contexts ranging from gene expression to neural activity in the brain. The possible roles of fluctuations for the dynamics and functions of such systems have thus received increasing attention~\cite{EloLev02, RajOud08, TahBra15, ArzTah24, FaiSel08,DoiLit16,WasKlo21}.
%THIS NEEDS TO BE CITED~\cite{[Elowitz et al., Science (2002); Raj, Oudenaarden, Cell (2008)], %[Stochastic cell division: Teheri-Agari et al., Current biology %(2015)], [Tissue mechanics: Arzash et al., arxiv (2023)], FaiSel08}%Hairbundel
An important tool to understand the collective dynamics of biological systems consisting of many interacting units
%(particles, neurons, genes, ...)
are low-dimensional descriptions of the effective dynamics at the \rot{mesoscopic or macroscopic scale}. Taking fluctuations properly into account can be necessary to establish a correct understanding of the system and can provide crucial insight into the origins of observed biological variability. 
\rot{At the mesoscopic scale, fluctuations naturally arise from the finite number of underlying units. These ``finite-size''} fluctuations may then be accounted for by stochastic low-dimensional models.
In the context of living systems, examples range from Langevin equations for gene regulation~\cite{KepEls01} to coupled molecular motors~\cite{GuePro11, TouSch11, MaKli14} to ecology~\cite{DobFre12}.

In neuroscience, neural-population models (also known as firing-rate or neural-mass models) such as the Wilson-Cowan model~\cite{WilCow72}  have been highly successful in describing various cortical phenomena such as oscillations~\cite{LuRin24}, multistability~\cite{LevBuz19}, and nonlinear response properties~\cite{AhmRub13}. Stochastic versions of neural-population models  have been critical for understanding fluctuating neural population dynamics such as metastability~\cite{WonWan06,ShpMor09,LevBuz19}, stochastic oscillations~\cite{DumNor14}, and response properties of cortical variability~\cite{KanOck17,HenAhm18}. Despite their mathematical tractability, however, these models are often phenomenological, lacking a clear link to an underlying \rot{finite-size} network of spiking neurons. 

A link to biophysical properties is maintained through bottom-up mean-field \rot{approaches}~\cite{Ama72,GerKis14,DecJir08,FauTou09,elBDes09,LaC22,Ock23}, where the macroscopic or mesoscopic dynamics is analytically derived from a microscopic model.
The resulting mean-field \rot{dynamics} are amenable to mathematical analysis~\cite{BruHak99,Bru00,Ger00,MonPaz15,SchLoe23} and enable efficient and accurate simulations of large spiking neural networks~\cite{NykTra00,CaiIye16,SchDeg17,AugLad17}. 
\rot{Recently, a bottom-up theory for the mesoscopic dynamics of finite-size populations of generalized integrate-and-fire neurons with escape noise has been developed~\cite{SchDeg17,SchLoe23}.
It} suggests an efficient multiscale modeling framework of cortical circuits~\cite{SchDeg17,RenLon20,WanSch22} grounded in realistic neuron dynamics~\cite{PozMen15,BilCai20}. 
However, this and other mean-field theories~\cite{BuiCow10,BuiCho13,SchDeg17,DumPay17,HeeSta21,KliKir22,SchLoe23} for finite-size neural networks 
%However, current mean-field \rot{theories for finite-size neural networks \cite{BuiCow10,BuiCho13,SchDeg17,DumPay17,HeeSta21,KliKir22,SchLoe23} 
typically assume homogeneous, all-to-all (full) connectivity. \rot{This assumption can be regarded as an approximation of a non-fully connected random network by its homogeneous mean connectivity while neglecting its quenched variability \cite{DegSch14,SchDeg17}. On the other hand, the role of random connectivity for intrinsically generated fluctuations of synaptic inputs has been extensively studied using dynamical mean-field theories \cite{SomCri88,MolSch92,BruHak99,Bru00, VanSom96,RenMor07,HelTet14,GoldiV21}, albeit only in the limit of infinitely large networks.
  
Both assumptions--homogeneous full connectivity ignoring quenched variability and infinite network size ignoring fluctuations--are severe limitations for describing biologically realistic, recurrent neural networks.} For example, rat barrel cortex contains only on the order of $N\sim 100$ to $N\sim 1000$ neurons per cell type per layer \cite{LefTom09} -- a population size for which finite-size fluctuations \rot{can be significant, as shown in computational models \cite{ComBru00,MazFon15,GigDec15,SchDeg17,SchGer20, PieSch22}}. Furthermore, the local connection probabilities $p$ in mouse visual cortex vary on a wide range ($0\lesssim p\lesssim 0.7$) with low average ($p\sim 0.1$) and  larger connection probabilities ($p\sim 0.5$) for similarly tuned neurons~\cite{KoHof11}. \rot{How such pronounced deviations from fully connected networks ($p=1$) affect the mesoscopic dynamics of finite-size neural populations is still poorly understood.}

\rot{Both finite population size ($N<\infty$) and quenched random connectivity (e.g.~$p<1$)} cause intrinsically generated (endogenous) fluctuations of the recurrent, synaptic input currents.
%, and may thus strongly affect the nonlinear population dynamics. However, their combined effects are still poorly understood. 
%Moreover, these 
\rot{These intrinsic fluctuations} act as multiplicative noise whose strength increases with population firing rates. On the other hand, neural variability in the cortex often decreases upon stimulation \cite{ChuByr10} -- a phenomenon that could be explained, among others, with a model %, 
where variability is generated by an external, additive noise \cite{HenAhm18}. To clarify the role of intrinsic fluctuations for such cortical phenomena under biologically realistic neuron numbers and connection probabilities requires a mean-field \rot{theory} that accounts for these different origins of fluctuations.

In a homogeneous, fully connected network, \rot{as arising under a mean-connectivity approximation,} the input fluctuations are common to all neurons. This is because  all neurons receive identical recurrent inputs and thus experience the same finite-size fluctuations of the population activity. These ``coherent'' fluctuations lead to stochastic population equations with finite-size noise of order $1/\sqrt{N}$. 
In contrast, in random networks with connection probability $p<1$, neurons no longer receive identical synaptic inputs, \rot{and hence input fluctuations become partially decorrelated. Consequently, the assumption of perfectly coherent fluctuations no longer holds, raising questions about the validity of the mean-connectivity approximation for finite-size networks. }

A mean-field description of 
%non-fully connected, 
\rot{random networks (including non-fully connected networks, $p<1$)} requires the treatment of fluctuations resulting from a ``quenched'' (i.e. temporally constant) random synaptic connectivity. Previous approaches to random networks considered the macroscopic limit $N\to\infty$. Depending on the concrete limiting procedure and network architecture, the recurrent fluctuations seen by different neurons either vanish~\cite{Ama72,Ger00,MonPaz15,Ock23} or become perfectly independent (``incoherent'') in this limit~\cite{SomCri88,VanSom96,BruHak99,Bru00, RenMor07,HelTet14,GoldiV21} 
%% JR: Addded Augustin & Ladenbauer here; I think their paper would deserve a more prominent place as they explicitly derive low-dimensional models for spiking networks
unless correlations are induced externally. 
A famous example is the mean-field theory for sparsely connected networks by Brunel and co-workers~\cite{BruHak99, Bru00}. 
In this theory, the neural population dynamics is deterministic and given by a nonlinear Fokker-Planck equation. The effect of incoherent fluctuations of the recurrent synaptic input \rot{caused by the random connectivity} appears as an additional contribution to the diffusion coefficient in this equation.  Although this approach \rot{relies on an annealing approximation that} neglects temporal correlations in the spike trains caused by the quenched disorder, the nonlinear Fokker-Planck equation has been successful to capture the influence of intrinsic fluctuations on the nonlinear population dynamics (for a treatment of temporal correlations, see e.g.~\cite{WieBer15}).  Using this theory, intrinsic fluctuations have been shown to crucially shape nonlinear response properties~\cite{SanHis20}, network oscillations~\cite{BruHak99, SchSch24} and multistability~\cite{MazFon15}. 

Even though dynamical mean-field theories for $N\to\infty$ have been used for real neural networks of finite size (see e.g.~\cite{EkeKra23}), they cannot describe finite-size fluctuations at the population level. Furthermore, finite-size networks are necessarily outside the sparse limit (i.e.~$p>0$) which entails a non-vanishing probability of shared recurrent inputs among neurons. Therefore the assumption of perfectly incoherent noise in dynamical mean-field theories cannot hold fully true for $N<\infty$. While finite-size extensions of the mean-field theory for sparsely connected networks have been proposed in~\cite{BruHak99,Bru00,MatGiu02}, low-dimensional neural population models that reveal and take into account the distinct effects of coherent and incoherent fluctuations are currently lacking.

In this paper, \rot{we consider finite-size networks with random connectivity, allowing} arbitrary connection probabilities $p\in[0,1]$. \rot{For these networks,} we derive and analyze a simple \rot{mesoscopic dynamics} that describes incoherent and coherent recurrent fluctuations. 
To this end, we study a network of Poisson neurons \rot{using an annealed approximation \cite{DerPom86,BruHak99} of the} quenched random connectivity. The annealed approximation neglects temporal correlations of incoherent fluctuations but captures the effects of \rot{random} connectivity and finite network size surprisingly well. \rot{In contrast to the case $N\to\infty$ of previous theories \cite{BruHak99,Bru00}, we need to treat the fluctuating input as a doubly stochastic process to account for both finite $N$ and random connectivity.
%In the presence of coherent finite-size fluctuations, this approximation allows us to treat the synaptic input as a doubly stochastic process, 
This description eventually yields} a three-dimensional Langevin equation \rot{for the mesoscopic dynamics}. We compare this \rot{dynamics} with a naive mean-field \rot{theory} in which the variance of the random connectivity is neglected, % 
corresponding to the ad hoc mean-field approximation used in previous \rot{studies}~\cite{DegSch14,SchDeg17}. While the focus of the paper is the Langevin dynamics for a homogeneous population with quenched or annealed random connectivity and common external noise, the theoretical framework can be extended to heterogeneous populations, where the resting potentials or external currents of the neurons exhibit quenched Gaussian disorder.  In summary, we present a simple theoretical framework for the effects of various sources of noise and quenched disorder on the neural population dynamics at the mesoscopic scale ($1\ll N<\infty$).

The paper is organized as follows. 
%The
We introduce the %
microscopic network model with quenched random connectivity and its annealed approximation 
%are introduced 
in Sec.~\ref{sec:model}. By reducing the annealed network to a mean-connectivity network with incoherent dynamical noise, we derive in Sec.~\ref{sec:derivation} a mesoscopic population model as a system of three coupled stochastic differential equations. In Sec.~\ref{sec:analysis}, 
we then analyze %
the mesoscopic model 
%is then analyzed 
with respect to fixed-point solutions and their stability, the linear response to dynamic stimuli and second-order statistics. 
In particular, we compare the 
%The 
obtained analytical results 
%are compared 
to the corresponding statistics of the quenched and annealed microscopic models using extensive simulations. 
%Our results are concluded and discussed 
We conclude and discuss our results 
in Sec.~\ref{sec:conclusions}. The extension to quenched Gaussian disorder as well as longer, detailed calculations are provided in the Appendix.

\section{Microscopic models}
\label{sec:model}

\subsection{Quenched network model}

At the microscopic level, we consider a random network of $N$ interacting Poisson neurons (also called nonlinear Hawkes processes). \rot{While our theory will be developed for general random connectivity with finite variance, we focus on the specific example of binary, non-full connectivity to study the effect of connection probabilities $p<1$. The Poisson} neurons fire spikes stochastically with conditional intensity $\lambda_i(t) = \phi(h_i(t^-))$, 
$i=1,\ldots,N$, %
where $h_i(t)$ is the input potential of neuron $i$ and $\phi$ is a non-negative hazard function. 
%of sigmoidal shape. This means that spikes 
% Spikes thus %
% occur 
% %conditionally 
% independently with 
% %conditional 
% firing probability $\mathbb{P}\{\text{neuron } i \text{ spikes in $(t,t+dt]$}|h_i(t)\}=\lambda_i(t)dt+o(dt)$. 
The dynamics of the input potentials is given by 
%the 
a %
system of coupled first-order equations with delay, %
\begin{align}
\tau\od{h_i}{t}  &= - h_i  + \muext(t) + \frac{w}{C}\sum\limits_{j=1}^N a_{ij} \dot Z_j(t-d),\label{eq:micro_main}
\end{align}
$i=1,\dotsc,N$. Here, $\tau$ is 
the time constant of the low-pass filter dynamics, $\muext(t)$ is an external drive and $\dot Z_i(t)=\sum_l\delta(t-t_{i,l})$ is the spike train  of neuron $i$ with $\{t_{i,l}\}$ being the individual spike times.
%with stochastic intensity
%\begin{equation}
%\label{eq:spike-train-intens}
%    \lambda_i(t)=\phi\bigl(h_i(t^-)\bigr).
%\end{equation}
%% JR: would be repeating ourselves
%If $\{t_{i,k}\}$ are the spike times of neuron $i$, the spike train is thus given by $s_i(t)=\sum_k\delta(t-t_{i,k})$.  
 The presence or absence of synaptic connections between neurons are described 
by the random adjacency matrix $[a_{ij}]$
%Furthermore, $a_{ij}\in\{0,1\}$ is the random adjacency matrix 
with fixed in-degree $C=\sum_{j=1}^{N} a_{ij}$ and elements $a_{ij}\in\{0,1\}$ that are marginally Bernoulli distributed with mean $\lrk{a_{ij}}=p=C/N$ representing the connection probability. This random connectivity can be constructed by choosing for each neuron $C$ presynaptic neurons randomly, and the resulting connectivity is fixed (``quenched'') in time. 
Furthermore, $d$ is the synaptic delay and $w=CJ$ is the total coupling  strength  with $J$ being the efficacy of a single synapse (i.e. $J/\tau$ is the jump size in millivolt of the postsynaptic potential caused by a single presynaptic spike). Since we focus on inhibitory networks in this paper, we assume that $J<0$.

As a concrete hazard function, we choose here a sigmoidal function in the form of an error function \cite{Ama72}: 
\begin{equation}
    \phi(h) = \phimax\Phi(\beta (h-\vartheta)),\label{eq:phi-erf}
\end{equation}
where $\Phi(x)= \lreckig{1+\erf\lrrund{x/\sqrt{2}}}/2$.
Although not essential for the general theory, this choice will allow us to analytically calculate the first and second moments of $\phi(h_i)$.
The parameter $\beta$ determines the steepness around the inflection point. Without loss of generality, we choose the position of the inflection point $\vartheta$ to be zero because we can always measure voltages with respect to the potential $\vartheta$. The sigmoidal shape with an upper limit $\phimax$ prevents the Poisson neurons to fire with arbitrarily high rates.
While the concave behavior and saturation of the sigmoidal hazard function for $h>0$ is technically useful, we are in the following mainly interested in dynamical regimes operating in the convex part of the hazard function $\phi(h)$, i.e. corresponding to mean inputs below the inflection point, $h<0$. This is because biologically realistic hazard functions of cortical neurons are typically convex \cite{JolRau06,PriFer08}.
As default parameters of the model, we choose $\tau= 20$ ms, $\phimax=100$ Hz and $d=0$ ms unless otherwise noted.

The aim of this paper is to derive a mesoscopic population model that generates population activities that statistically match the population activities obtained from a microscopic network simulation. We define the population activity $A_N(t,\Delta t)$ with respect to a time discretization with time step $\Delta t$ as the total number of spikes per neuron and time step: 
\begin{equation}
\label{eq:empir-popact}
    A_N(t;\Delta t)=\frac{\Delta Z(t)}{N\Delta t}.
\end{equation} 
Here, $\Delta Z(t)$ denotes the total number of spikes in the time interval $(t,t+\Delta t]$. 
%The population activity represents the main mesoscopic observable. 

\subsection{Mean-connectivity network}
\label{sec:mean-connectivity-net}

\begin{figure}[t]
  \centering
    \includegraphics[width=\linewidth]{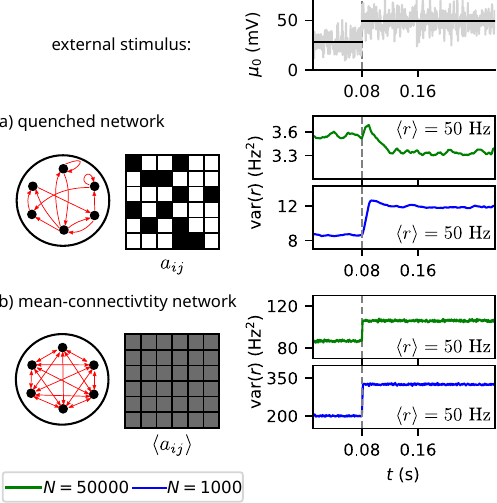} 
  \caption{Response of the population rate variance to a change of the mean stimulus strength in the original, quenched random network~(a) and in a corresponding fully-connected, mean-connectivity network that only accounts for the mean synaptic weight~(b). Top: Noisy external stimulus $\muext(t)$ whose mean increases at $t=0.08$ s in a step-like fashion from $\mu_0=28$ mV to $\mu_0=50$ mV; the noise strength $\sigmaext^2 = 1$ mV$^2$ is kept fixed. (a) Schematic of a quenched network with fixed adjacency matrix $a_{ij}$ (left) and the trial-averaged variance of the population firing rate as a function of time (right) for different network sizes (green line: $N=5\cdot 10^4$, blue line: $N=10^3$, in both cases $C=100$). (b) Same for a mean-connectivity network, Eq.~\eqref{eq:1st-order-net-micro}, as an ad hoc approximation of the quenched network, where the quenched connectivity $a_{ij}$ is replaced by the average connectivity $\langle a_{ij}\rangle$. 
  %Note the different scaling of the y-axis in a) and b). %\gruen{Thus, while the mean stationary population rates after the step (as indicated in the right panels) are equal in both models, the variances of the population rate strongly differ.} 
  While the mean stationary population rates after the step are equal in both models \rot{(measured rates are provided in the right panels)}, the variances of the population rate strongly differ (note the different scaling of the $y$-axis).
  \rot{Other parameters: $\beta=5$ mV$^{-1}$, $w=-1$ mVs.}
%  affected by the mean-connectivity approximation.}
  %(a) Top: Illustration of the original network with quenched random connectivity. Bottom: Variance of the stochastic population rate under a white-noise external stimulus $\muext(t)$, whose mean $\mu_0$ has a step change as depcted in the bottom panel. (b) As in (a) but for the corresponding mean-field model, where the connectivity has been replaced by the average weighted graph. .... MUSTER, $N=1000$ is missing
  }
  \label{fig:intro}
\end{figure}

A major obstacle in deriving population models analytically is the quenched randomness of the connectivity \blau{$J_{ij}:=J a_{ij}$. A standard mean-field approach to tackle this problem is to neglect fluctuations in the recurrent synaptic input by scaling the synaptic weights inversely proportional to the network size, $J_{ij}=w_{ij}/N$, where the coupling strength $w_{ij}$ is assumed to be of order $1$. Then, in the limit $N\to\infty$, the synaptic input $N^{-1}\sum_jw_{ij}\dot Z_j$ converges to the deterministic input $\lrk{w_{ij}}\langle\dot Z_j\rangle$ \cite{LaC22,Ock23}, where $\lrk{\cdot}$  denotes the average over both quenched disorder and Poisson noise.
In our case, this corresponds to a network where the quenched connectivity 
%$w_{ij}=\frac{w}{p}a_{ij}$ 
$w_{ij}$ % not necessary to give definition here
(Fig.~\ref{fig:intro}a, left) is replaced by the mean connectivity 
%\lrk{w_{ij}}=\frac{w}{p}\lrk{a_{ij}}=w$ 
$\lrk{w_{ij}}=NJ\lrk{a_{ij}}\rot{=CJ}=w$ 
%% JR: More connected to defs. previous 
(Fig.~\ref{fig:intro}b, left). Importantly, the mean connectivity effectively yields a fully connected network with homogeneous coupling $w$. In the limit $N\to\infty$, the synaptic input is thus not affected by the variance 
\begin{equation}
    \sigma_w^2=w^2(1-p)/p
    \label{eq:sigma_w}
\end{equation} 
of the disordered couplings $w_{ij}$ but only by the mean $\lrk{w_{ij}}$.

%While this approach is exact in the limit $N\to\infty$ with $1/N$-scaling of the synaptic weights, 
The replacement of all elements of the adjacency matrix $a_{ij}$ with their mean $p$ has been used ad hoc also for finite network size  $N<\infty$~\cite{DegSch14,SchDeg17}. However, this ad hoc approach is no longer exact for finite $N$ but must be regarded as an approximation that retains the mean synaptic connectivity $\lrk{J_{ij}} \rot{= pJ}$ but ignores all higher-order cumulants of the random variables $J_{ij}$. In the following, we will refer to this effective, fully connected network} as the \emph{mean-connectivity network}, whereas the original model, Eq.~\eqref{eq:micro_main}, will be referred to as the \emph{quenched network} \rot{(Table \ref{sum_micro})}. 
Replacing the connection strengths $w_{ij}$ with their means $w$, we obtain from Eq.~\eqref{eq:micro_main} the dynamics of the mean-connectivity network:
\begin{equation}
\label{eq:1st-order-net-micro}
    \tau dh_i(t)=\lreckig{-h_i+\muext(t)}dt+\frac{w}{N}\sum_{j=1}^NdZ_j(t-d),
\end{equation}
where $dZ_j(t)\sim \text{Pois}(\phi(h_j(t))dt)$ are the conditionally independent Poisson increments of the spike count of neuron $j$. 
The mean-connectivity network of Poisson neurons, Eq.~\eqref{eq:1st-order-net-micro}, is a microscopic model that admits an exact reduction to a mesoscopic model for the total spike count $dZ(t)=\sum_{i=1}^N dZ_i(t)$. To see this, we note that the sum of independent Poisson numbers in Eq.~\eqref{eq:1st-order-net-micro} is a Poisson number with mean $Nr(t-d)dt$, where 
\begin{equation}
\label{eq:stoch-rate-def}
    r (t)=\frac{1}{N}\sum_{i=1}^N\phi\bigl(h_i(t)\bigr).
\end{equation} 
is the stochastic population rate. Furthermore, because all neurons are driven by the same synaptic input, 
their input potentials $h_i(t)$ coincide (or converge after an initial transient). 
The common input potential $\bar h(t)$ thus obeys the mesoscopic dynamics  
\begin{equation}
\label{eq:1st-order-net-meso}
    \tau d\bar h(t)=\lreckig{-\bar h+\muext(t)}dt+\frac{w}{N}dZ(t-d),
\end{equation}
where now $dZ(t)\sim \text{Pois}(N\phi(\bar h(t))dt)$ describes the increment of the total spike count of the population. Equation~\eqref{eq:1st-order-net-meso} describes the mesoscopic dynamics as it only involves the population activity $A_N(t,\Delta t)=\int_t^{t+\Delta t}dZ(t)/(N\Delta t)$.

Simulations of the quenched and the mean-connectivity network yield similar results for the stationary mean firing rates and mean input potentials. However, the mean-connectivity network may differ drastically compared to the quenched network when it comes to the non-stationary response (linear response function) and the fluctuations (second-order statistics). 
To illustrate the failure of the mean-connectivity approximation, let us consider the variance of the stochastic population rate $r(t)$. This variance represents the part of the mesoscopic neural variability (variance of the population activity $A_N(t,\Delta t)$) that is caused by the rate variability rather than the Poisson spiking noise. It can be measured from population activity data (see Appendix, Sec.~\ref{Sec:App:Estimation of the rate variance from population activities}), or directly computed from a simulated time series of $r(t)$ in our model. 
A non-vanishing variance of the population rate can occur for two reasons in our model: first, because of intrinsic finite-size fluctuations, and second, because of externally injected common noise. Here, we model the common external noise as a Gaussian white noise (more precisely, $\mu(t)=\bar\mu(t)+\sqrt{\tau\sigmaext^2}\hat\zeta(t)$, where $\bar\mu(t)$ is the mean stimulus and $\hat\zeta(t)$ is a standard Gaussian white noise process).
%obeying $\lrk{\hat\zeta(t)\hat\zeta(t')}=\delta(t-t')$). 

Interestingly, simulations of the mean-connectivity network (Eq.~\eqref{eq:1st-order-net-micro} or \eqref{eq:1st-order-net-meso}) strongly overestimate the population-rate variance of the quenched network by more than one order of magnitude (Fig.~\ref{fig:intro}, note the different scale of the y axes). \rot{We stress that this difference is not an effect of a difference in firing rates: The stationary mean population rates $\lrk{r}$ are the same in both networks (e.g., the numerical values of the firing rates at high stimulus amplitudes are approximately equal as indicated in Fig.~\ref{fig:intro}a,b).}
In addition to this large quantitative difference, we also observe a qualitatively different response to fast stimulus changes \rot{
when %if 
the network is large ($N=5\cdot 10^4$, green lines in Fig.~\ref{fig:intro})}: When the mean stimulus strength jumps from $28$ mV to $50$ mV, the variance of the population rate exhibits a small but significant decrease in the quenched network (Fig.~\ref{fig:intro}a, right). This suppression of variability is in marked contrast to the prediction by the mean-connectivity network, which exhibits instead an increase in the variance (Fig.~\ref{fig:intro}b, right). \rot{This qualitatively different behavior of the population-rate variance in response to a step stimulus is, however, not observed for smaller network size ($N=10^3$, but otherwise same parameters and similar firing rates; blue lines in Fig.~\ref{fig:intro}).}
%\rot{when network is large so that the variance induced by the external fluctuations dominate over the intrinsic finite-size fluctuations}. 

How can we understand non-stationary responses and neural variability of mesoscopic variables theoretically? In this paper, we present a ``second-order'' mesoscopic mean-field theory that also captures second-order statistics and nonstationary responses quantitatively, and explains non-trivial, qualitative phenomena such as the suppression of variability observed in Fig.~\ref{fig:intro}a.  In particular, the explanation of the failure of the mean-connectivity approximation with respect to the stationary variances in Fig.~\ref{fig:intro}b and its resolution by the ``second-order'' mean-field theory will be presented towards the end of the paper in Sec.~\ref{sec:variances}.

\subsection{Annealed network model}
\label{sec:annealing}

\begin{figure}[t]
  \centering
   \includegraphics[width=\linewidth]{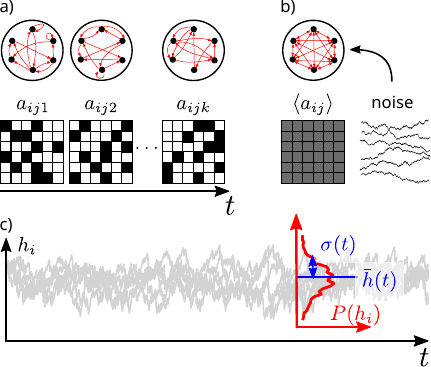} 
  \caption{Schematic illustration of the annealed network and its approximation by a mean-connectivity network with additional noise. (a)~In the annealed network model, Eq.~\eqref{eq:micro_annealed}, the random connectivity is re-sampled at each spike time resulting in an adjacency matrix $a_{ijk}$ that varies at each spike $k$. (b) The annealed model can be mapped to a mean-connectivity network with additional noise, Eq.~\eqref{eq:annealed-network-model}. The network is all-to-all connected with an average connectivity $\langle a_{ij}\rangle$. Additionally, each neuron is subject to an independent noise that captures the random connectivity. (c)~Example trajectories of the input potentials in the mean-connectivity network with additional noise (gray). The distribution $P$ of the input potentials follows a Gaussian distribution with a time-dependent mean $\bar{h}(t)$ and a time-dependent variance $\sigma^2(t)$. The second-order mean-field model~\eqref{eq:meso_main} aims to describe the stochastic dynamics of these two mean-field variables.}
  \label{fig:illu}
\end{figure}

The quenched connectivity causes heterogeneity of neural firing 
%(e.g. spatial and temporal correlations), 
with spatial and temporal correlations which often 
%makes 
make a direct mean-field treatment infeasible. Here, we follow a different approach based on a dynamical resampling of the connectivity (``annealing'') at each spike~\cite{DerPom86,BruHak99} (Fig.~\ref{fig:illu}a). This procedure largely ignores the part of the spatial and temporal correlations that is caused by the quenched connectivity, but leads to a tractable model of a fully connected neural network with additional dynamical noise that, as we will see, retains and captures the main statistical features of the recurrent synaptic input to a neuron (Fig.~\ref{fig:illu}b). We expect that the annealed approximation is better in dense networks of strongly correlated neurons. In this case, the temporal correlations in the spike input of a given neuron are partially retained upon randomization of the presynaptic neurons. In contrast, in sparse networks, the membrane potentials are independent among neurons, and therefore, the resampling of presynaptic neurons in time destroys the temporal correlations in the presynaptic spike train of a fixed presynaptic neuron.

In the annealed approximation, the connections $a_{ij}$ are re-sampled at each spike occurring in the network. Thus, the time-varying adjacency matrix can be written as $[\hat a_{ijk}]$, where the additional index $k$ labels the $k$-th spike occurring in the network. Then, for a fixed $k$, the $N\times N$-adjacency matrix $[\hat a_{ij}]_k$ is still a random matrix with fixed in-degree, and therefore has independent rows but dependent elements $\hat a_{ijk}$ within a given row $i$. However, the dependence caused by the fixed-in-degree constraint is immaterial for the recurrent inputs in the annealed approximation. This is because only the $j_k$-th column of the matrix $[\hat a_{ij}]_k$, which has independent elements, matters for the  transmission of the $k$-th spike, where $j_k$ is the index of the neuron that fired the $k$-th spike, and because of the re-sampling of the adjacency matrix. The latter implies that for different spike times, the columns are independent. From the perspective of the recurrent inputs, we can thus equally well re-sample the elements of the adjacency matrix as independent Bernoulli variables $a_{ijk}\sim \text{Ber}(p)$ with success probability $p=C/N$. Thus, our model, Eq.~\eqref{eq:micro_main}, changes in the annealed approximation to
\begin{align}
\tau dh_i(t)  &= [- h_i  + \muext(t)]dt + \rot{\frac{1}{N}} dY_i(t-d),\label{eq:micro_annealed}\\
\frac{dY_i}{dt}(t)&=\rot{\frac{w}{p}}\sum\limits_{j=1}^N\sum_l a_{ijk_{j,l}}\delta(t-t_{j,l})\label{eq:dY}
\end{align}
where the index $k_{j,l}$ delivers the location of the $l$-th spike time $t_{j,l}$ of neuron $j$ within the sequence of all spikes in the network. The quantity $Y_i(t)$ is the integrated synaptic input received by neuron $i$ until time $t$. We refer to this model as the \emph{annealed network} \rot{(Table \ref{sum_micro})}. We note that in simulations of the annealed network, it is not necessary to re-sample the entire adjacency matrix at each spike but 
%it is sufficient to re-sample
only the column that corresponds to the neuron that fired the spike. In other words, whenever a neuron fires a spike, this spike is transmitted independently to postsynaptic neurons with probability $p=C/N$. 

\rot{By generalizing from binary weights $w_{ij}=\frac{w}{p}a_{ij}$ to an arbitrary connectivity matrix $w_{ij}$, the annealed network can also be defined for such general connectivity through dynamic resampling, which produces time-dependent synaptic weights $w_{ij}(t)$. In the specific case of binary synapses, $w_{ij}(t_{j,l})=\frac{w}{p}a_{ijk_{j,l}}$ as in Eq.~\eqref{eq:dY}.}

Apart from the random connectivity, another source of quenched disorder that can be included in our theory is heterogeneity in the external current $\mu_0$. For the sake of clarity, we will begin our analysis without heterogeneity in $\mu_0$ and will defer the straightforward extension to heterogeneity to a brief section in the Appendix (Sec.~\ref{sec:hetero}).

\begin{table}%[H] add [H] placement to break table across pages
\rot{
\caption{\label{sum_micro} Summary of microscopic network models.}
\begin{ruledtabular}
\begin{tabular}{lcc}
Network&Equation&Connectivity\\
\hline
Quenched&\eqref{eq:micro_main}& random, fixed in time,\\
&&$\lrk{w_{ij}}=w$, $\var{w_{ij}}=\sigma_w^2$\\
Mean-connectivity& \eqref{eq:1st-order-net-micro}& mean-matched all-to-all,\\
&&$\lrk{w_{ij}}=w$, $\var{w_{ij}}=0$\\
%\hline
Annealed&\eqref{eq:micro_annealed}, \eqref{eq:dY}&random, dynamic\\
&&(resampling in time),\\
&&$\lrk{w_{ij}(t)}=w$, $\var{w_{ij}(t)}=\sigma_w^2$
\end{tabular}
\end{ruledtabular}
}
\end{table}

\section{Derivation of the mesoscopic model}
\label{sec:derivation}

\subsection{A mean-connectivity network with additional noise that accounts for second-order connectivity statistics}

The first step of the derivation is a temporal coarse-graining of the annealed model leading to an effective, fully connected network with homogeneous weights (mean-connectivity network) and additional noise. The additional noise will be the crucial difference to the mean-connectivity network discussed in Fig.~\ref{fig:intro}. To coarse-grain, we integrate Eq.~\eqref{eq:micro_annealed} over a small time step of length $\Delta t$:
\begin{equation}
\label{eq:coarse-h}
    \tau \Delta h_i(t)  = \int_t^{t+\Delta t}[- h_i(s)  + \muext(s)]\,ds + \rot{\frac{1}{N}}\Delta Y_i(t-d).
\end{equation}
Here, $\Delta h_i(t)=h_i(t+\Delta t)-h_i(t)$ is the increment of the input potential, \rot{$\Delta Y_i(t)=\sum_{j=1}^N w_{ij}(t)\Delta Z_j(t-d)$ is the synaptic input  received by neuron $i$ during the time step $\Delta t$ and $\Delta Z_j(t)=1$ if neuron $j$ has a spike during that time step and $\Delta Z_j(t)=0$ otherwise. In the annealed approximation, the time-varying coupling strengths $w_{ij}(t)$ are independent random variables (across time steps and synapses) with mean $w$ and variance $\sigma_w^2$. In our specific example, the random variables are binary taking the value $w/p$ with probability $p$ and $0$ otherwise. However, the derivation does not hinge upon binary connectivity but applies more generally to any distribution of coupling strengths with finite variance $\sigma_w^2<\infty$.
}

% Consider the total number of spikes in the population during the same time step, $\Delta Z(t)=\sum_{i=1}^N\Delta Z_i(t)$,} as it corresponds to 
% the desired mesoscopic population activity via Eq.~\eqref{eq:empir-popact}. 
The synaptic input $\Delta Y_i(t)$ can be regarded as the result of a doubly stochastic process: On the one hand, given the total spike count $\Delta Z(t)=\sum_{i=1}^N\Delta Z_i(t)$, each $\Delta Y_i(t)$ is the sum of $\Delta Z(t)$ independent random variables. Hence, the conditional mean and covariance are
\begin{align*}
  &\langle\Delta Y_i(t)|\Delta Z(t)\rangle=\rot{w}\Delta Z(t),\\%\label{eq:dy-mean}\\
  &\cov{\Delta Y_i(t), \Delta Y_j(t)|\Delta Z(t)}=\rot{\sigma_w^2}\Delta Z(t)\delta_{ij},\label{eq:dy-var}
\end{align*}
where $\delta_{ij}$ denotes the Kronecker delta. If $N$ is large such that $p\Delta Z(t)\gg 1$, the synaptic input $\Delta Y_i(t)$ for a given $\Delta Z(t)$ is approximately Gaussian distributed. Under this condition, we can write
\begin{equation}
  \label{eq:gaussianY_Delta_Z}
  \Delta Y_i(t)\approx \rot{w}\Delta Z(t)+\rot{\sigma_w}\sqrt{\Delta Z(t)}n_i(t),
\end{equation}
where $n_i(t)\sim \mathcal{N}(0,1)$ are independent standard normal random variables. The first term in Eq.~\eqref{eq:gaussianY_Delta_Z} is proportional to the total number of spikes or the mesoscopic population activity and thus corresponds to the input in an effective mean-connectivity network. This term is common to all neurons. In contrast, the second term is independent across neurons  and only arises when \rot{$\sigma_w>0$. Specifically,} it captures the individual differences of the synaptic inputs in a non-fully connected (``diluted'') network \rot{with $p<1$}. 

On the other hand, $\Delta Z(t)$ is itself a Poisson random variable with mean $Nr (t)\Delta t$. From Eq.~\eqref{eq:gaussianY_Delta_Z}, or, alternatively, from the law of total expectation and total covariance, we find  that the synaptic inputs $\Delta Y_i(t)$ exhibit the following mean and covariance for a given stochastic population rate $r(t)$:
\begin{align*}
    \lrk{\Delta Y_i(t)\mid r(t)} &= \rot{wN}r(t)\Delta t\\
    \cov{\Delta Y_i(t), \Delta Y_j(t)\mid  r(t)}&= \rot{\lrrund{w^2+\sigma_w^2\delta_{ij}}N}r(t)\Delta t. 
\end{align*}
The last equation shows that pairwise correlations of the synaptic input to different neurons are non-zero for \rot{finite $N$}. These correlations arise from the common fluctuations of the first term in Eq.~\eqref{eq:gaussianY_Delta_Z} representing the population activity.

The covariance structure of the synaptic inputs $\Delta Y_i(t)$ is preserved if their representation by Eq.~\eqref{eq:gaussianY_Delta_Z} is replaced by
% The first term in Eq.~\eqref{eq:gaussianY_Delta_Z} has a clear convergence to the population activity in the continuum limit $\Delta t \to 0$, whereas the second term is more complicated. There, we have a product of the random variables of the square root $\Delta Z$ and and $n_i$. We want to replace this product with a simpler term that still contains the same statistics. It is sufficient to replace $\Delta Z(t)$ with the mean value because the replacement process
 \begin{equation*}
  \label{eq:gaussianY}
  \Delta \hat Y_i(t)= \rot{w}\Delta Z(t)+\rot{\sigma_w}\sqrt{\rot{N}r(t)\Delta t}n_i(t).
\end{equation*}
This representation allows to take the temporal continuum limit. 
%has the same conditional mean $\lrk{\Delta \hat Y_i(t)\mid r(t)}$ and covariance $\cov{\Delta \hat Y_i(t), \Delta\hat Y_j(t)\mid  r(t)}$ as in Eq.~\eqref{eq:gaussianY_Delta_Z}. We can therefore represent $\Delta Y_i(t)$ with $\Delta\hat Y_i(t)$ if we ignore higher moments.
Substituting $\Delta \hat Y_i(t-d)$ for $\Delta Y_i(t-d)$ in Eq.~\eqref{eq:coarse-h}, dividing by $\Delta t$ and taking the limit $\Delta t\to 0$, we obtain
% we can now approximate Eq.~\eqref{eq:micro_annealed} by
% \begin{multline}
%   \tau \Delta h_i(t)=
%   [-h_i(t)+ \muext(t)]\Delta t+
%   w\frac{\Delta Z(t-d)}{N}\\+w\sqrt{\frac{1}{C}\lrrund{1-\frac{C}{N}}r(t-d)\Delta t}n_i(t-d),
% \end{multline}
% where $\Delta h_i(t):=h_i(t+\Delta t)-h_i(t)$ is the increment of the input potential.
% Taking the  continuum limit $\Delta t\rightarrow 0$, we obtain 
\begin{multline}
  \label{eq:annealed-network-model}
  \tau\frac{dh_i}{dt}=-h_i+\muext(t)  +\frac{w}{N}\sum_{j=1}^N \dot Z_j(t-d)+\rot{\sigma_w\sqrt{\frac{r(t-d)}{N}}}\zeta_i(t).
\end{multline}
Here, we have used that the population activity, Eq.~\eqref{eq:empir-popact}, can be rewritten as
\begin{equation}
\label{eq:popact-spikes}
    A_N(t,\Delta t)=\frac{1}{N\Delta t}\int_t^{t+\Delta t}\sum_{i=1}^N \,dZ_i(t'),
\end{equation}
where the spike trains $\dot Z_i(t)$ have stochastic intensities $\phi(h_i(t^-))$. Furthermore, $\zeta_i(t)$ are  independent Gaussian white noises obeying $\langle\zeta_i(t)\zeta_j(t')\rangle=\delta_{ij}\delta(t-t')$. 

Equation~\eqref{eq:annealed-network-model} represents an effective, fully connected network with rescaled synaptic efficacy $w/N=pJ$ and additional dynamic noise whose intensity is proportional to the variance $\sigma_w^2$ of the coupling strengths $w_{ij}$. Therefore, the dynamics is the same as the mean-connectivity network, Eq.~\eqref{eq:1st-order-net-micro}, but with an additional noise term that captures the fluctuations caused by the random connectivity. This representation also reveals the distinct effects of finite network size ($N<\infty$) and random dilution ($p<1$) on the recurrent synaptic fluctuations. First, the stochastic input spikes from all neurons in the network contribute a common synaptic input current proportional to the population activity (third term on the right-hand side of Eq.~\eqref{eq:annealed-network-model}), which is stochastic for finite $N$ and thus exposes the coherent part of the synaptic fluctuations. More precisely, the population activity has conditional mean $r(t)$ (given all input potentials $h_i(t^-)$, $i=1,\dotsc,N$) and finite-size fluctuations of order $N^{-1/2}$ because the spike trains $\dot Z_i(t)$ are conditionally Poisson (given $h_i(t^-)$) with conditional mean $r(t)$. 

In contrast, the last term of Eq.~\eqref{eq:annealed-network-model}  contains a Gaussian white noise $\zeta_i(t)$ that is independent for each neuron, and thus represents the incoherent part of the synaptic fluctuations. This incoherent noise captures the effect of the random connectivity ($\sigma_w^2>0$), and, specifically, the random dilution ($p<1$). It therefore vanishes in the limit of fully connected, homogeneous networks ($p\to 1$, $\sigma_w^2\to 0$), in which case the mean-connectivity network is recovered.

In the following, we assume a large network size $N$ such that a diffusion approximation of the shot noise can be made: noting that $\Delta Z(t)$ is conditionally Poisson with mean and variance equal to $Nr(t)\Delta t$, the population activity can be approximated for large $N$ as
% \begin{gather}
% \label{eq:popact-whitenoise}
%     A_N(t;\Delta t)\approx\frac{1}{\Delta t}\int_{t}^{t+\Delta t} A(t')\,dt',\\
%     A(t)= r(t) + \sqrt{\frac{r(t)}{N}}\eta(t),\nonumber
% \end{gather}
% \begin{equation}
% \label{eq:popact-whitenoise}
%     A_N(t;\Delta t)\approx\frac{1}{\Delta t}\int\limits_{t}^{t+\Delta t} A(t')\,dt',\quad
%     A(t)= r(t) + \sqrt{\frac{r(t)}{N}}\eta(t),    
% \end{equation}
\begin{equation}
\label{eq:popact-whitenoise}
    A_N(t;\Delta t)\approx\frac{\int_{t}^{t+\Delta t} A(t')\,dt'}{\Delta t},\quad
    A(t)= r(t) + \sqrt{\frac{r(t)}{N}}\eta(t),    
\end{equation}
where $\eta(t)$ is a common Gaussian white noise with $\langle \eta(t)\eta(t')\rangle =\delta(t-t')$. In this form, the finite-size fluctuations of order $1/\sqrt{N}$ become explicit. We note that the diffusion approximation of the population activity, Eq.~\eqref{eq:popact-whitenoise}, can in principle become negative. However, if $N$ is sufficiently large such that $Nr(t)\Delta t\gg 1$, the probability that such event happens is negligibly small.

\subsection{Second-order mesoscopic mean-field dynamics}
\label{sec:mesot-dyn}

\rot{
All-to-all uniformly connected networks with independent noise, such as the mean-connectivity network with additional noise derived in the previous section, 
have been well studied. In particular, in the limit $N\to\infty$, the population dynamics can be described by a Fokker-Planck equation \cite{BruHan06}, while for finite $N$, the stationary statistics of the fluctuations has been calculated using linear response theory \cite{LinDoi05}. However, despite earlier theoretical approaches \cite{BruHak99,MatGiu02}, a derivation of stochastic population dynamics that captures both finite-size fluctuations and nonstationary dynamics remains challenging.
In our setting, this derivation is facilitated by the absence of resets in the Poisson neuron model. To obtain a self-consistent description, we follow a standard two-step procedure:} first, we consider the dynamics of the input potentials
\begin{equation}
    \tau \od{h_i}{t}  = -h_i+f(t)+g(t)\zeta_i(t), \label{eq:effective_OU}
\end{equation}
driven by fixed, time-dependent input signals $f(t)$ and $g(t)$, and derive the dynamics of the mean and variance conditioned on these inputs. We then derive 
the dynamics of the stochastic rate $r(t)$ for fixed input signals $f$ and $g$ such that it depends on these input signals only through the conditional mean and variance and an additional colored noise. In the second step, in turn, the input signals will be expressed in terms of the stochastic rate $r(t)$ according to Eqs.~\eqref{eq:annealed-network-model}, \eqref{eq:popact-spikes} and \eqref{eq:popact-whitenoise}:
\begin{equation}
\label{eq:f-g}
    \begin{aligned}
    f(t) &= \muext(t)+w\lreckig{r(t-d) + \sqrt{\frac{r(t-d)}{N}}\eta(t-d)},\\
    g(t) &= \rot{\sigma_w\sqrt{\frac{r(t-d)}{N}}}
\end{aligned}
\end{equation}
This step will close the system for the conditional mean and variance and the colored noise because the stochastic rate depends itself only on these variables. In this way, we will obtain a self-consistent mean-field description for the annealed model, Eq.~\eqref{eq:annealed-network-model}, as detailed in the following. In view of Eq.~\eqref{eq:f-g}, we also note that the conditioning on the input signals $f$ and $g$ will allow for the interpretation of the mean-field variables at time $t$ as conditional averages given the history of the noise $\eta(t'-d)$, the stochastic rate $r(t'-d)$ and the external input $\muext(t')$ for $t'<t$.

In the first step, when $f$ and $g$ are fixed, the input potentials $h_i$ represent $N$ independent, time-inhomogeneous Ornstein-Uhlenbeck processes for which the ensemble mean $\bar{h}$ and ensemble variance $\sigma^2$ obey the dynamics
\begin{equation}
\label{eq:condit-moments-f-g}
    \tau \od{\bar{h}}{t}  = -\bar{h} + f(t),\quad
    \tau \od{\sigma^2}{t}  = -2\sigma^2 +\frac{g^2(t)}{\tau}.
\end{equation}
% \begin{equation}
% \label{eq:condit-moments-f-g}
%     \begin{aligned}
%     \tau \od{\bar{h}}{t}  &= -\bar{h} + f(t)\\
%     \tau \od{\sigma^2}{t}  &= -2\sigma^2 +\frac{g^2(t)}{\tau}.
% \end{aligned}
% \end{equation}
% To  close the mean-field dynamics self-consistently, we express $r(t)$ and thus the functions $f$ and $g$ in terms of $\bar h$ and $\sigma^2$. 
These variables are the conditional mean and variance described above. To relate the stochastic rate to these variables, we first consider the limit $N\to\infty$. In this limit, the stochastic rate, Eq.~\eqref{eq:stoch-rate-def}, converges to its conditional mean
\begin{align}
    F\left(\bar{h}(t), \sigma^2(t)\right)&:=\langle\phi(h_i(t))\mid \bar h(t),\sigma^2(t)\rangle\nonumber\\
    &=\int_{-\infty}^{\infty}\phi(h) g_{\bar{h}, \sigma^2}(h)\,dh\label{eq:Integrated_transfer_function}
\end{align}
where $g_{\bar h,\sigma^2}(h)=\exp[-(h-\bar h)^2/(2\sigma^2)]/\sqrt{2\pi\sigma^2}$. Note that the conditional mean depends on the functions $f$ and $g$ only through the conditional mean $\bar h(t)$ and conditional variance $\sigma^2(t)$.
Next, we consider the case of finite $N$. In this case, the stochastic rate $r(t)$ has finite-size fluctuations of order $1/\sqrt{N}$, hence we rewrite Eq.~\eqref{eq:stoch-rate-def} as 
\begin{equation}
    r(t) = \left[F\left(\bar{h}(t), \sigma^2(t)\right) +\frac{1}{\sqrt{N}}\xi(t)\right]_+\label{eq:Integrated_transfer_function-1}
\end{equation}
with $\xi(t)=\sqrt{N}\left[\frac{1}{N}\sum_{i=1}^N\phi(h_i(t))-F\left(\bar{h}(t), \sigma^2(t)\right)\right]$. \blau{The rectification bracket $[\cdot]_+=\max(0,\cdot)$ could be inserted because $N^{-1}\sum_i\phi(h_i)$ is always a non-negative quantity. Including the rectification is necessary for a Gaussian approximation of $\xi$ to enforce a non-negative population rate $r(t)$, albeit negative values are extremely rare events for biologically relevant network sizes $N$. 
%For example, such events have never occurred in our extensive simulations.
} 
For fixed input functions $f$ and $g$, the second term $F\left(\bar{h}(t), \sigma^2(t)\right)$ is deterministic, and hence, for large $N$, the variable $\xi(t)$ is an approximately Gaussian, colored noise with mean $0$ and auto-covariance function $c_\xi^{f,g}(t,t')=\cov{\phi(h_i(t)),\phi(h_i(t'))}$ (again taken with fixed functions $f$ and $g$). 
Although there exist explicit expressions for the auto-covariance function of a nonlinear transformation of a Gaussian process with given auto-covariance function (see e.g.~\cite{MusGer19}), these expressions are in the form of infinite series or double integrals, and are  not directly usable for the derivation of mesoscopic dynamics. Therefore, we follow a simpler ad hoc approach here:
we assume that $\phi(h_i(t))$ has approximately the same temporal correlation structure as $h_i(t)$. Therefore, we make the heuristic approximation $c_\xi^{f,g}(t,t')\approx \frac{c_\xi^{f,g}(t,t)}{\sigma^2}c_h^{f,g}(t,t')$, where $c_h^{f,g}$ denotes the auto-correlation function of the process Eq.~\eqref{eq:effective_OU}.
Because $\xi(t)$ is Gaussian and $h_i(t)$ has an exponential auto-correlation function with correlation time $\tau$, we hence model the colored noise as an Ornstein-Uhlenbeck process of the form
\begin{equation*}
    \tau\frac{d\xi}{dt}=-\xi+\sqrt{2\tau\sigma_\lambda^2(t)}\zeta(t).
\end{equation*}
Here, $\zeta(t)$ is Gaussian white noise with $\lrk{\zeta(t)}=0$ and $\lrk{\zeta(t)\zeta(t')}=\delta(t-t')$, and 
\begin{align*}
\sigma_\lambda^2(t)&=\lrk{\phi(h_i(t))^2\mid \bar h(t),\sigma^2(t)}-\lrk{\phi(h_i(t))\mid \bar h(t),\sigma^2(t)}^2\nonumber\\
    &=G\left(\bar{h}(t), \sigma^2(t)\right)
\end{align*}
is the conditional variance of the stochastic intensity $\lambda(t)=\phi(h_i(t))$ given by
\begin{equation*}
\label{eq:Integrate_square_transfer_function}
    G\left(\bar{h}, \sigma^2\right):=\int_{-\infty}^\infty \phi^2(h) g_{\bar{h}, \sigma^2}(h)\,dh-F^2\left(\bar{h}, \sigma^2\right).
\end{equation*}
We note again that the conditional variance depends on the functions $f$ and $g$ only through the conditional mean $\bar h(t)$ and conditional variance $\sigma^2(t)$.

We now proceed with the second step of the mean-field derivation, where we close the system in the mean-field variables $\bar h$, $\sigma^2$ and $\xi$ self-consistently. 
%In the first step, we have expressed the stochastic rate $r(t)$ in terms of the dynamics of these mean-field variables (cf. Eq.~\eqref{eq:Integrated_transfer_function-1}). In the second step, we can thus close the system by relating 
To this end, the functions $f$ and $g$ are related back to the stochastic mean-field rate $r(t)$ using Eq.~\eqref{eq:f-g}. Inserting this equation into Eq.~\eqref{eq:condit-moments-f-g} we finally arrive at the mesoscopic mean-field \rot{(MF)} dynamics
\begin{align}
\label{eq:meso_main}
\begin{aligned}
\tau\od{\bar{h}}{t} &= -\bar{h} + \muext(t) + w\lreckig{r(t-d) + \sqrt{\frac{r(t-d)}{N}}\eta(t)}\\
\tau\od{\sigma^2}{t} &= -2\sigma^2 + \frac{\rot{\sigma_w^2}}{\tau N} r(t-d)\\
\tau\od{\xi}{t} &= -\xi + \sqrt{2\tau G\left(\bar{h}(t), \sigma^2(t)\right)}\zeta(t),    
\end{aligned}
\end{align}
where $r(t) =  [F\left(\bar{h}(t), \sigma^2(t)\right) + \xi(t)/\sqrt{N}]_+$. The population activity $A_N(t,\Delta t)$ associated with this mesoscopic dynamics is given by Eq.~\eqref{eq:popact-whitenoise}.
Because we consider the population mean $\bar{h}(t)$ and population variance $\sigma^2(t)$, we call Eq.~\eqref{eq:meso_main} the 2nd-order mean-field (MF2) \rot{dynamics}. 

The \rot{MF2 dynamics captures the effects of both finite $N$ and random connectivity}. Firstly, the \rot{finite-size effect manifests itself in the multiplicative noise term proportional to $\sqrt{r/N}$ in the equation for $\bar h$.} The finite-size noise increases with larger firing rates, reflecting the Poisson noise in the underlying microscopic model, and it vanishes in the large-$N$ limit.
Secondly, the \rot{effect of random connectivity becomes manifest in a non-vanishing variance $\sigma^2$, which is driven by the term proportional to $\sigma_w^2/N$. 
Note that in our case of binary random connectivity, the variance of the synaptic weights is $\sigma_w^2=w^2(1-p)/p$ (cf. Eq.~\eqref{eq:sigma_w}). Thus, for full connectivity, $p=1$, the variance $\sigma^2$ vanishes. It also vanishes for $N\to\infty$ unless $C$ is kept constant (sparse limit) in which case $\sigma_w^2/N=w^2(1/C-1/N)$ remains of order $1$.

A non-vanishing variance $\sigma^2$ (in our case $p<1$)}, in turn, impacts the population rate $r$ through the averaged nonlinearity $F$, Eq.~\eqref{eq:Integrated_transfer_function}.
However, the population average, Eq.~\eqref{eq:stoch-rate-def}, is not perfectly equal to $F$ for finite $N$ and $\sigma^2>0$. The imperfect averaging leads to finite-size fluctuations $\xi(t)$ \rot{-- a third effect caused by the combination of} both finite network size and random connectivity. This effect \rot{vanishes when $N\to\infty$ or $\sigma_w^2=0$ ($p=1$) because} the noise strength is proportional to $\sqrt{G/N}$ and $G$ vanishes for $\sigma^2=0$.

If we had neglected the dispersion of the input potentials due to the random connectivity, i.e. assuming $\sigma^2\equiv 0$, we would have arrived at a first-order MF \rot{theory} \rot{(MF1)} with only one equation for the mean $\bar{h}(t)$:
\begin{align}
\tau\od{}{t}\bar{h} &= -\bar{h} + \muext(t) + w\lreckig{r(t-d) + \sqrt{\frac{r(t-d)}{N}}\eta(t)}\nonumber\\
r(t) &= \phi\bigl(\bar{h}(t)\bigr).\label{eq:naive_main}
\end{align}
Here, we have used that, for $\sigma=0$, the population transfer function $F(\bar{h},\sigma^2)$ simplifies to the transfer function of a single Poisson neuron  $\phi(\bar{h})=F(\bar{h}(t), 0)$ and that $G(\bar{h}, 0) = 0$. 
\rot{Theoretical predictions of MF1, Eq.~\eqref{eq:naive_main}, can always be obtained from the corresponding result of MF2 by taking} the limit $p\to 1$ while keeping the network size $N=C/p$ and the coupling strength $w=JpN$ constant. 

While the \rot{MF1 dynamics, Eq.~\eqref{eq:naive_main}, still captures finite-size effects through the noise term, it does not describe the effect of randomness in the connectivity $\sigma_w^2>0$ when $p<1$. In fact, MF1 precisely corresponds to the mean-connectivity approximation that underlies the assumption of full connectivity made in previous finite-size theories \cite{BuiCow10,BuiCho13,DegSch14,SchDeg17,DumPay17,HeeSta21,KliKir22,SchLoe23}. This can be seen by applying the diffusion approximation of the Poissonian shot noise for large $N$, Eq.~\eqref{eq:popact-whitenoise},  to the exact mesoscopic dynamics, Eq.~\eqref{eq:1st-order-net-meso}, of the mean-connectivity network, Eq.~\eqref{eq:1st-order-net-micro}, which recovers MF1. Thus,} by comparing simulations of the microscopic models (quenched and annealed networks) with MF1 and MF2, we will thus be able to judge the mean-connectivity approximation used in previous studies~\cite{DegSch14,SchDeg17} and the correction for the effect of random connectivity ($p<1$) given by the 2nd-order MF theory.

In the limit $N\to\infty$, the \rot{MF2 theory}, Eq.~\eqref{eq:meso_main}, converges to either a one- or a two-dimensional, deterministic \rot{dynamics} depending on whether the connectivity is dense or sparse. In the dense limit, where $p$ is kept constant and, hence, $C=pN\to\infty$, the variance $\sigma^2$ as well as the fluctuations $N^{-1/2}\eta$ and $N^{-1/2}\xi$ vanish, leaving a \blau{one-dimensional} deterministic firing-rate model. This case is equivalent to the large-$N$ limit of the 1st-order MF \rot{theory and can thus neither capture finite-size nor $p<1$ effects}. In contrast, in the sparse limit, where $C$ is kept constant and, hence, $p=C/N\to 0$, the noise terms $N^{-1/2}\eta$ and $N^{-1/2}\xi$ vanish but the variance $\sigma^2$ does not. The result is a two-dimensional, deterministic mean-field dynamics for the mean and variance of the input potentials, 
\rot{\begin{align}
\label{eq:meso_sparse_limit}
\begin{aligned}
\tau\od{}{t}\bar{h} &= -\bar{h} + \muext(t) + wr(t-d) \\
\tau\od{}{t}\sigma^2 &= -2\sigma^2 + \frac{w^2}{\tau C} r(t-d),
\end{aligned}
\end{align}
where $r(t) = F(\bar{h}(t), \sigma^2(t))$. This sparse connectivity limit is} similar to the classical model of Amari \cite{Ama72}. \rot{While it captures the effect of random connectivity, it cannot describe finite-size effects and hence second-order statistics caused by finite-size fluctuations. In the following, we will refer to the dynamics in the sparse limit, Eq.~\eqref{eq:meso_sparse_limit}, in short as \emph{MFsparselim}.}

\rot{The effects described by the three population dynamics MF1, MF2 and MFsparselim for different parameter regimes are summarized in Table~\ref{tab:MF-models} (Appendix, Sec.~\ref{sec:tables}).}

\subsubsection{Functions $F$ and $G$ for an error-function nonlinearity}

For the error function nonlinearity, Eq.~\eqref{eq:phi-erf}, the functions $F$ and $G$ can be calculated explicitly as (see appendix, Sec.~\ref{sec:App:Calculation of the function F and G})
\begin{align}
&F(h, \sigma^2) = \phimax\Phi\left(\betaeff(\sigma^2) h\right)\label{eq:F_formula},\\
&G(h,\sigma^2)=\phimax^2 \Phi\left(\betaeff(\sigma^2)h\right)- 2\phimax^2T\Big(\betaeff(\sigma^2)h, \betaeff(2\sigma^2)h\Big)\label{eq:G_formula}
\end{align}
with $T$ being the Owen T-function and 
\begin{equation*}
    \beta_\text{eff}(\sigma^2) = \frac{\beta}{\sqrt{1+\beta^2\sigma^2}}
\end{equation*}
being the effective steepness of the sigmoidal nonlinearity at the population level. Because $\betaeff(\sigma^2) \leq \betaeff(0) = \beta$, the effective nonlinearity that governs the population dynamics has a reduced steepness compared to the the single-neuron nonlinearity \rot{(Fig.~\ref{fig:first_order_and_stability_in_main_text}a)} and this reduction is caused by the strictly positive spread $\sigma>0$ of the membrane potentials in a non-fully connected random network.

\section{Analysis of the mesoscopic model}
\label{sec:analysis}

The mesoscopic  \rot{dynamics MF2}, Eq.~\eqref{eq:meso_main}, allows us to  calculate and analyze the first- and second-order statistics of mesoscopic quantities such as the stochastic population rate $r(t)$, the population-averaged input potential $\bar h(t)$ and the population activity $A_N(t,\Delta t)$. \rot{We will mainly focus on the second-order statistics, the analysis of which will be based on a linearization of the dynamics about the fixed points of the noiseless systems. Therefore, we will first calculate the fixed points and their stability. Although not the focus of the paper, this analysis will also yield approximations for first-order statistics including the stationary mean population activity and the linear response of the mean population activity to time-dependent external stimuli.}

\subsection{Stationary mean population activity}
\label{sec:main:Stationary mean population activity}

We first calculate the fixed-points $\bar h(t)=\bar h_0$, $\sigma^2(t)=\sigma_0^2$  and $\xi(t)=0$ of the deterministic system when the noise terms in Eq.~\eqref{eq:meso_main} are set to zero and the external drive is constant, $\muext(t) =\mu_0$. In the following, we assume an inhibitory network ($w<0$) balanced with a positive external drive $\mu_0>0$. In this case, the system exhibits a unique fixed point as shown in the Appendix, Sec.~\ref{sec:unique_FP}.  If the fluctuations are sufficiently small and the fixed point is stable, the quantity $r_0:=F(h_0,\sigma_0^2)$ then provides an approximation  of the mean stationary population activity (or equivalently, the mean stationary firing rate of neurons). At equilibrium, \rot{the input potential $h_0$ is given by the} fixed-point equation
\begin{equation}
\frac{\bar{h}_0-\mu_0}{w} = F\lrrund{\bar{h}_0,\frac{w(1-p)}{2\tau pN}\left(\bar{h}_0-\mu_0\right)},\label{eq:fixpoints_one_line}
\end{equation}
\rot{(see appendix, Sec.~\ref{sec:Fixed point solutions})}.
%Note that both sides of the equation correspond to the stationary firing rate $r_0$. 
Furthermore, the fixed points of the MF1 are obtained by setting $p=1$ in Eq.~\eqref{eq:fixpoints_one_line} corresponding to setting $\sigma_0^2=0$ in the sigmoidal function $F(h_0,\sigma_0^2)$. The solutions for both \rot{MF1 and MF2} can be determined graphically \rot{(Fig.~\ref{fig:first_order_and_stability_in_main_text}a)}: 
%While the left-hand side is identical for the two MF models and is described by a line, the nonlinear function $F$ on the right-hand side of Eq.~\eqref{eq:fixpoints_one_line} differs in its second argument $\sigma_0^2$. 
For MF1, $\sigma_0^2=0$ and therefore the sigmoidal function $F(h_0,0)$ is steeper compared to MF2, for which $\sigma_0^2$ is strictly positive. 

\rot{While the fixed-points can be obtained numerically, for the analytical theory we use the approximations
\begin{align}
\label{eq:approximate_form_fixedpoints_r}
    r_0&\approx -\frac{\mu_0}{w},\\
    \bar h_0 &\approx \Phi^{-1}\left(-\frac{\mu_0}{w\phimax}\right)\sqrt{\beta^{-2}+\sigma_0^2}     \label{eq:approximate_h}\\
    \sigma_0^2 &\approx -\frac{w\mu_0 (1-p)}{2\tau N p}\label{eq:sigma0-approx}
\end{align}
for stimuli $0 < \mu_0 < -w\phimax$, where the firing rates are low ($0< r<\phimax/2$), see appendix,  Sec.~\ref{sec:Fixed point solutions}. The approximation holds for large $\beta \gg 1/\sigma_0^2$ and for $-\mu_0/w\gg (1-p)/(\tau C)$ and $-\mu_0/w> 0.023\phimax$. The fixed-point solutions yield theoretical predictions for the stationary first-order statistics such as the stationary mean firing rate $\lrk{r}$, mean membrane potential $\lrk{\bar h}$ and mean membrane-potential spread $\lrk{\sigma^2}$. However, we postpone the discussion of these predictions to the Appendix~\ref{sec:Fixed point solutions}, because our main focus is the second-order statistics.
}

\subsection{Stability of the fixed-point solutions}
\begin{figure}[!t]
  \centering
\includegraphics[width=\linewidth]{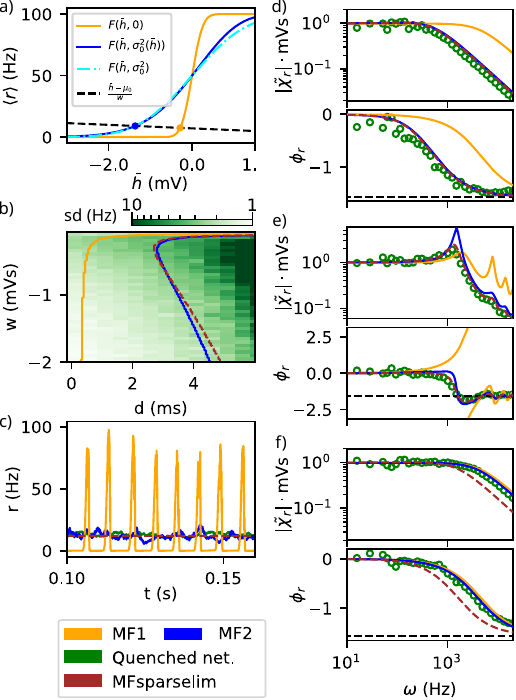}
  \caption{
  %BILDVORSCHLAG! f) hat quenched data von $p=1$, bis $p=0.9$ fertig ist (deshalb grau gemacht!!)
  \rot{Analysis of the deterministic mean-field dynamics. (a) Graphical solution of the fixed point equation \eqref{eq:fixpoints_one_line}, here illustrated for $\mu_0=5$ mV and $w=-0.7$ mVs. Solid lines: function $F(\bar h,\sigma^2(\bar h))$ for MF1 (orange, $\sigma^2(\bar h)\equiv 0$) and MF2 (blue, $\sigma^2(\bar h)$ as in Eq.~\eqref{eq:fixpoints_one_line}). Black dashed line: left-hand side of Eq.~\eqref{eq:fixpoints_one_line}. Cyan dash-dotted line: $F$ with the second argument fixed at the fixed-point value $\sigma_0^2$. (b) $(d, w)$-phase diagram with Hopf bifurcation lines (onset of oscillations). Bifurcation lines of MF2 and MFsparselim  explain stable asynchronous activity and oscillatory instability.  The heat map shows the standard deviation of the population rate $r(t)$ in the quenched network ($p=0.1$) to indicate the presence of oscillations. Hopf bifurcations, calculated from Eq.~\eqref{eq:char-eq}, are displayed for MF1 (orange line), MF2 (blue line) and MFsparselim (brown line). (c) Sample trajectories of population rate $r(t)$ for quenched network (green), MF1 (orange), MF2 (blue) and MFsparselim (brown) with $d=1$ ms.  (d)-(f) Linear response of the time-dependent mean population rate (``rate susceptibility'') to a weak modulation of the external current $\mu(t)$ for quenched network (green circles), MF1 (orange), MF2 (blue) and MFsparselim (dashed brown). The MF theories are given by  Eq.~\eqref{eq:chi_r_linear}, where $p$ is set to $1$ for MF1, and the sparse limit $N\to\infty$, $pN=C=\text{const}$ is taken for MFsparselim. For each panel, top and bottom plots show the amplitude $\vert \tilde{\chi}_r\vert$ and the phase $\phi_r$ of the susceptibility, respectively. Dashed horizontal line: $\phi_r = -\pi/2$. 
  (d) $p=0.1$ ($C=100$), $d=0$; (e) $p=0.4$ ($C=400$), $d=1\,$ms; (f) $p=0.95$ ($C=950$), $d=0$. Other parameters unless otherwise indicated: $N=1000$, $w=-1$ mVs, $\beta=5$ mV$^{-1}$, $\mu_0=10$ mV, $\sigmaext=0$.}}
  \label{fig:first_order_and_stability_in_main_text}
\end{figure}

In the presence of a non-zero transmission delay $d>0$, the fixed-point solution of the mean-field model may become unstable, leading to an oscillatory regime, see e.g.~\cite{BruHak08}.
%for an analysis of this regime in networks of LIF neurons in the sparse limit. 
For such an instability to correctly predict oscillatory behavior of our microscopic network model,  
%depends on an accurate description of the recurrent fluctuations in the mean-field theory.
our mean-field theory must accurately describe the recurrent fluctuations in the network.
%Comparing simulations of the microscopic network models with simulations of the first- and second-order mean-field \red{theories} \red{(Fig.~\ref{fig:first_order_and_stability_in_main_text}c)}, we observe that for the chosen parameters the 1st-order MF model wrongly predicts strong oscillations when the actual population activity of the network is nearly constant corresponding to an asynchronous firing regime.   
%Indeed, we observe in simulations that an actually non-oscillatory population activity of the network is not well described by the first-order mean-field model, which incorrectly predicts strong oscillations (Fig.~\ref{fig:Asynchronous_vs_oscillations}a). 
%In contrast, the 2nd-order theory correctly reproduces the constant population activity. 
%For the given set of parameters (see caption), only the 1st-order mean-field model exhibits strong oscillations, while the 2nd-order mean-field model and the microscopic models remain in the asynchronous state. 

%This behavior 
The onset of oscillations 
can be understood by a linear stability analysis.
To this end, 
we linearize 
the noiseless system 
%is linearized 
around the fixed point. Small deviations of the mesoscopic variables,
%from the fixed point, 
$\delta\bar h=\bar h(t)- \bar h_0$, $\delta\sigma^2=\sigma^2(t)-\sigma_0^2$, obey the linearized dynamics 
\begin{align}
\od{}{t} \Vect{X}(t) = \Mat{T} \Vect{X}(t) + \Mat{W} \Vect{X}(t-d),\label{eq:linearized_system-2}
\end{align}
where we introduced the deviation vector $\Vect{X}(t)=\lrrund{\delta\bar{h}(t),\delta\sigma^2(t),\xi(t)}^T$ and the matrices
\begin{gather*}
\Mat{T}= -\frac{\text{diag}(1,2,1)}{\tau},\quad \Mat{W}= \lrrund{\frac{w}{\tau}, \frac{w^2\left(1-p\right)}{\tau^2 pN}, 0}^T\Vect{L}^T
\end{gather*}
with $\Vect{L}^T = (F_h,F_\sigma, N^{-1/2})$.
Here, $F_h$ and $F_\sigma$ denote the partial derivatives of $F(h,\sigma^2)$ with respect to $h$ and $\sigma^2$ at the fixed point (exact and approximate expressions for $F_h$ and $F_\sigma$ are given in the Appendix \ref{sec:App:approx_Fh_Fs}). 
% Alte Gleichungen:
% \begin{subequations}\label{eq:meso_linearized_for_stability_analysis}
% \begin{align}
%     \tau\od{}{t}\delta h &= -\delta h + w\delta r(t-d)\label{eq:meso_linearized_for_stability_analysis_h}\\
%     \tau\od{}{t}\delta\sigma^2 &= -2\delta\sigma^2 + \frac{w^2(1-p)}{\tau C} \delta r(t-d)\label{eq:meso_linearized_for_stability_analysis_sigma}\\
%     \tau\frac{d\xi}{dt}&=-\xi\label{eq:meso_linearized_for_stability_analysis_xi}\\
%     \delta r(t) &= \left(F_h \delta h(t) + F_\sigma \delta \sigma^2(t) + \frac{1}{\sqrt{N}}\xi(t)\right)\label{eq:meso_linearized_for_stability_analysis_r},
% \end{align}
% \end{subequations}
We note that any perturbation in $\xi$ exponentially decays to zero with a time constant $\tau$. 
%For the linear stability analysis it 
It 
is therefore sufficient to look for solutions of the form $\Vect{X}(t)=e^{\lambda t}\lrrund{\hat{h}(\lambda),\hat{\sigma}(\lambda),0}^T$ with some constant parameter $\lambda$. The sign of the real part of $\lambda$ then determines the stability of the fixed point; in particular, if $\Re(\lambda) >0$ the fixed-point is unstable. As shown in Appendix~\ref{sec:App:Stabilization of the asynchronous state}, the parameter $\lambda$ must satisfy the characteristic equation
\begin{align}
\lambda \tau &= -1 + w{F}_h e^{-\lambda d} + \frac{w{F}_h \hat{F}_\sigma e^{-2\lambda d}}{2+\lambda \tau -\hat{F}_\sigma e^{-\lambda d}},
\label{eq:char-eq}
\end{align}
where $\hat{F}_\sigma := w^2 \frac{1-p}{\tau C} F_\sigma$. 

To find the onset of oscillations, we look for a Hopf bifurcation, at which $\lambda = i\omega$ with a frequency $\omega\in\R\setminus\{0\}$. Imposing such purely imaginary values of $\lambda$ in Eq.~\eqref{eq:char-eq} allows us to find the ``Hopf boundary'' of the oscillatory phase in the parameter space. In the following, we investigate the stability depending on the transmission delay $d$ and the coupling strength $w$ \rot{(Fig.~ \ref{fig:first_order_and_stability_in_main_text}b)}. In the $(d, w)$-space, we can see that the bifurcation lines of the noiseless systems significantly differ between 2nd-order and first-order MF dynamics. For sufficiently strong coupling strength, the 1st-order MF dynamics exhibits a Hopf boundary at small values of $d$, predicting oscillations already at small delays. In contrast, the 2nd-order theory predicts an oscillation onset for much larger delays. Interestingly, 
%for \rot{random} connectivity with sufficiently small $p$, 
MF2 predicts that a stronger coupling strength $w$ requires a larger delay to enable oscillations, whereas the first-order model requires a smaller delay. 
\rot{We note that the bifurcation line obtained for MFsparselim is close to MF2.
The bifurcation lines qualitatively explain the presence or absence of large noisy oscillations in the mean-connectivity and random network models (Fig.~ \ref{fig:first_order_and_stability_in_main_text}b,c). 
Furthermore, the simulations of the MF dynamics indicate that the Hopf bifurcation is supercritical: varying the delay, the variance of the population rate $r$ changes continuously at the bifurcation and no bistability between fixed point and limit cycle is observed. 
}

In conclusion, our stability analysis of MF2 shows that connectivity disorder leads to a stabilization of the non-oscillatory stationary state. In particular, the disordered connectivity leads to incoherent fluctuations of the recurrent input, which flattens the effective nonlinearity of the macroscopic dynamics and thus changes the stability properties of the fixed point.

\subsection{Linear response of the mean activity}

We now analyze how the network responds to a small, time-dependent  perturbation of the external current, $\muext(t)=\mu_0+\mu_1(t)$ by calculating the  linear response function \cite{LedBru11}.
To this end, we linearize our mesoscopic mean-field \rot{dynamics}, Eq.~\eqref{eq:meso_main}, around the fixed point and obtain a generalization of Eq.~\eqref{eq:linearized_system-2}:
\begin{align}
\od{}{t} \Vect{X}(t) = \Mat{T} \Vect{X}(t) + \Mat{W} \Vect{X}(t-d) + \Vect{M}(t) + \Mat{B}\Vect{\zeta}(t).\label{eq:linearized_system}
\end{align}
Here, the vector $\Vect{X}(t)$ represents again the deviation from the fixed point as in the previous section and the three-dimensional vector $\Vect{M}(t)$ represents a general small perturbation to the system. Furthermore, the last term represents the finite-size noise, where
%\begin{gather}
%\Mat{B} = \PM{\frac{w}{\tau} \sqrt{\frac{r_0}{N}}&0\\0&0\\0&\sqrt{\frac{2G_0}{\tau N}}},\quad \nonumber\Vect{\zeta}(t)= \PM{\eta(t)\\ \zeta(t)}
%\end{gather}
\begin{gather}
\Mat{B} = \PM{\frac{w}{\tau} \sqrt{\frac{r_0}{N}}&0\\0&0\\0&\sqrt{\frac{2G_0}{\tau}}},\quad \nonumber\Vect{\zeta}(t)= \PM{\eta(t)\\ \zeta(t)}
\end{gather}
and $G_0 = G(\bar{h}_0, \sigma_{0}^2)$. In the linearized dynamics we neglected terms of order $O(1/N)$
and higher.
In the Fourier domain, the solution  reads
\begin{equation}
    \tilde{\Vect{X}}(\omega) = \tilde{\Mat{\chi}}(\omega)\left(\tilde{\Vect{M}}(\omega)+\Mat{B}\tilde{\Vect{\zeta}}(\omega)\right),\label{eq:tildeX}
\end{equation}
where we introduced the susceptibility matrix
\begin{equation}
\label{eq:chi}
    \tilde{\Mat{\chi}}(\omega)= \left[i\omega\Id_3 -\Mat{T}-\Mat{W}e^{-i\omega d}\right]^{-1}.
\end{equation}
Here, $\Id_3$ denotes the $3\times 3$ identity matrix and the tilde notation $\tilde f(\omega)$ for a function $f(t)$ denotes its Fourier transform, i.e.~$\tilde{f}(\omega)=\int_{-\infty}^\infty f(t) e^{-i\omega t}\,dt$. The explicit expressions for the elements of $\tilde{\Mat{\chi}}$ are given in the appendix, Eq.~\eqref{eq:App:chi_expicit}.

The susceptibility matrix provides the response of the ensemble mean $\lrk{\Vect{X}(t)}$ of the deviation from the fixed point of the full nonlinear system to an infinitesimally  small perturbation $\Vect{M}(t)$ in the Fourier domain via the relation $\lrk{\tilde{\Vect{X}}(\omega)} = \tilde{\Mat{\chi}}(\omega)\tilde{\Vect{M}}(\omega)$. Similarly, the response of the ensemble-averaged firing rate  $\langle r(t)\rangle=F(h_0,\sigma_0^2)+\delta r(t)$ is given
%by its Fourier transform as 
in the Fourier domain by
$\tilde{\delta r}(\omega)
%=\Vect{L}^T\langle\tilde{\Vect{X}}(\omega)\rangle
=\Vect{L}^T\tilde{\Mat{\chi}}(\omega)\tilde{\Vect{M}}(\omega)$. For an
external stimulus $\muext(t)=\mu_0+\mu_1(t)$, the perturbation is $\tilde{\Vect{M}}(\omega)=(\tilde\mu_1(\omega)/\tau,0,0)^T$. In this case, we 
%thus 
then 
have $\tilde{\delta r}(\omega)=\tilde{\chi}_r(\omega)\tilde{\mu}_1(\omega)$, where we introduced the rate susceptibility
\begin{equation}
    \tilde{\chi}_r(\omega)=\frac{1}{\tau}\left[F_h\tilde{\Mat{\chi}}_{11}(\omega)+F_\sigma\tilde{\Mat{\chi}}_{21}(\omega)\right].\label{eq:chi_r_linear}
\end{equation}
Here, we used that $\tilde{\chi}_{31}\equiv 0$.

%We can compare the prediction of Eq.~\eqref{eq:chi_r_linear} with simulations of the susceptibility (Fig.\ref{fig:chi_r}). The theory for the mesoscopic models is no approximation and simulations converge exactly on theory which has been confirmed in comparison to simulations, Fig~\ref{fig:chi_r} only shows the theory for the mesocopic models. 
The rate susceptibility is in general complex-valued with absolute value $\vert\tilde \chi_r(\omega)\vert$ and phase $\phi_r(\omega)$. 
%An 
A straightforward 
interpretation 
%of the susceptibility 
is that the system driven by a sinusoidal external input modulation $\mu_1(t)=\epsilon\sin(\omega t)$  with small amplitude $\epsilon$ will respond with a sinusoidal rate modulation  with amplitude $\epsilon\vert\tilde \chi_r(\omega)\vert$ and a phase shift $\phi_r(\omega)$. For all models studied here, the amplitude decreases for high frequencies with a power law with exponent -1 \rot{(Fig.~\ref{fig:first_order_and_stability_in_main_text}d-f)}. The MF2 theory agrees very well
\rot{with the quenched microscopic network}, 
%with each other. This is 
in contrast to the MF1 \rot{theory} which massively overestimates the amplitude at high frequencies. The quenched network shows a slightly smaller amplitude than predicted by the 2nd-order MF \rot{theory}, but is still reasonably well approximated by the latter. While the amplitude vanishes at low frequencies as expected,  the phase shift approaches  $-\frac{\pi}{2}$ in the limit of large frequencies \rot{(Fig.~\ref{fig:first_order_and_stability_in_main_text}d-f)}. We can understand this behavior analytically by 
considering
%analysing 
the limits of Eq.~\eqref{eq:chi_r_linear}. For $\omega=0$ we obtain
\begin{equation}
    \label{eq:chir_lowfreq}
    \tilde\chi_r(0) = \frac{F_h}{1-wF_h-\frac{w^2}{2\tau}F_\sigma \left(\frac1{C}-\frac1{N}\right)},
\end{equation}
which is real-valued and has therefore a phase zero. The amplitude in the low-frequency limit is approximately the negative inverse of the coupling strength $-w^{-1}$, under the condition that $1\ll wF_h$ and $C$ large. Similarly, for large frequencies we find asymptotically
\begin{equation}
    \label{eq:chir_highfreq}
    \tilde\chi_r(\omega) \sim -i\frac{F_h}{\tau}\cdot \frac1{\omega},\qquad \omega\to\infty.
\end{equation}
In this limit, the phase converges to $-\frac{\pi}{2}$ and the amplitude follows a power law $\omega^{-1}$ scaled by $F_h/\tau$. The different values for the derivative $F_h$ explain the difference between the MF1 and the  MF2 \rot{theory}. To understand this difference analytically, we use the following approximation for the slope of the transfer function at the fixed point (see Appendix \ref{sec:App:approx_Fh_Fs}): 
\begin{equation}
\label{eq:Fh-approx}
    F_h\approx\frac{\phimax Q\lrrund{-\frac{\mu_0}{w\phimax}}}{\sqrt{2\pi\lrrund{\beta^{-2}+\sigma_0^2}}},
\end{equation}
where $\sigma_0^2$ is given by Eq.~\eqref{eq:sigma0-approx}. Furthermore, we introduced the non-dimensional function
\begin{equation}
\label{eq:Q-func}
    Q(x) := \exp\left[-\frac12 \Phi^{-1}\left(x\right)^2\right]
\end{equation}
defined on the interval $[0,1]$. 
This function has an inverted-U shape and is symmetric with respect to $x=\frac{1}{2}$ (a graph of $Q$ is shown in Fig.~\ref{fig:var_r_with_external_noise_and_Fh}c, dotted lines). Equation \eqref{eq:Fh-approx} clearly shows that the factor $F_h$ increases with the connection probability $p$ because $\sigma_0^2$ decreases with $p$, cf. Eq.~\eqref{eq:sigma0-approx}. The MF1 theory corresponding to the mean-connectivity network (without additional 
%i.e. 
noise) is obtained by setting $p=1$ in Eq.~\eqref{eq:sigma0-approx}. In this case, the dispersion $\sigma_0^2$ vanishes, and hence the slope $F_h$ is maximized.  
Therefore, the MF1 and MF2 \rot{theories} predict the same power-law behavior 
\rot{of the rate susceptibility with $\omega$ at large frequencies}, but with a higher magnitude for the 1st-order model \rot{(Fig.~\ref{fig:first_order_and_stability_in_main_text}d)}. With increasing connection probability $p$ 
%in the system 
the susceptibility approaches the 1st-order MF theory. The low-frequency limit is in both mesoscopic models almost independent of $p$ as the low-frequency limit is roughly given by $-w^{-1}$.
Simulations of the annealed network network (not shown in the figure) match the predictions of the MF2 \rot{theory} while the quenched network model has slightly lower amplitude and slightly lower phase for intermediate frequencies. Overall, the MF2 theory yields a strongly improved prediction compared to the MF1 \rot{theory}.

The synaptic delay $d$ appears in $\tilde\chi_r$ only via the complex exponential in Eq.~\eqref{eq:chi}. On the other hand, the delay does not influence the low-frequency limit and the high-frequency asymptotics, Eqs.~\eqref{eq:chir_lowfreq} and \eqref{eq:chir_highfreq} of the response, respectively. For intermediate frequencies the complex exponential causes a modulation and a peak in the amplitude as a function of $\omega$ \rot{(Fig.~\ref{fig:first_order_and_stability_in_main_text}e)}. Simulations of the annealed network model perfectly confirm \rot{(not shown in figure)} the prediction of the 2nd-order MF \rot{theory}, whereas the quenched network model has a less pronounced peak at a lower frequency. \rot{The susceptibility of MFsparselim fits well with the quenched network for small or medium connection probability,  $p=0.1$ and $p=0.4$ (Fig.~\ref{fig:first_order_and_stability_in_main_text}d, e). For larger $p$, however, there is a noticeable discrepancy for MFsparselim, while MF1 and MF2 have a good match with the quenched network (at $p=0.95$, see Fig.~\ref{fig:first_order_and_stability_in_main_text}f). 
%While MFsparselim and MF1 do not fit well for high connectivity and sparse connectivity respectively, the predictions of MF2 are good in both cases.
In marked contrast to MF1 and MFsparselim, the predictions of MF2 are good over the entire range of connection probabilities $p$.}

\subsection{Power spectral density}

\begin{figure}[t]
  \centering
  \includegraphics[width=\linewidth]{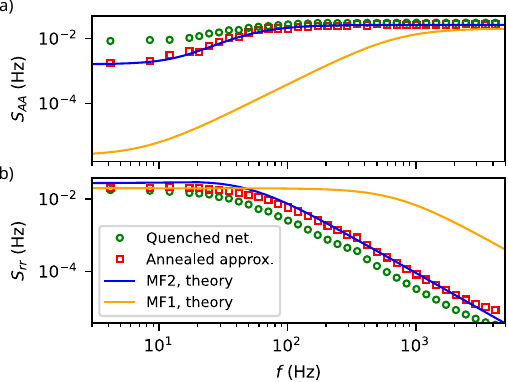}
  \caption{\rot{Power spectral densities of the population activity and population rate. a) Power spectrum of $A_N(t;\Delta t)$ for quenched (green circles) and annealed network (red squares) and 1st- and 2nd-order theory (orange and blue lines, Eqs.~\eqref{eq:S_AA_1st} and \eqref{eq:S_AA}, respectively). b) Same for the power spectrum of $r(t)$ (theory: Eq.~\eqref{eq:S_rr})} Parameters: $w=-1$ mVs a) $N=500$, $p=0.07$, $\mu(t)=\mu_0=10$~mV, \rot{$\sigmaext=0$}, \rot{$\Delta t = 0.1$~ms}.}
  \label{fig:PSD}
\end{figure}

%$=\lrrund{\bar h(t),\sigma(t),N^{-1/2}\xi(t)}^T$

%So far, we have analyzed the first-order statistics of the network dynamics such as the (time-dependent) mean firing rate and the mean input potential. 
Now we turn to the second-order statistics of the stationary dynamics. We assume that the external stimulus $\mu(t)$ is a stationary process with given mean $\mu_0$ and power spectral density $S_{\mu\mu}(\omega)$. We begin with the power spectrum of 
%the mean-field state vector $\Vect{X}(t)$ (cf.~Eq.~\eqref{eq:linearized_system-2}). This matrix involves, e.g., the power spectrum of the average input potential $\bar h(t)=\bar h_0+X_1(t)$ but also cross spectra such as between $\bar h(t)$ and $\sigma^2(t)=\sigma_0^2+X_2(t)$.  Furthermore, we are interested in the power spectra of the average intensity $r(t)$ and 
the population activity $A(t)$ \rot{and population rate $r(t)$.} 
%, as well as in the variance of $r(t)$. The latter is studied in the subsequent section. 
\rot{This statistics has been calculated analytically using linear-response theory for finite-size networks with uniform full connectivity \cite{LinDoi05, DegSch14}. In a similar fashion, we here derive the power spectrum via a linearization of our 2nd-order MF dynamics. To this end, we start with} the power spectral density matrix $\Mat{S}(\omega)$ \rot{of the mean-field state vector $\Vect{X}(t)$ (cf.~Eq.~\eqref{eq:linearized_system-2})} defined via the relation 
\begin{equation}
\label{eq:psd-def}
     \left\langle \tilde{\Vect{X}}(\omega)\tilde{\Vect{X}}^*(\omega') \right\rangle=2\pi\Mat{S}(\omega)\delta(\omega-\omega')
\end{equation}
where $^*$ denotes the conjugate transpose of a matrix. Using Eq.~\eqref{eq:tildeX} for the linearized system, we find the following  approximation for $\Mat{S}(\omega)$ describing weak fluctuations around the fixed points (see Appendix, Sec. \ref{sec:App:Variance and power spectral density in the stationary state}):
\begin{multline*}
        \Mat{S}_{ij}(\omega)=\frac1{\tau^2}\left[S_{\mu\mu}(\omega) + \frac{w^2r_0}{N}\right]\tilde{\chi}_{i1}(\omega)\tilde{\chi}_{j1}^*(\omega)\\
    +\frac{2G\left(\bar{h}_0, \sigma_{h, 0}^2\right)}{\tau}\tilde{\chi}_{i3}(\omega)\tilde{\chi}_{j3}^*(\omega).
\end{multline*}
The matrix $\Mat{S}$ gives us the (cross-) spectral densities between $\bar{h}$, $\sigma^2$ and $\xi$. For the power spectral densities of the population activity $A(t)$, we use the linear approximation
\begin{equation}
    \tilde{A} \approx \tilde{r} + \sqrt{\frac{r_0}{N}}\tilde{\eta},\quad \tilde{r} \approx F_h\tilde{h} + F_\sigma \widetilde{\sigma^2}+\frac{1}{\sqrt{N}}\tilde{\xi}.
    \label{eq:r_tilde_linear}
\end{equation}
\rot{which yields approximate expressions for  the power spectrum of the stochastic population rate $r(t)$,}
\begin{multline}
    S_{rr}(\omega) \approx F_h^2 S_{11}(\omega) +F_{\sigma}^2 S_{22}(\omega) + \frac1{N}S_{33}(\omega)+\\
    2F_hF_\sigma\mathfrak{Re}(S_{12}(\omega)) + 2\frac{F_h\mathfrak{Re}(S_{13}(\omega))+F_\sigma\mathfrak{Re}(S_{23}(\omega))}{\sqrt{N}},\label{eq:S_rr}
\end{multline}
and of the population activity $A(t)$:
\begin{equation}
    S_{AA}(\omega) \approx S_{rr}(\omega) + \frac{r_0}{N} + 2\sqrt{\frac{r_0}{N}}\mathfrak{Re}(S_{r\eta}(\omega)).\label{eq:S_AA}
\end{equation}
The cross spectral density $S_{r\eta}$ that appears here can be calculated using Eq.~\eqref{eq:r_tilde_linear} and \eqref{eq:tildeX}:
\begin{equation*}
    S_{r\eta}(\omega) \approx \frac{w}{\tau} \sqrt{\frac{r_0}{N}} \left[F_h \tilde{\chi}_{11}(\omega) + F_\sigma \tilde{\chi}_{21}(\omega)\right].
\end{equation*}
%The linear theory for the power spectrum of $A(t)$, Eq.~\ref{eq:S_AA}, is compared to simulations of the microscopic and mesoscopic models (Fig. \ref{fig:PSD}). 
In Fig.~\ref{fig:PSD}, we compare the \rot{theoretical predictions, Eqs.~\eqref{eq:S_rr} and \eqref{eq:S_AA},  to the corresponding power spectra obtained from simulations of the quenched and annealed network.
In all cases,} the power spectra \rot{of the population activity} display a pronounced trough at low frequencies \rot{(Fig.~\ref{fig:PSD}a)}. 
The \rot{MF1 theory} severely underestimates the observed low-frequency power, whereas the \rot{MF2 theory} matches well the microscopic model under the annealed approximation. \rot{Nonetheless, even with this marked improvement over MF1, the annealed network’s low‐frequency power remains lower than that of the quenched network, exposing a limitation of the annealed approximation. The discrepancy of the power spectra of quenched and annealed networks is caused by the partial elimination of the effect of the single-neuron auto-correlations on the recurrent input when applying the annealing (i.e. the permanent resampling of connectivity). We note that the elimination of temporal correlations only concerns the incoherent part of the recurrent input, as expressed by the white noise in Eq.~\eqref{eq:annealed-network-model}. In contrast, for finite $N$, the coherent part of the recurrent input fluctuations (finite-size fluctuations of population activity) still contains effects of spike-train auto-correlations, even in the annealed network. 
Finally, we mention that we have not included MFsparselim in Fig.~\ref{fig:PSD}  because its stationary dynamics is characterized by a deterministic equilibrium without any fluctuations and a vanishing power spectrum.}

\subsection{Variances}
\label{sec:variances}

\begin{figure}[t]
  \centering
\includegraphics[width=\linewidth]{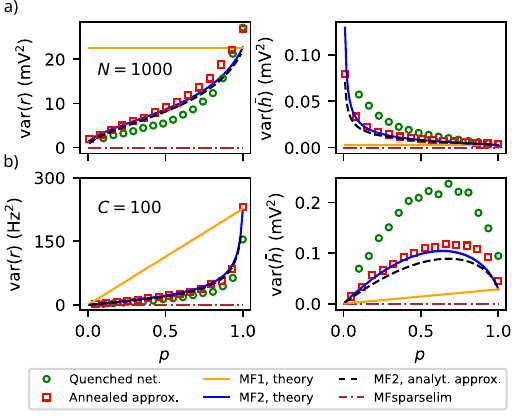}
%\caption{Simulations for mean and variances of the poullation firing rate $r(t)$ for different connection probabilities $p=C/N$. Symbols show the simulation results of the mesoscopic or microscopic models, solid lines are the corresponding linear theories for the mesoscopic models, $\mu_0=10$ mV and $w=-1$ mVs. \textbf{First row}: $N=1000$ neurons is fix and $p$ is changed by changing the number of synapses $C$. \textbf{Second row}: $C=100$ is fix and $N=C/p$ is chosen accordingly. \textbf{a)} and \textbf{d)} variance of the mean firing rate for the quenched microscopic model (green), the annealed microscopic model (red), the 2nd-order (blue) and 1st-order (orange) mean-field models. The variances increase with increasing $p$, execpt for the 1st-order mean-field model which is independent of $p$, for $N$ constant. \textbf{b)} and \textbf{e) } Stationary values of the (population averaged) mean firing rate  \textbf{c) } and \textbf{f) } variance of the population mean of the input potentials}
\caption{ 
%MF$_{N\to\infty}^{p\to 0}$ 
%MFsparselim (preferred), 
%MF$_\text{sparse limit}$, MF$_\text{sparselim}$, MFsparse 
Stationary variance of the population rate and mean input potential for varying connection probability $p$. (a) The number of neurons is fixed to $N=1000$ (hence $C=1000p$). (b) The in-degree is fixed to $C=100$ (hence $N=100/p$). Left panels of (a) and (b): Variance of the population rate $r(t)$ (Symbols: simulations of microscopic networks as indicated in the legend; orange and blue solid lines: full theory Eq.~\eqref{eq:variance_r_linear_with_finite_size} and Eqs.~\eqref{eq:App:sig13_sig23}; black dashed line: analytical approximation for the variances based on the 2nd-order MF theory, Eq.~\eqref{eq:sigma_rr_approx}, \eqref{eq:var-r-Cconst} and Eqs.~\eqref{eq:sigma11_approximation}, \eqref{eq:var-hbar-Cconst} for $\var{r}$ and $\var{\bar h}$, respectively; \rot{dash-dotted line: MFsparselim.}) Right panels of (a) and (b): Same as left panels but for variance of $\bar h$. Parameters: $\mu_0=10$ mV, $w=-1$ mVs, $\sigmaext=0$ mV, $\beta = 5$ mV$^{-1}$}
  \label{fig:variance_vs_p}
\end{figure}

\begin{figure}
    \centering
\includegraphics[width=\linewidth]{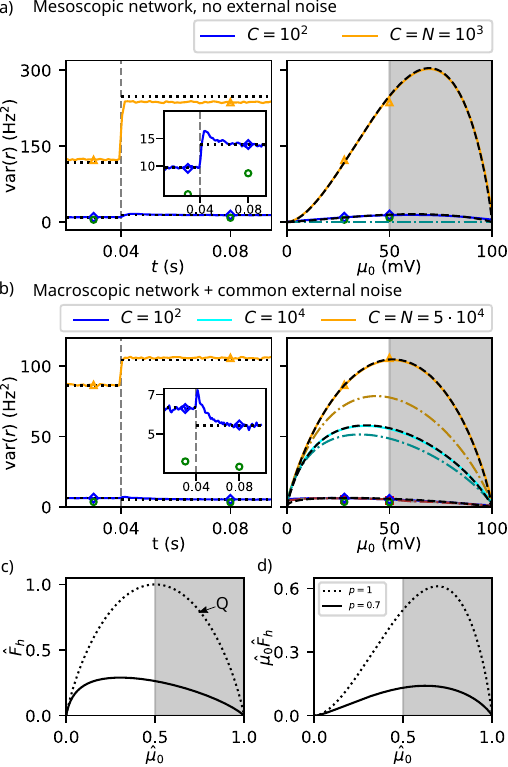}
    \caption{Variance of the stochastic population rate $r(t)$ for varying mean external stimulus $\mu_0$. (a) Moderate-size network of $N=10^3$ neurons whose inputs exhibit internally generated, finite-size noise but no external noise, $\sigmaext=0$. \rot{Different in-degree $C$ are encoded by color}. Left: Response to a step stimulus, where $\mu_0$ jumps from $28$~mV to $50$~mV at time $t=0.04$~s. Blue/orange line: simulations of \rot{MF2} for $C=100$ ($p=0.1$) and $C=1000$ ($p=1$), the latter ($p=1$) equivalent to MF1 or the mean-connectivity approximation of the former ($p=0.1$). Black dotted: corresponding theory for the stationary variances, Eq.~\eqref{eq:variance_r_linear_with_finite_size}. Gray dashed: stimulus onset. Inset: magnified data for $C=100$. \rot{Symbols represent stationary variances of network simulations and theory, see description of right panel.} Right: Stationary variance of $r(t)$ vs.~mean stimulus strength $\mu_0$. Symbols: simulations for $C=100$ of quenched network (green circles),  
    %annealed network (red squares),  
    MF1 (orange triangles), and MF2  (blue diamonds). Solid lines: full MF1 (orange) and MF2 theory (blue). Black dashed: corresponding analytical approximations, Eq.~\eqref{eq:var-r-Cconst}; \rot{dash-dotted: MFsparselim}. The gray dashed line marks the border of the region of interest $\mu_0\le -w\phimax/2$ where the hazard function at the fixed point, $\phi(\bar h_0)$ is convex. 
    (b) Analogous to (a) for a much larger network ($N=5\cdot10^4$) with negligible internally generated finite-size noise but with additional external noise $\sigmaext^2=1$~mV$^2$. The connection probability is varied by different values of $C$ \rot{(encoded by color)}. Orange line: common MF1 theory or mean-connectivity network. \rot{Dash-dotted lines: MFsparselim.}
    %(color-coded by $C$.} 
    (c) Slope $\hat{F}_h$ vs.~mean stimulus $\hat{\mu}_0$. Dotted: $p=1$ corresponding to the function $Q$, solid line: $p=0.7$. Gray dashed as in (a),(b) right. (d) Same for the product $\hat\mu_0\hat{F}_h$. \rot{(c), (d): $N=1000$, $\hat{\beta}=500$ (equivalent to $\beta=5$ mV$^{-1}$ in (a), (b)).}}
\label{fig:var_r_with_external_noise_and_Fh}
\end{figure}

The stationary variances of $r(t)$ and $\bar{h}(t)$ can be calculated by standard methods for the linearized system in the absence of synaptic delay, $d=0$ (see Appendix, Sec.~\ref{sec:App:Variance and power spectral density in the stationary state}). Furthermore, we assume that the external stimulus $\muext(t)$ has mean $\mu_0$ and white Gaussian fluctuations of strength $\sigmaext^2$ as in Sec.~\ref{sec:mean-connectivity-net}.  
%The result is 
Under these assumptions, one eventually obtains
% a system of linear equations for the elements of the covariance matrix $\Mat{\Sigma}$. Because of the symmetry of $\Mat{\Sigma}$, we only have six equations for the $3\times 3$ matrix. The elements of the covariance matrix relating to the variables $\bar{h}$ and $\sigma^2$ are
% \begin{multline}
%     \PM{\sigma_{11}\\ \sigma_{12}\\ \sigma_{22}} = \frac{-2}{E}\left(\frac{w^2r_0}{\tau^2N}+\frac{\sigma_\text{ext}^2}{\tau}\right) \PM{ (\Gamma_{11}+\Gamma_{22})\Gamma_{22}-\Gamma_{12}\Gamma_{21}\\ -\Gamma_{22}\Gamma_{21}\\ \Gamma_{21}^2} \\
%     -\frac{4\sigma_{13}}{E} f_{13}(\Mat{\Gamma}) -\frac{4\sigma_{23}}{E} f_{23}(\Mat{\Gamma})
% \end{multline}
% where $\Gamma_{ij}$ are the elements of the matrix $\Mat{\Gamma}:=\Mat{T}+\Mat{W}$. Furthermore, the functions $f_{13}$ and $f_{23}$ are given in the appendix, Eq. ....
% The first two terms correspond to  compared to a system~\eqref{eq:linearized_system} where the variable $\xi$ is set to zero. We expect these terms to be small for the number of neurons $N$ large. The terms are fully written in the Appendix, Sec \ref{sec:App:Variance and power spectral density in the stationary state}. We use the abbreviation
% \begin{equation}
%     E = 4\left(\Gamma_{12}\Gamma_{21}-\Gamma_{11}\Gamma_{22}\right)\left(\Gamma_{11}+\Gamma_{22}\right).
% \end{equation}
% From the linearized system, we obtain the following approximation for the variance of the stochastic mean-field rate
\begin{multline}
    \label{eq:variance_r_linear_with_finite_size}
    \var{r} \approx F_h^2 \sigma_{11} + F_\sigma^2 \sigma_{22} +2F_hF_\sigma \sigma_{12}\\
    +\frac{2F_h}{\sqrt{N}}\sigma_{13}+\frac{2F_\sigma}{\sqrt{N}}\sigma_{23} + \frac{1}{N} G(\bar h_0,\sigma^2_0).
\end{multline}
where $\sigma_{ij}=\lrk{X_i(t)X_j(t)}$ are  the elements of the covariance matrix of the state vector $\Vect{X}(t)$. In particular, $\sigma_{11}$ yields the variance of the population-averaged input potential $\bar h(t)$. Approximate analytical expressions for $\sigma_{ij}$ can be calculated using the linearized system, Eq.~\eqref{eq:linearized_system}, and are given in the appendix, Sec.~\ref{sec:App:Variance and power spectral density in the stationary state}. Our theory,  Eq.~\eqref{eq:variance_r_linear_with_finite_size}, agrees well with microscopic simulations (Fig.~\ref{fig:variance_vs_p} and \ref{fig:var_r_with_external_noise_and_Fh}), discussed in more detail below.
%as discussed in the following (Fig.~\ref{fig:variance_vs_p} and \ref{fig:var_r_with_external_noise_and_Fh}). 
In general, we find a good quantitative match between the theory and simulations of the annealed network. While, in the case of quenched random connectivity, the theory again exhibits quantitative deviations, our 2nd-order MF theory still captures the parameter dependence of the variance qualitatively and yields a much better quantitative prediction than the 1st-order MF theory. \rot{Again, MFsparselim without any noise in the external input deterministically converges into an equilibrium with zero variance.}

\subsubsection{Dependence on the connection probability}

Which connection probability causes the largest variability in our network model? To address this question, we analyze the variances of the  population rate $r$ and the average input potential $\bar h$ as a function of $p$ (Fig.~\ref{fig:variance_vs_p}). To avoid a trivial dependence through the mean firing rate, we vary $p$ such that the mean firing rates remain roughly unchanged. According to Eq.~\eqref{eq:approximate_form_fixedpoints_r}, a constant rate can be achieved by keeping the coupling strength $w=NpJ$ constant. In an experimental setting, where the synaptic efficacy $J$ is the basic physiological parameter, this constraint could be realized in two different scenarios, either by fixing the network size $N$ and changing the synaptic efficacy $J$ such that $Jp=\text{const.}$, or by fixing both the number of input connections $C=Np$ and the synaptic efficacy $J$. 

In the first scenario with constant $N$, 
%when $N$ is constant, 
we find that the variance of the rate $r(t)$ increases with connection probability in microscopic network simulations  (Fig.~\ref{fig:variance_vs_p}a, left), although the variance of the input potential $\bar h$ decreases  (Fig.~\ref{fig:variance_vs_p}a, right). \rot{MF1} cannot explain this behavior because the model equation \eqref{eq:naive_main} does not depend on the parameter $p$ when $w$ is fixed and $N$ is constant. 
%At all connection probability, 
Thus, for all $p\in[0,1]$, 
the variance predicted by the 1st-order MF \rot{theory} corresponds to the 
%all-to-all-connected 
fully connected case $p=1$, i.e. to the mean-connectivity network. In contrast, the dependence on $p$ of the variances measured from simulations of the annealed network are quantitatively well reproduced by the 2nd-order MF theory, Eq.~\eqref{eq:variance_r_linear_with_finite_size}. The 2nd-order theory also captures the variances of the quenched network qualitatively. The quantitative agreement is also reasonably good, however, we find that the rate variance is overestimated and the variance of the population-averaged input potentials is underestimated by the annealed network. Nevertheless, the prediction of variances by the 2nd-order theory yields a significant improvement over the 1st-order MF theory. 

In order to gain an analytical understanding, we use \rot{the following explicit approximations of the variances (see Appendix, Sec.~\ref{sec:App:Variance and power spectral density in the stationary state}):
\begin{equation}
    \label{eq:variances_first_approximation}
    \var{\bar h}\equiv \sigma_{11} \approx  \frac{\frac{-w\mu_0}{2\tau N}+\frac{\sigma_{\text{ext}}^2}{2}}{(1 - w  F_h)},\quad \var{r}\approx F_h^2\var{\bar h}.
\end{equation}
In the last approximation, we have used Eq.~\eqref{eq:variance_r_linear_with_finite_size} and the fact that,} for the parameters considered in this study, the variance of $r$ is vastly dominated by the first term in that equation. By contrast, the other terms in Eq.~\eqref{eq:variance_r_linear_with_finite_size} are small enough to be ignored. 
\rot{Equation~\eqref{eq:variances_first_approximation} reveals that an increasing slope $F_h$ decreases the variance of $\bar h$ in an inhibitory network ($w<0$). Intuitively, this effect can be understood by the increase of the effective leakage $(1-wF_h)/\tau$ in the linearized dynamics of $\bar h$ which dampens the input fluctuations. In contrast, the variance of the population rate can increase with increasing slope because the quadratic prefactor $F_h^2$ can overcompensate the decrease of $\var{\bar h}$. To make this argument more clear in the theoretical discussion, we further assume that $-w F_h \gg 1$ (which holds for our parameter setting). Under this assumption, the variance of the population-averaged input potentials becomes inversely proportional to $F_h$,}
\begin{equation}
    \label{eq:sigma11_approximation}
    \var{\bar h} \approx  \lrrund{\frac{\mu_0}{2\tau N} + \frac{\sigmaext^2}{-2w}}\frac{1}{F_h}.
\end{equation}
\rot{In turn, the quadratic factor $F_h^2$ due to the conversion from variance of $\bar h$ to variance of $r$ leads effectively to a linear dependence of $\var{r}$ on $F_h$,}
\begin{equation}
    \label{eq:sigma_rr_approx}
    \var{r} \approx \left(\frac{\mu_0}{2\tau N} + \frac{\sigmaext^2}{-2 w}\right)F_h . 
\end{equation}
The variance formulas \eqref{eq:sigma11_approximation} and \eqref{eq:sigma_rr_approx} share a common factor that represents the sum of two contributions to the variability. The first term of the sum corresponds to the intrinsically generated variability caused by finite-size fluctuations of order $\sqrt{r_0/N}$ in the mesoscopic MF \rot{dynamics}, Eq.~\eqref{eq:meso_main}. We recall that the dependence of these fluctuations on the firing rate $r_0=-\mu_0/w$, arises from the Poisson spiking noise at the microscopic level. Thus, the Poisson property that the variance scales proportionally to the mean causes finite-size fluctuations that increase with the mean stimulus strength $\mu_0$ (see below). We emphasize that this contribution to the variability is a clear finite-size effect that vanishes in the limit $N\to\infty$.
In contrast, the second term of the sum in Eqs.~\eqref{eq:sigma11_approximation} and \eqref{eq:sigma_rr_approx} represents externally generated variability proportional to the variance of the common external stimulus.

For constant $N$, the $p$-dependence in both formulas, Eq.~\eqref{eq:sigma11_approximation} and \eqref{eq:sigma_rr_approx}, is solely contained in the slope $F_h$ of the transfer function.
% Inserting our previous approximations for $\sigma_0^2$, Eq.~\eqref{eq:sigma0-approx}, into Eq.~\eqref{eq:Fh-approx}, this dependence is given by
% \begin{equation}
%     \label{eq:Fh_approx_p}
% F_h\approx 
%     \frac{\phimax Q\lrrund{-\frac{\mu_0}{w\phimax}}}{\sqrt{2\pi\lrrund{\beta^{-2}+\frac{-w\mu_0(1-p)}{2\tau Np}}}}.
%     %\sigma_{rr} \approx -\frac{wr_0}{2\tau N} F_h.
% \end{equation}
% \begin{equation}
%     \label{eq:sigma_rr_approx}
%     \var{r} \approx 
%     \left(\frac{-w r_0}{2\tau N} + \frac{\sigmaext^2}{-2 w}\right)  \sqrt{\frac{-\tau N p} {\pi w\mu_0 (1-p)} } \phimax Q\lrrund{-\frac{\mu_0}{w\phimax}}.
%     %\sigma_{rr} \approx -\frac{wr_0}{2\tau N} F_h.
% \end{equation}
As discussed in Eq.~\eqref{eq:Fh-approx} above, the slope $F_h$ is a monotonously increasing function of $p$. The slope dependence thus explains the increase of the variance of the population rate (Fig.~\ref{fig:variance_vs_p}a left) and the decrease of the variance of the mean input potential (Fig.~\ref{fig:variance_vs_p}a right), respectively. 

In the second scenario with constant $C$ and varying population size $N=C/p$, we observe different dependencies of the variances on $p$ compared to the first scenario with constant $N$ (Fig.~\ref{fig:variance_vs_p}b). In simulations of microscopic networks and 2nd-order MF \rot{dynamics}, the rate variance increases supralinearly  with $p$ and the variance of the mean input potential $\bar h$ exhibits a non-monotonic behavior with a maximum inside the range $0<p<1$. Again, the 1st-order MF theory does not capture these behaviors. Note that there is a good qualitative agreement between 2nd-order MF theory and microscopic network simulations. As in the first scenario, however, the variance of the average input potential in the quenched network is underestimated by the annealed network, which, in turn, is quantitatively well matched by the 2nd-order MF theory.

To gain a theoretical understanding of these observations, we use our analytical approximation  Eq.~\eqref{eq:sigma_rr_approx}. For constant $C$, the approximation  can be rewritten as
% \begin{equation}
%     \var{r}\approx  \left(\frac{\mu_0p}{2\tau C} + \frac{\sigmaext^2}{-2 w}\right)\frac{\phimax Q\lrrund{-\frac{\mu_0}{w\phimax}}}{\sqrt{2\pi\lrrund{\beta^{-2}+\frac{-w\mu_0(1-p)}{2\tau C}}}}.
%     \label{eq:var-r-Cconst}
% \end{equation}
\begin{equation}
    \var{r}\approx\frac{\phimax^2\hat\beta}{\sqrt{8\pi}}\lrrund{\frac{\hat \mu_0p}{\hat\tau C}+\hat\sigma_\text{ext}^2}\hat F_h.
    \label{eq:var-r-Cconst}
\end{equation}
where we have introduced the dimensionless parameters $\hat \mu_0=-\mu_0/(w\phimax )$, $\hat{\sigma}_\text{ext}^2=\sigmaext^2/(w\phimax)^2$, $\hat\beta=-\beta w \phimax$, $\hat\tau=\tau \phimax$, as well as the dimensionless slope of the transfer function
\begin{equation}
    \hat{F}_h := \sqrt{2\pi}\frac{F_h}{\beta \phimax}\approx\frac{Q(\hat\mu_0)}{\sqrt{1+\frac{\hat\beta^2 (1-p)}{2\hat\tau C}\hat\mu_0}}
    \label{eq:Fh-dimensionless}
\end{equation} 
(cf. Eq.~\eqref{eq:Fh-approx}).
The expression for the rate variance in Eq.~\eqref{eq:var-r-Cconst} is a product of two monotonously increasing functions of $p$ -- a linear function (the prefactor of $\hat F_h$) and a strictly convex function ($\hat F_h$). Hence, the variance of the population rate is a monotonously increasing, strictly convex function of $p$ explaining our observation in Fig.~\ref{fig:variance_vs_p}b (left). In contrast, the variance of the mean input potential, Eq.~\eqref{eq:sigma11_approximation}, can be rewritten for constant $C$ and dimensionless parameters as
\begin{equation}
\label{eq:var-hbar-Cconst}
       \var{\bar h} \approx \frac{\sqrt{2\pi}w^2\phimax^2}{2\hat \beta}\left(\frac{\hat\mu_0p}{\hat\tau C} + \hat\sigma_\text{ext}^2\right)\frac{1}{\hat F_h}.
\end{equation}
% \begin{equation}
% \label{eq:var-hbar-Cconst}
%        \var{\bar h} \approx \left(\frac{\mu_0p}{2\tau C} + \frac{\sigmaext^2}{-2 w}\right) \frac{\sqrt{2\pi\left(\beta^{-2} + \frac{\vert w\vert \mu_0 (1-p)}{2\tau C}\right)} } {\phimax Q\left(\frac{\mu_0}{-w\phimax}\right)},
% \end{equation}
%\begin{equation}
%\label{eq:var-hbar-Cconst}
%       \var{\bar h} \approx \frac{\mu_0 \sqrt{2\pi}}{2\tau C \phimax} \frac{p\sqrt{\beta^{-2} + \frac{\vert w\vert \mu_0 (1-p)}{2\tau C}} } {Q\left(\frac{\mu_0}{-w\phimax}\right)},
%\end{equation}
This expression is a product of a monotonously increasing and a monotonously decreasing function of $p$ (the prefactor of $1/\hat F_h$ and $1/\hat{F}_h$, respectively). The product may therefore exhibit a non-monotonic behavior. A closer inspection of Eq.~\eqref{eq:var-hbar-Cconst} reveals a maximum at an intermediate connection probability 
%\begin{equation}
%    p_\text{max}= \frac{2}{3} \left(1+\frac{2\tau C}{\beta^2 \vert w\vert \mu_0}\right),
%\end{equation}
\begin{equation*}
    p_\text{max} = \frac{2}{3}\left(1+\frac{2\tau C}{-w\mu_0}\left(\beta^{-2} - \frac{\sigma_\text{ext}^2}{4}\right)\right)
\end{equation*}
%provided that $4\tau C/(\beta^2(-w)\mu_0)<1$. In the limit of a steep single-neuron transfer function ($\beta\to\infty$) and in the absence of external noise, the maximum is attained at $p_\text{max}=2/3$. Such maximum is indeed verified by simulations  (Fig.~\ref{fig:variance_vs_p}b, right).
in the interval $(0,1)$ if $-2/3 < \frac{2\tau C}{-w\mu_0}\left(\beta^{-2}-\sigmaext^2/4\right)<1/2$. In the limit of a steep single-neuron transfer function ($\beta\to\infty$) and in the absence of external noise, the maximum is attained at $p_\text{max}=2/3$. Such maximum is indeed verified by simulations  (Fig.~\ref{fig:variance_vs_p}b, right).
Furthermore, the variance $\text{var}(\bar{h})$ approaches zero for $p\to 0$ and approaches a non-vanishing value  for $p\to 1$.

In contrast to the 2nd-order theory, the 1st-order MF theory cannot explain the strictly convex and non-monotonic behavior of the variances of the population rates and input potentials, respectively (Fig.~\ref{fig:variance_vs_p}b). In the second scenario with constant $C$, the 1st-order MF theory is obtained from the above formulas by fixing the ratio $p/C$ and letting $p\to 1$. Thus, the common factor $\bigl(\frac{\hat \mu_0p}{\hat\tau C}+\hat\sigma_\text{ext}^2\bigr)$ in Eqs.~\eqref{eq:var-r-Cconst} and~\eqref{eq:var-hbar-Cconst} remains unchanged, whereas the second factor becomes independent of $p$. Because the common factor  is an increasing linear function of $p$, the variances are thus also increasing linear functions, in line with simulations of the 1st-order mean-field model (but in contradiction to microscopic simulations and 2nd-order MF theory).

\subsubsection{Dependence on the mean stimulus strength}

We started our paper with network simulations pointing out marked quantitative and qualitative discrepancies of the mean-connectivity network with respect to the stimulus dependence of the rate variability (Fig.~\ref{fig:intro}). In particular, we observed a strong overestimation of the rate variance by the mean-connectivity network compared to the original quenched network. 
\rot{At first sight, it may appear counter-intuitive that more homogeneous connectivity leads to a stronger variance in the population rate. This apparent paradox results from the fact that in mean-connectivity networks, finite-size fluctuations are `directly' picked up by the steep single-unit transfer function that also determines the population rate, $F(h,\sigma^2)=\phi(h)$ for $\sigma^2=0$, whereas the effect of these fluctuations at the population level is attenuated for networks with heterogeneous connectivity due to due to the flattening of the effective population rate transfer function; the difference in the population rate variance can then be seen from Eq.~\eqref{eq:sigma_rr_approx}. We will make this argument more explicit below. We also note that the large variance of the population rate in the mean-connectivity network (or MF1) can be attributed to an increased high-frequency power (Fig.~\ref{fig:PSD}b).}

Furthermore, for a noisy external stimulus and large network size ($N=50000$), we observed a suppression of the rate variability by an increase in the mean stimulus strength in the quenched network. In contrast, the mean-connectivity network incorrectly predicted an increase in rate variability rather than a suppression.
As we will see now, these observations can also be well explained by our 2nd-order MF theory. 

%To this end, we 
Let us %
study two opposing scenarios: first, we consider a small (``mesoscopic'') network without noise in the external stimulus ($\sigmaext=0$). In this case, the rate variability solely originates intrinsically from the finite-size fluctuations. Second, we consider a large (``macroscopic'') network with a noisy external stimulus ($\sigmaext>0$). In this case, the rate variability is purely externally generated.    
In both cases, we find that the first-order MF theory largely overestimates the variance of the population rate in the quenched and annealed network, whereas the 2nd-order MF theory, Eq.~\eqref{eq:variance_r_linear_with_finite_size}, correctly predicts the magnitude of the rate variance (Fig.~\ref{fig:var_r_with_external_noise_and_Fh}a,b). Because the 1st-order MF \rot{dynamics} is the exact MF \rot{dynamics} for the mean-connectivity network, this large deviation explains our observation in Fig.~\ref{fig:intro}.

Furthermore, in the two different scenarios, we find a different monotonicity of the rate variance if we restrict ourselves to the biologically interesting range where the fixed point is in the convex part of the sigmoidal hazard function, $\bar h_0<0$ and $r_0<\phimax/2$, i.e. below the inflection point of $\phi(h)$ (cf. Fig.~\ref{fig:first_order_and_stability_in_main_text}a). The upper boundary of this range given by the inflection point corresponds to a single-neuron firing rate at half maximum, $\phimax/2$, and hence an external stimulus $\mu_0^*=-w\phimax/2$ (Fig.~\ref{fig:var_r_with_external_noise_and_Fh}a,b right, dashed vertical line).
In the first scenario without external noise, the finite-size induced rate variability increases monotonically for $\mu_0<\mu_0^*$ for both 1st- and 2nd-order MF theory. In contrast, in the second scenario with external noise and without finite-size fluctuations, the variance of the rate shows a non-monotonic behavior with a maximum below $\mu_0^*$ if $p<1$. 
%\red{We also note that the models with full connectivity have a much shorter relaxation time to the stationary values than those with $p<1$.}

These observations can be understood analytically using the approximation Eq.~\eqref{eq:var-r-Cconst},~\eqref{eq:Fh-dimensionless}.
% Introducing the dimensionless parameters $\hat \mu_0=-\mu_0/(w\phimax )$, $\hat{\sigma}_\text{ext}^2=\sigmaext^2/(w\phimax)^2$, $\hat\beta=-\beta w \phimax$, $\hat\tau=\tau \phimax$, as well as the dimensionless slope of the transfer function
% \[\hat{F}_h := \sqrt{2\pi}\frac{F_h}{\beta \phimax}\approx\frac{Q(\hat\mu_0)}{\sqrt{1+\frac{\hat\beta^2 (1-p)}{2\hat\tau Np}\hat\mu_0}}\] 
% (cf. Eq.~\eqref{eq:Fh-approx}), we can rewrite our simplified approximation Eq.~\eqref{eq:sigma_rr_approx} for the rate variance as
% \begin{equation}
%     \var{r}\approx\frac{\phimax^2\hat\beta}{\sqrt{8\pi}}\lrrund{\frac{\hat \mu_0}{\hat\tau N}+\hat\sigma_\text{ext}^2}\hat F_h.
% \end{equation}
Let us first discuss how the factor $\hat F_h$ in Eq.~\eqref{eq:var-r-Cconst} depends on $\hat\mu_0$.
For $p=1$, i.e. in the 1st-order MF theory, $\hat{F}_h$ is exactly given by the function $Q(\hat\mu_0)$ displayed by the dotted line in Fig.~\ref{fig:var_r_with_external_noise_and_Fh}c. This function, and thus $\hat{F}_h$, grows monotonically for $\hat{\mu}_0>0$ until it reaches a maximum at $\hat\mu_0^*=-\mu_0^*/(w\phimax)=0.5$, which corresponds to the upper bound $\phimax/2$ of the biologically relevant dynamical range.  For $p<1$, the factor $\hat F_h$ is the quotient  of the symmetric function $Q$ and an increasing function, which shifts the maximum of $\hat F_h$ to a value $<1/2$ (Fig.~\ref{fig:var_r_with_external_noise_and_Fh}c, solid line). Hence, $\hat F_h$ exhibits a non-monotonic behavior in the biologically relevant range $0<\mu_0<\mu_0^*$. 
Besides the shift of the maximum to smaller values of the mean stimulus strength $\hat \mu_0$, the maximum also decreases when the connection probability is lowered from $p=1$ to smaller values. This decrease is a consequence of dividing the function $Q(\hat\mu_0)$ by the square-root term that is strictly larger than one for $p<1$. Importantly, $\hat F_h$ can decrease by a large factor if the single-neuron transfer function is steep, i.e. $\hat\beta\gg 1$, which explains the drastic decrease of the rate variance from 1st- to 2nd-order MF theory (Fig.~\ref{fig:var_r_with_external_noise_and_Fh}a,b, where $\hat\beta=500$, \rot{corresponding to $\beta=5$mV$^{-1}$ as in Fig.~\ref{fig:intro}}).

The behavior of $\hat F_h$ just described explains the variance of the population rate for an infinitely large network with external noise (second scenario, Fig.~\ref{fig:var_r_with_external_noise_and_Fh}b). 
In fact, the prefactor of $\hat F_h$ in Eq.~\eqref{eq:var-r-Cconst} becomes independent of $\hat\mu_0$ in the limit when $N=C/p\to \infty$, and hence the dependence on $\hat\mu_0$ is fully captured by $\hat F_h$. In particular, the stark difference in magnitude of $\var{r}$ between the 1st- and 2nd-order MF prediction, as well as the qualitative response of the rate variability (suppression or amplification) to increasing stimuli, is explained by how the monotonicity and magnitude of the slope of the population transfer function $\hat F_h$ varies with $p$.

In contrast, in the first scenario when $\sigmaext=0$ and $N=C/p$ is finite, Eq.~\eqref{eq:var-r-Cconst} tells us that the variance of the population rate is proportional to
\[\hat\mu_0\hat F_h=\frac{\hat\mu_0}{\sqrt{1+\frac{\hat\beta^2 (1-p)}{2\hat\tau C}\hat\mu_0}}Q(\hat\mu_0).\]
The right-hand side is a  product of a monotonically increasing function of $\hat\mu_0$ and the function $Q(\hat\mu_0)$, which is maximized at $\hat\mu_0^*=1/2$. Therefore, the maximum of $\hat\mu_0\hat F_h$ is shifted to values $\hat\mu_0>1/2$ for all $p\in[0,1]$ (Fig.~\ref{fig:var_r_with_external_noise_and_Fh}d). As a consequence, the variance of the population rate $\var{r}$ is a monotonically increasing function of the mean stimulus strength in the biologically relevant range $0\le \hat\mu_0\le 1/2$ (Fig.~\ref{fig:var_r_with_external_noise_and_Fh}a right, $\mu_0<50$~mV), and thus an increase of the mean stimulus strength always increases the rate variability in the absence of external common noise (Fig.~\ref{fig:var_r_with_external_noise_and_Fh}a left).

To keep the arguments simple, we studied here two clear-cut cases -- one with and another without external noise.  According to our theory, Eq.~\eqref{eq:sigma_rr_approx}, the rate variance  is a sum of two terms, where each term on its own corresponds to one of these cases. In general, the theory also applies to the mixed case, where the rate variance is the sum of externally and intrinsically induced variability. For example, in the simulations of Fig.~\ref{fig:intro}, we also compared two networks of size $N=1000$ and $N=50000$ but the external common noise was present in both cases. Nonetheless, the mesoscopic case of $N=1000$ with external noise can still be qualitatively understood by the limit case $\sigmaext=0$ because the finite-size fluctuations dominate over the external noise. \rot{Lastly, we remark that as $\hat\mu_0\to0$, e.g.~for strongly inhibition-dominated networks, the population rate variance decreases for both the mean-connectivity network (MF1) and MF2 in both cases as can be seen from Fig.~\ref{fig:var_r_with_external_noise_and_Fh}c,d.}

\section{Conclusions}
\label{sec:conclusions}

In this work, we set out to derive a stochastic population dynamics for \rot{finite-size} networks of spiking neurons that correctly accounts for fluctuations at the mesoscopic scale when the \rot{connectivity is random -- as in the case of non-full connectivity ($p<1$)}. While such regime is biologically highly relevant, to our knowledge it has not yet been systematically investigated within the frameworks of previously developed mean-field theories~\cite{BruHak99,Bru00}. %In particular, the combined effects of quenched disorder and stochastic neuronal firing on the population dynamics have been largely unclear. 
The low-dimensional, stochastic \rot{population} model derived here allows one to study specifically how the combination of quenched disorder and stochastic neuronal firing affects the mesoscopic activity of finite-size neuronal populations. We thus hope to close an important gap in the use cases of simplified descriptions of networks and to provide the foundations for a mean-field modeling framework of neural variability at the mesoscopic population level.
%that in general have proven to be extremely helpful for a deeper understanding of their dynamics.

%For simplicity, we studied random networks of Poisson neurons.
%with fixed in-degree.
%, i.e.~an identical number of presynaptic connections, and chose a specific transfer function (or $f-I$ curve) $\Phi(h)$ that relates the instantaneous firing rate to the intensity or input current $h$. 
Based on a Poisson assumption for neural firing and a ``dynamically annealed'' description of the connectivity, we approximated the recurrent input to each neuron by its (population) mean, a coherent fluctuating part shared among all neurons, and an individually fluctuating contribution reflecting the connectivity disorder. Eventually, we obtained three coupled stochastic differential equations (SDEs) that describe the evolution of the mean input potential $\bar h$ within the population, the input potential's variance $\sigma^2$, as well as a colored noise $\xi$ in the population rate  that vanishes in both limits $N\to\infty$ \blau{and $p\to 1$. 
Our theory thus goes beyond previous, deterministic mean-field dynamics for first- and second-order cumulants \cite{Ama72} or pseudo-cumulants \cite{GoldiV21} that similarly captured the quenched variability of synaptic weights but assumed the macroscopic limit $N\to\infty$.}
%While our specific choice of the transfer function enabled us to analytically evaluate the effect of a finite variance in the input currents on the population rate, the reduction of the spiking network dynamics to a low-dimensional system of coupled SDEs should remain valid for arbitrary choices of the transfer function, albeit probably more difficult to compute. The assumption of fixed in-degree allowed us to assume that the mean input current as well as the strengths of the fluctuating parts are identical among neurons, but in practice we observed that a purely random Erdös-Réniy network shows almost identical statistics as those described in the previous sections for networks with fixed in-degree (see also Appendix~\ref{}).

The analysis of the new ``second-order'' mean-field (MF2) dynamics showed that the quenched connectivity disorder for $p<1$ has a drastic effect on the neural variability, the response to time-dependent stimuli, and the stability of the network in biologically relevant regimes. Specifically, our main findings with regard to the effects of a finite connectivity $0<p<1$ and finite network size $N<\infty$ are the following: (i) A finite variance in the input currents ($\sigma^2>0$) caused by the quenched random connectivity with finite $C=pN$ leads to an effective broadening of the neuron transfer function, but the stationary mean firing rate remains rather unaffected \rot{(Figs.~\ref{fig:first_order_and_stability_in_main_text}a, \ref{fig:Fixpoint_properties}b)}. (ii) In the presence of synaptic delays, the stability properties of fixed points of the dynamics and locations of bifurcations can change considerably. In particular, we showed that the connectivity disorder can stabilize networks in regimes where a ``first-order'' mean-field (MF1) theory corresponding to a mean-connectivity approximation would predict oscillatory dynamics (Fig.~\ref{fig:first_order_and_stability_in_main_text}b,c). (iii) The population-rate response to high-frequency stimuli and (iv) the variance of the population rate are significantly lower than predicted by the MF1 theory (Fig.~\ref{fig:first_order_and_stability_in_main_text}d,e and \ref{fig:variance_vs_p}). Moreover, (v) in the presence of shared noise due to common external input, an increase in the mean external input can actually lead to a \emph{decrease} of the variance of the population rate (Figs.~\ref{fig:intro} and \ref{fig:var_r_with_external_noise_and_Fh}) -- an effect which again is not captured in the MF1 theory.

Overall, we provided simple analytical explanations for the effects of finite connectivity and finite network size in terms of the slope of the population transfer function and the multiplicative character of the finite-size noise. In particular, we found that a decreasing connection probability $p$ lowers the slope $F_h$ through an increase of the membrane potential variance $\sigma^2$ and that, at finite network size $N$, the neural variability has a rate-dependent contribution that is proportional to the product of mean stimulus strength (or firing rate) and the slope of the transfer function (Fig.~\ref{fig:var_r_with_external_noise_and_Fh}d, Eq.~\eqref{eq:sigma_rr_approx}).

Of note, our mesoscopic approach differs from previous theoretical studies that focussed on how quenched random connectivity shapes microscopic dynamics \cite{SomCri88,KadSom15,MusGer19,DahGru19}.
Neural networks with quenched disorder can display a wide repertoire of dynamical states even under homogeneous random connectivity \cite{DahGru19}, as considered here.
In particular, inhibition-dominated networks operating near the edge of chaos produce rich single-neuron dynamics that may support complex computations \cite{HenVog14,DahGru19}.
% A particularly interesting dynamical state can be found in inhibition-dominated networks that operate close to the edge of chaos.
% This dynamics near criticality has been shown to produce rich single-neuron dynamics, which are potentially beneficial for neural computations \cite{HenVog14,DahGru19}.
Importantly, this critical dynamical state refers to the microscopic dynamics, as characterized by broad distributions of pairwise correlations \cite{HenVog14,DahGru19} and neuronal time scales \cite{ShiZer25}. 
The annealed approximation underlying our theory cannot reproduce these microscopic features.
However, this limitation is not problematic for our mesoscopic description of the coarse-grained population activity.
%While the annealed approximation underlying our theory cannot reproduce these microscopic features, this limitation is not problematic for our mesoscopic description of the coarse-grained population activity.
As already noted in \cite{DahGru19}, the mesoscopic population activity is largely unaffected by the critical dynamics at the microscopic scale. % (apart from finite-size effects). 
We have furthermore shown here that, at the mesoscopic scale, the annealed approximation captures surprisingly well both the dynamics and their fluctuations across the full range of connection probabilities.
% The close agreement between 
% %the 
% our 
% MF2 theory 
% %(and thus the annealed approximation) 
% and simulations of quenched networks indicates 
% %that:
% %(i) 
% that fluctuations arise predominantly from spiking noise rather than chaotic rate dynamics, and
% that 
% %(ii) 
% the influence of random connectivity on these fluctuations is well described by an annealed model that preserves connectivity statistics  at any moment in time.

% three main assumptions: 1) annealed network, 2) single population, 3) neurons modeled by simple first-order Poisson models
To derive a simple mean-field model with fluctuations that highlights the effects of finite connectivity and finite size, we made several simplifications. In particular, we used (i) an annealed approximation of the quenched random connectivity, (ii) only one cell type (single population), and (iii) a first-order Poisson model of the spiking-neuron dynamics. In the following, we will briefly discuss these simplifications and possible extensions.

In our ``dynamically annealed'' approximation, we assumed that each presynaptic spike would be randomly distributed among all neurons according to the connection probability $p$, independent of the actual, fixed (quenched) synaptic connectivity matrix of a given network. A systematic comparison of our theory with simulations of quenched networks showed that our MF2 theory captures reliably the behavior of quenched networks \blau{with respect to the main findings described above. Deviations of the annealed network (from the quenched network) and, correspondingly, the MF2 \rot{dynamics}, are noticeable for several statistics, though. However, these deviations were considerably smaller than the respective deviations of the mean-connectivity network and the corresponding MF1 \rot{dynamics} used in previous studies~\cite{DegSch14,SchDeg17,KulRan20}. 

The remaining deviations of the annealed network can most likely be attributed to the neglect of temporal correlations in the recurrent input caused by the annealing, i.e. the incessant resampling of the connections in time. This explanation would be consistent with the underestimation of the low-frequency power by the annealed network which we observed in Fig.~\ref{fig:PSD}. For infinitely large, sparsely connected networks, a proper account of temporal correlations can, in principle, be obtained through a self-consistent treatment of the auto-correlations of the recurrent fluctuations \cite{KadSom15,WieBer15}, which typically adds low-frequency power \cite{WieBer15}. Such self-consistent theory would lead to colored noise rather than white noise for the incoherent fluctuations in Eq.~\eqref{eq:annealed-network-model}. However, how to build a stochastic, dynamical population model for finite-size, random networks that accounts for temporal correlations in a self-consistent manner is an open theoretical problem which goes beyond the scope of the present study. }  %\red{If this explanation of the deviation of the annealed approximation to the quenched network holds true, one might wonder whether applying the annealed approximation to a modified quenched system with less incoherent temporal correlations in the recurrent input will yield a better match. So to speak, if we consider a system with less temporal correlations to begin with, the annealed approximation can destroy less. We expect that introducing a reset to a membrane potential $h_\text{reset}$ upon spiking [...] A full systematic investigation or even finding a mesoscopic theory for the system with reset is beyond the scope of this paper. Nonetheless, as a simple start we show here simulations the networks with reset for the variances at different connection probabilities $p$ (Fig.~\ref{fig:Var_IF}), similar to Fig. YYY in ZZZ without reset. Indeed, we find that the annealed approximation fits well with the variances of the quenched network, much better than for the system without reset (Fig. YYY). Notice that the qualitative behavior of the variance }

%\red{As the deviations between the quenched and the annealed network are possibly explained by the decreased temporal correlations between two individual neurons in the annealed network, one expects a better match for networks with reset. The reset of an leaky integrate-and-fire (LIF) neuron whitens the power spectral density resulting in weaker temporal correlations for the quenched network as well, closing the gap between quenched and annealed. Fig.~\ref{fig:Var_IF} ...}

\begin{figure}
    \centering
\includegraphics[width=\linewidth]{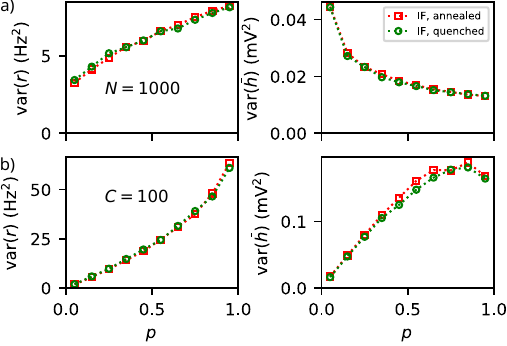}
    \caption{\rot{Variance of $r$ and $\bar h$ for a network of LIF neurons with a rectified linear hazard function $\phi(h)=\beta_\text{lin}[h-\theta_\text{lin}]_+$ and a post-spike reset to $h_\text{reset}$ as a function of connection probability $p$. The LIF network behaves qualitatively similar to the Poisson network without reset in this paper (cf.~Fig.~\ref{fig:variance_vs_p}). Variances are calculated from network simulations of quenched (green) and annealed (red) networks for (a) $N=1000$ and (b) $C=100$ for different connection probabilities $p$. The rectified linear hazard rate was matched both by value and slope of the single neuron transfer function Eq.~\eqref{eq:phi-erf} at a reference point $h_\text{ref} = -0.25$ mV. Parameters: $h_\text{reset}=-15$~mV, $\beta_\text{lin} \approx 91.32$~mV$^{-1}$, $\theta_\text{lin}\approx -0.37$~mV, $w=-1$ mVs, $\mu_0=10$~mV}.}
\label{fig:Var_IF}
\end{figure}

Interestingly, perhaps not surprisingly, the comparison of our theory with simulations of ``dynamically annealed'' networks shows that  it can be considered almost exact for the annealed case. One may ask whether a biological realization of such a ``dynamically annealed'' network, where our theory becomes exact, might actually exist. A possible origin of such randomly distributed spikes might be the probabilistic vesicle release: A fully connected network where synapses would transmit a spike with probability $p$ would be an exact realization of the system we describe by our approximation. 
\blau{Thus, the annealed network model may also be regarded as a simple caricature of a spiking neural network with probabilistic synaptic transmission.
Another type of annealing occurs in networks with full connectivity but dynamic synapses (short-term synaptic plasticity). Recently, such dynamically changing networks have also been reduced to low-dimensional Langevin dynamics based on a similar Poisson neuron model \cite{SchGer20,PieSch22}.

For the sake of a systematic and thorough development of the theory, we here began with the case of one population corresponding to a single neuron type; especially, we focused on the case of an inhibitory network. However, our theory suggests a straightforward extension to multiple populations. Cortical networks typically consist of several cell types and are generally described as excitatory-inhibitory networks that are, in the simplest case, organized as two coupled populations. Rate models of such two-population} E-I networks remain a popular tool to describe network dynamics up to this day, whether to characterize inhibition stabilization in cortical networks~\cite{SanAki20} or spatial patterns of  oscillatory activity observed in motor cortex~\cite{KulRan20,KanRan23}. It will certainly be worthwhile to investigate how the combined effects of \rot{random} connectivity and network size we began to explore here may shape the more complicated dynamics of networks that comprise more than one population. \blau{Furthermore, a multi-population extension 
would offer the possibility to model specific cortical microcircuits, such as canonical circuit models \cite{HahKum22} or cortical column models with multiple interneuron types \cite{BilCai20}, on the mesoscopic scale, which accounts for fluctuations.
%suggests a powerful modeling framework for cortical microcircuits on the mesoscopic scale.
}
%% JR: SanAki20 = https://elifesciences.org/articles/54875
%%     KanRan23 = https://elifesciences.org/articles/81446

%\tilo{extension to more realistic neuronal dynamics such as integrate-and-fire}
\blau{To develop the theory, we used Poisson neurons based on a one-dimensional dynamics for the membrane potential of each neuron. While a Poisson model neglects spike-history effects such as neuronal refractoriness, it allowed us to derive a low-dimensional, closed set of mean-field equations for the first- and second-order cumulants. 
%$\bar{h}(t)$ and $\sigma^2(t)$ and one further fluctuation variable $\xi(t)$. 
We expect that in regimes where the neuron's firing is Poissonian, similar observations can be made in more biophysically grounded models of spiking networks. While this is beyond the scope of this work, it would certainly be worthwhile to systematically investigate the effects described here in simulations of such networks. 

The main difference of our Poisson model compared to more realistic integrate-and-fire (IF) models is the absence of a reset mechanism (in fact, with reset, our model would be equivalent to a leaky IF model with escape noise \cite{Ger00,YuTai22,SchLoe23,Ock23,DumHen24}). While the resetting in IF neurons captures refractoriness, it significantly complicates the population dynamics. Specifically, the population dynamics of IF neurons is usually infinite dimensional, represented by partial differential equations \cite{BruHak99,DumHen24}, integral equations \cite{GerKis14}, infinite systems of SDEs \cite{KniMan96,MatGiu02}, or, in the case of finite $N$, stochastic versions of these \cite{MatGiu02,SchDeg17,SchLoe23,VinBen23}. In principle, it seems possible to carry over our theory to these population equations by again using the same annealed approximation. It would be interesting to study how well the annealed approximation works for neurons with refractoriness.
One may speculate that similar second-order mean-field theories may provide a quantitatively correct account for networks e.g.~of generalized integrate-and-fire neurons in the presence of finite connectivity where effective first-order mean-field theories have already proven to be quantitatively precise for fully connected networks~\cite{SchDeg17,KulRan20,SchLoe23}.

\rot{While a systematic analysis of the effect of  resets is beyond the scope of this paper, we present here some first  simulation results. We consider a quenched network, in which the input potential is reset to $h_\text{reset}$ after each spike. With reset, the single-neuron hazard rate no longer requires an upper bound, and we therefore adopt a rectified linear hazard rate \cite{Ock23}. 
As an example, we repeat the simulations from Fig.~\ref{fig:variance_vs_p} with reset. The results (Fig.~\ref{fig:Var_IF}) show that the variances of the population rate $r$ and mean input potential $\bar h$ as a function of the connection probability $p$  behave qualitatively similarly to the case without reset (Fig.~\ref{fig:variance_vs_p}). Specifically, for fixed $N$, the variance of $r$ increases with $p$, while the variance of $\bar h$ decreases. For fixed $C$, the variance of $r$ increases, and the variance of $\bar h$ exhibits a maximum. Importantly, the annealed approximation now matches the quenched network much more closely  than in the absence of reset. This aligns with our earlier explanation that discrepancies between the quenched network and the annealed approximation arise from neglecting temporal correlations in the incoherent recurrent input due to annealing. The reset reduces the enhanced low-frequency power of the model without reset, effectively whitening the spike-train power spectra and thus reducing temporal correlations in the input. In other words, there are less  temporal correlations to be removed by the annealing, leading to a better fit.}

%In this study, we characterized the effect of finite connectivity for networks of Poisson neurons, where the firing of individual units does not rely on a biophysical description of the neuron dynamics like it is the case for integrate-and-fire models. 
For simulations and concreteness, we made some further specific modeling choices: For example,} we studied random networks with fixed in-degree, i.e.~an identical number of presynaptic connections, and chose a specific transfer function (or $f-I$ curve) $\phi(h)$ that relates the instantaneous firing rate to the intensity or input current $h$.  While our specific choice of the transfer function also enabled us to analytically evaluate the effect of a finite variance in the input currents on the population rate, the reduction of the spiking network dynamics to a low-dimensional system of coupled SDEs should remain valid for arbitrary choices of the transfer function, albeit probably more difficult to compute. The assumption of fixed in-degree allowed us to assume that the mean input current as well as the strengths of the fluctuating parts are identical among neurons, but in practice we observed that a purely random Erd\H{o}s-Réniy network shows almost identical statistics as those described in the previous sections for networks with fixed in-degree.
%(see also Appendix~\ref{App:Fixed in-degree and Erdős-Rényi connectivity are equivalent for an annealed network}).

Here, we showed how to derive low-dimensional stochastic models that describe fluctuations in a complex, biological model system, namely a finite-size spiking neural networks with \rot{random} connectivity.
Already in the simple case of one population and Poissonian spiking, we found highly non-trivial effects such as the non-monotonic dependence of the variance of the recurrent input on connection probability (Fig.~\ref{fig:variance_vs_p}b) and the weak suppression of variability by external stimuli (Fig.~\ref{fig:var_r_with_external_noise_and_Fh}b). Thus, even though a single population and Poisson spiking dynamics may be too simple to quantitatively model certain phenomena in cortex (such as a stronger suppression of variability \cite{ChuByr10}), the fact that our MF2 theory explains non-trivial fluctuation effects demonstrates the power of the approach as a proof of principle. 
Importantly, with the biological extensions 
%to multiple populations and biologically more realistic, spiking-neuron dynamics, as 
discussed above, we believe that our theory may pave the way for a useful modeling framework for variability in cortical circuits at the mesoscopic population level.

%%%%%%%%%%%%%%%%%%%%%%%%%%%%%%%%%%%%%%%%%%%%%%%%%%%%%%%%%%%%%%%%%%%%%%%%%%%%%%%%%%%%%%%%%%%%%%%%%%%%%%

\section{Acknowledgment}
We are grateful to Jakob Stubenrauch for useful comments on the manuscript.

\section{Data availability}

The code that supports the findings of this article is openly
available \cite{GreRan25_repo}.

% \section{Author contributions}

% %J.R. and T.S. conceptualization and supervision. N.G. simulation, data analysis and visualization. N.G., J.R. and T.S. formal analysis. N.G. and T.S. writing - original draft. N.G., J.R. and T.S. writing - review and editing.

% J.R.~and T.S.: conceptualization, methodology and supervision. N.G.: simulation, data analysis and visualization. N.G.~and T.S.: formal analysis, writing - original draft. N.G.,~J.R., and T.S.: writing - review and editing.

% %N.G was jointly supervised by J.R. and T.S..
\appendix

\section{Mesoscopic dynamics in the presence of heterogeneous external drive}
\label{sec:hetero}

An additional source of disorder in the microscopic models is the heterogeneity in the parameters. For Eq.~\eqref{eq:micro_main}, we  include heterogeneity in the external drive as follows: 
\begin{equation}
    \label{eq:muextsplit}
    \mu_{\text{ext},i}(t) = \muext(t)+\hat\mu_i.
\end{equation}
The random variables $\hat\mu_i$ for each neuron $i$ have mean zero and are independently and identically distributed with probability density $\rho(\hat\mu)$. They can be interpreted as heterogeneity of the external drives, the resting potentials or the thresholds $\vartheta$ of the different neurons. Importantly, the variables $\hat\mu_i$ are constant in time, and thus represent quenched disorder.
The derivation of the microscopic annealed network \eqref{eq:annealed-network-model} does not explicitly depend on the value of $\muext(t)$ and we therefore obtain the same results with $\muext(t)$ replaced with Eq.~\eqref{eq:muextsplit}. Specifically, Eq.~\eqref{eq:effective_OU} is modified as follows
\begin{equation*}
    \tau \od{h_i}{t}  = -h_i+\hat\mu_i+f(t)+g(t)\zeta_i(t). \label{eq:effective_OU2}
\end{equation*}
Splitting off the heterogeneous part from $h_i(t)$ yields a new variable $x_i(t)=h_i(t)-\hat\mu_i$ that obeys
\begin{align*}
%    h_i(t) &=: x_i(t) + \mu_i\label{eq:hi_sum_xi_mui}\\ 
    \tau \dot{x}_i &= -x_i + f(t)+g(t)\zeta_i(t).%\label{eq:annealed-separated_heterogeneity}
\end{align*}
The last equation is of the same form as Eq.~\eqref{eq:effective_OU} and can be dealt with accordingly. For the self-consistent closure, the probability density of the input potentials $h_i$ is needed for the calculation of $\lrk{\phi(h_i(t))}$. At each time point we assume $h_i(t)=x_i(t)+\hat\mu_i$ is the sum of two independent random variables and thus its density is the convolution $g_{\bar x, \sigma_x^2}\ast \rho$. For simplicity, we assume that the heterogeneity is Gaussian distributed $\hat\mu_i\sim\mathcal{N}(0, \sigma_\mu^2)$ with a given fixed variance $\sigma_\mu^2$. In this case, we have a Gaussian probability for the distribution $h_i$ with mean $\bar x(t)$ and variance $\sigma_x^2+\sigma_\mu^2$.
We therefore have
\begin{align*}
%\label{eq:meso_with_mu_i}
\begin{aligned}
\tau\od{\bar{x}}{t} &= -\bar{x} + \muext(t) + w\lreckig{r(t-d) + \sqrt{\frac{r(t-d)}{N}}\eta(t)}\\
\tau\od{\sigma_x^2}{t} &= -2\sigma_x^2 + \frac{w^2\left(1-p\right)}{\tau pN} r(t-d)\\
\tau\od{\xi}{t} &= -\xi + \sqrt{2\tau G\left(\bar{x}(t), \sigma^2(t)\right)}\zeta(t)    
\end{aligned}
\end{align*}
where
\begin{align*}
r(t) &= F\left(\bar{x}(t), \sigma^2(t)\right) + \frac{1}{\sqrt{N}} \xi(t)\\
\sigma^2(t) &= \sigma_x^2(t) + \sigma_\mu^2.
\end{align*}
In other words, we obtain the very same mean-field dynamics as in the case without heterogeneity, Eq.~\eqref{eq:meso_main}, but with an increased variance $\sigma^2$. \blau{See also the classical mean-field theory of Amari \cite{Ama72} for a similar treatment of heterogeneity.}
%%The first-order mean-field model (..) can also be extended in the same way where Eq.~\eqref{eq:meso_with_mu_i} is reduced with $\sigma_x^2(t)= 0$ and $\xi(t)= 0$.
%The effects of a heterogeneity in the drive is illustrated in Fig.~\ref{fig:mu_i}, where a sample trajectory is shown. The model Eq.~\eqref{eq:meso_with_mu_i} accurately captures the mean and the variance of the input potentials $h_i$ of the annealed microscopic model with a heterogeneity in the drive. We can contrast that result with the 2n-order mean-field model without the effect of external drive which underestimates the variance and overestimates the mean.

% \begin{figure*}[t]
%   \centering
% \includegraphics[width=\linewidth]{Figure_mu_i.pdf}
%   \caption{Time series of \textbf{a)} the mean input potential $\bar h$ and \textbf{b)} the variance of the input potentials $\sigma^2$ for a heterogenous external drive with $\sigma_\mu^2=4$ mV$^2$. Trajectory in black is the 2nd-order mean-field model which does not account for the heterogeneity. Dotted lines are the respective theoretical values of the fixed points. Parameters: $w=1$ mVs, $p=0.1$, $\mu_0=10$ mV}
%   \label{fig:mu_i}
% \end{figure*}

\section{\rot{Fixed point solutions}}\label{sec:Fixed point solutions}
At equilibrium, \rot{the sparse limit, Eq.~\eqref{eq:meso_sparse_limit},} gives
\begin{equation}
        \label{eq:fixed_points_noiseless}
         \bar h_0 =\mu_0 + wr_0,\quad   \sigma_0^2 =\frac{w^2}{2\tau C}r_0,\quad r_0=F(\bar h_0,\sigma_0^2)
\end{equation}
which yields the fixed-point equation \eqref{eq:fixpoints_one_line} for $h_0$.

\begin{figure*}[t]
  \centering
\includegraphics[width=0.95\linewidth]{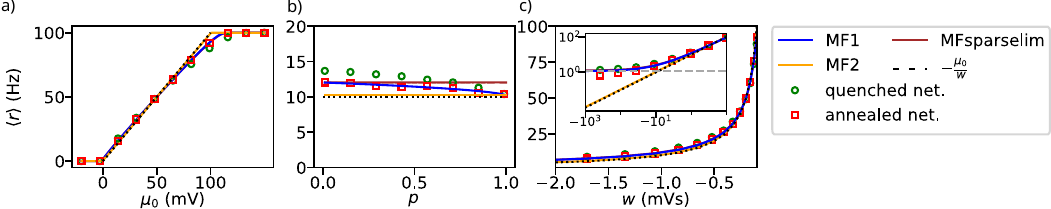}
  \caption{Mean stationary population rate of the quenched and annealed networks and fixed-point solutions of the noiseless mean-field dynamics.  Population averaged firing rates as a function of (a) the external input $\mu_0$, (b) the connection probability $p$ for $C=100$ constant, and (c) the coupling strength $w$. Orange (blue) line: MF1 (MF2), brown line: MFsparselim; red squares (green circles): annealed (quenched) network simulation; dotted line: analytical approximation \eqref{eq:approximate_form_fixedpoints_r}. The inset in (c) shows the same data, but for a larger range of $w$ in a double logarithmic scaling. Gray dashed line: 2nd-order theory $r_\infty$ for the firing rate  in the limit $w\to-\infty$ (Eq.~\eqref{eq:rate-inf}).  Note: MFsparselim is identical to MF2 in (a) and (c). Parameters: $N=1000$, $\beta=5$~mV$^{-1}$ (a) $p=0.1$ and $w=-1$~mVs, (b) $\mu_0=10$~mV, $w=-1$~mVs, (c) $p=0.1$, $\mu_0=10$~mVs.}
  \label{fig:Fixpoint_properties}
\end{figure*}

Because we assume an inhibitory network ($w<0$), the linear function on the left-hand side of Eq.~\eqref{eq:fixpoints_one_line} has a negative slope. If we choose $w$ such that the intersection occurs for values $\bar h_0$ lower than the inflection point $\vartheta=0$ of the hazard function $\phi(h)$ (corresponding to the biologically relevant convex part of $\phi(h)$), then from the graphical solution it is clear that the 2nd-order \rot{MF dynamics} exhibits a fixed-point with lower mean input potential $\bar h_0$ and higher population firing rate $r_0$ compared to the 1st-order MF \rot{dynamics}. This effect %is 
becomes 
stronger 
%the 
for smaller connection probability 
%$p$ is.
$p$.

The mean stationary firing rates of the quenched and annealed microscopic models are well captured by the numerical solution of the fixed-point equation \eqref{eq:fixpoints_one_line} \rot{(Fig.~\ref{fig:Fixpoint_properties}a-c)}. As a function of the external drive $\mu_0$, we observe a strikingly linear dependence between the saturation regimes at low and high values of $\mu_0$ \rot{(Fig~\ref{fig:Fixpoint_properties}a)}. 
%, we see that the fixed point values of the MF models follow the simulation results of the microscopic models with only minor deviations. The 2nd-order MF model fits perfectly on the network with dynamic rewiring, the 1st-order MF model less well. We notice that for $\mu_0\in[0;100]$ mV all models and network simulations have approximately a linear behavior. This behavior is most pronounced in the 1st-order MF model. For $\mu_0<0$, the firing rate is zero as the inhibitory network with negative drive is silent and for $\mu_0>100$ mV the firing rate is $\phimax= 100$ Hz as ...
Furthermore, we find that the firing rate is roughly constant when the connection probability $p$ is varied while $N$ is fixed \rot{(Fig~\ref{fig:Fixpoint_properties}b)}. The weak dependence on $p$ is slightly better predicted by the 2nd-order MF theory than by the 1st-order MF theory, which, by definition, has no dependence on $p$. 
%The 1st-order model has no explicit dependency on the connection probability $p$ and the fixed point of the firing rate is therefore indifferent of $p$ if $N$ is set constant . The 2nd-order MF model and the microscopic network simulations do depend on $p$, albeit the mean stationary activity only weakly decreases with increasing $p$. Again, 2nd-order MF model fits perfectly with the simulations of the network with dynamic rewiring. 
We also observe a power-law behavior of the mean firing rate in both microscopic simulations and fixed-point solutions if the coupling strength is not too strong \rot{(Fig.~\ref{fig:Fixpoint_properties}c)}. 
We note that, in general, the relative error of the 1st-order MF \rot{dynamics} (and  hence the mean-connectivity network) with respect to the stationary mean population rate is rather small compared to the much larger error  with respect to the second-order statistics reported in Fig.~\ref{fig:intro}. 

\subsection{Analytical approximation of fixed points}
The linear dependence on $\mu_0$, the weak dependence on $p$ and the power-law dependence on $w$ can be explained with our MF theory as follows.
Firstly, we notice that for a sufficiently large steepness $\beta\gg 1/\sigma_0^2$ of the single-neuron transfer function $\phi$, the 2nd-order MF dynamics remains practically unaltered if we take the limit $\beta\to\infty$. In this limit, the single-neuron transfer function tends to $\phimax\theta(\bar h)$, where $\theta$ denotes the Heaviside step function. Indeed, for the values of $\beta$ considered in this paper, taking the limit $\beta\to\infty$ does not noticeably alter the effective nonlinearity $F$ and hence the 2nd-order MF dynamics. The limiting value of the fixed point can be calculated analytically from the graphical representation  \rot{(Fig.~\ref{fig:first_order_and_stability_in_main_text}a)} which yields an approximate analytical expression for the firing rate in the 1st-order MF \rot{dynamics}, %if $\vert w\vert$ is not too large: 
\begin{equation}
\label{eq:approximate_form_fixedpoints_r_app}
    r_0\approx\begin{cases}
        0,&\quad \mu_0 \le 0\\
        -\frac{\mu_0}{w},&\quad 0 < \mu_0 < -w\phimax\\
        \phimax,&\quad \mu_0\ge -w\phimax.
    \end{cases}
\end{equation}
This analytical expression predicts \rot{that $r_0$ is inversely proportional to the} coupling strength in the balanced regime $0<\mu_0<-w\phimax$. Indeed, the numerical fixed-point solution for the 1st-order MF \rot{dynamics} exhibits an excellent quantitative agreement with this prediction \rot{(Fig.~\ref{fig:Fixpoint_properties}c)}.

Secondly, we note from the geometry of the graphical solution of the fixed points that the intersection point of the 1st-order MF \rot{dynamics} serves as a rough approximation for the firing rate $r_0$ of the 2nd-order MF \rot{dynamics} as well \rot{(Fig.~\ref{fig:first_order_and_stability_in_main_text}a)} unless $w$ is extremely negative as discussed below. 
Indeed, the horizontal stretching of the transfer function in the 2nd-order theory mainly affects the value of the fixed point $\bar h_0$ rather than the firing rate $r_0$. More precisely, a sufficient condition for the validity of our approximation is that $-\mu_0/w\gg (1-p)/(\tau C)$ and $-\mu_0/w> 0.023\phimax$ (see Appendix \ref{sec:App:Stationary mean population activity}).
We expect that the approximation~\eqref{eq:approximate_form_fixedpoints_r} slightly underestimates the firing rates if the intersection is below the inflection point ($\bar h_0<0$) and slightly overestimates the firing rates if the intersection is above the inflection point ($\bar h_0>0$). 
Our approximation Eq.~\eqref{eq:approximate_form_fixedpoints_r} well explains the piecewise linear behavior of the firing rate as a function of the external input as observed in \rot{Fig.~\ref{fig:Fixpoint_properties}a}. Furthermore, Eq.~\eqref{eq:approximate_form_fixedpoints_r} also explains why the firing rates are roughly independent of the connection probability $p$ \rot{(Fig.~\ref{fig:Fixpoint_properties}b)}, and why we see the power-law $w^{-1}$ even in non-fully connected networks \rot{(Fig.~\ref{fig:Fixpoint_properties}c)} unless $w$ is extremely negative (this case is treated below).
In general we note that, compared to the 1st-order \rot{MF dynamics}, the 2nd-order MF \rot{dynamics} shows a better agreement with the firing rate in microscopic simulations, in particular, it fits the annealed network almost perfectly.

Because closed-form analytical approximation of the fixed points will be crucial for a qualitative---if not quantitative---discussion of the linear analysis around fixed points in the following sections, we provide here also approximations for the fixed point values $\bar h_0$ and $\sigma_0^2$. Using our approximation for the firing rate, Eq.~\eqref{eq:approximate_form_fixedpoints_r}, we obtain from Eq.~\eqref{eq:fixed_points_noiseless} an approximation for the dispersion: 
\begin{equation}
\label{eq:sigma0-approx_app}
    \sigma_0^2 \approx -\frac{w\mu_0 (1-p)}{2\tau N p}
\end{equation}
if $0<\mu_0<-w\phimax$. To obtain a corresponding approximation for $\bar h_0$ in the 2nd-order MF theory, we solve the fixed point equation $r_0=F(\bar h_0,\sigma_0^2)$ (cf. Eq.~\eqref{eq:fixed_points_noiseless}) for $\bar h_0$  using our approximations for $r_0$ and $\sigma_0^2$ resulting in
\begin{equation}
    \label{eq:approximate_h_app}
    \bar h_0 \approx \Phi^{-1}\left(-\frac{\mu_0}{w\phimax}\right)\sqrt{\beta^{-2}+\sigma_0^2}.
\end{equation}

\subsection{Limit of strongly inhibitory networks}
Finally, the regime of strong inhibitory coupling reveals a qualitative difference between the the 1st- and 2nd-order mean-field prediction. In the limit  $w\to-\infty$, the right-hand side of Eq.~\eqref{eq:fixpoints_one_line} is asymptotic to 
\[\phimax\Phi\left(-\sqrt{\frac{2\tau pN(\bar{h}_0-\mu_0)}{(1-p)w}}\right),\] and thus the mean stationary firing rate  in a strongly inhibitory network, is given by the solution $r_\infty$ of the limiting equation
\begin{equation}
  \label{eq:rate-inf}
  r_\infty =\phimax \Phi\left(-\sqrt{\frac{2\tau pNr_\infty}{1-p}}\right).
\end{equation}
The graphical solution of this equation clearly shows a vanishing firing rate in the 1st-order MF theory ($p\to 1$), whereas the limiting equation has a non-vanishing solution for $p<1$ (2nd-order MF theory, \rot{Fig.~\ref{fig:Fixpoint_properties}c}, inset). 
%In the limit $p\to 1$ or in the limit $N\to\infty$ at fixed $p>0$, this relation implies that $r_0\to \\phimax\Phi(-\infty) =0$. Thus, in the case of full connectivity or in the dense limit, the firing rate vanishes for a strongly inhibitory networks. The fixed point equations for the 2nd-order mean-field theory with $p=1$ is equivalent to the fixed point equation for the 1st-order mean-field theory. 
The non-vanishing, limiting firing rate in the 2nd-order MF theory is in good agreement with simulations of the annealed and quenched microscopic model. In contrast, the 1st-order MF theory generally underestimates the firing rates in strongly inhibitory networks: for $w\to-\infty$, the 1st-order MF \rot{dynamics} has a rate that converges to zero like $-\mu_0/w$ as predicted by Eq.~\eqref{eq:approximate_form_fixedpoints_r}.

\subsection{Condition for the validity of the firing-rate approximation~\eqref{eq:approximate_form_fixedpoints_r}}
\label{sec:App:Stationary mean population activity}

In Eq.~\eqref{eq:approximate_form_fixedpoints_r}, a simple approximation for the fixed-point firing rate is given, namely $r_0\approx -\mu_0/w$, which we used for an inhibitory network ($w<0$) in the non-saturated regime ($0<\mu_0\le-w\phimax$). This approximation corresponds to the solution of the 1st-order MF theory in the limit $\beta\to\infty$. We argued that under certain sufficient conditions, the approximation is good for the 2nd-order MF \rot{dynamics} as well. Here we provide a justification for these conditions. For biological reasons, we make a slightly stronger assumption on the working point of our system: we assume that the fixed point is in the convex part of the transfer function, i.e. $h_0<0$ and $r_0<\phimax/2$,  as drawn in \rot{Fig.~\ref{fig:first_order_and_stability_in_main_text}a}. 
%Second, we assume that the fixed point is not located in the tail of the transfer function $F$ corresponding to negligibly small firing rates. Specifically, we require that $h_0>-a\sigma_0$ with $a=2$.

For the firing-rate approximation to be applicable also to the 2nd-order MF \rot{dynamics}, we require that the relative difference between the firing rates of the 1st- and 2nd-order MF theory is small:
\begin{equation}
\label{eq:rate-error}
    r_0^{\text{2nd}} - r_0^{\text{1st}} \ll r_0^{\text{1st}}.
\end{equation}
Here, $r_0^{\text{1st}}=-\mu_0/w$ and $r_0^{\text{2nd}}=F(h_0,\sigma_0^2)$. Furthermore, we have used that $r_0^{\text{2nd}}>r_0^{\text{1st}}$ because of our assumption, $h_0<0$ \rot{(Fig.~\ref{fig:first_order_and_stability_in_main_text}a)}. To estimate the left hand side of \eqref{eq:rate-error} from above, we want to obtain an upper bound approximation of the rate $r_0^{\text{2nd}}=(h_0-\mu_0)/w$. To this end, we lower bound the input potential $h_0$ (note that $w<0$). Because $h_0$ is expected to be on the order of the width $\sigma_0$ of the transfer function, we write the lower bound as $-a\sigma_0<h_0$, where $a$ is a positive number that is still unknown. This lower bound of $h_0$ yields an upper bound estimate for the mean firing rate: $r_0^{\text{2nd}}<(-a\sigma_0-\mu_0)/w$. Hence, the condition $\eqref{eq:rate-error}$ is surely met if
\begin{equation*}
    \frac{-a\sigma_0 - \mu_0}{w} - \frac{-\mu_0}{w} = -a\frac{\sigma_0}{w} \ll -\frac{\mu_0}{w},
\end{equation*}
hence $a\sigma_0 \ll \mu_0$. Unfortunately, we do not know the exact value of $\sigma_0$ for the 2nd-order model. However, if our approximation holds true, we can self-consistently use our approximation for $\sigma_0^2$, Eq.~\eqref{eq:sigma0-approx}, and the condition becomes
\begin{equation}
    \label{eq:App:Approx-condition-1}
    a \sqrt{\frac{-w(1-p)}{2\tau\mu_0 C}} \ll 1.
\end{equation}
The larger the value of $a$, the more conservative this sufficient condition becomes. However, to obtain a less conservative condition, we could lower the value of $a$, as long as $h_0>-a\sigma_0$. We check for this latter condition in the same manner as above, by self-consistently using the approximation of $h_0$ in Eq.~\eqref{eq:approximate_h}:
\begin{align}
    h_0 \approx \Phi^{-1}\left(\frac{-\mu_0}{w\phimax}\right)\sigma_0 &> - a\sigma_0\ ,\nonumber
\end{align}
where we have used the limit $\beta\to\infty$. Eventually, this leads to
\begin{align}
    \frac{-\mu_0}{w\phimax} &> \Phi(-a).\label{eq:App:Approx-condition-2}
\end{align}
The last step was possible, because $\Phi$ is strictly monotonously increasing. If there exists $a>0$ such that both Eq.~\eqref{eq:App:Approx-condition-1} and Eq.~\eqref{eq:App:Approx-condition-2} are fulfilled, we consider the approximations for the fixed points to be valid. For the parameter choice used in our simulations, $a=2$ was sufficient. The numerical evaluation of the error function yields $\Phi(-2)\approx 0.023$. Fixing $a=2$ yields the sufficient condition reported in Sec.~\ref{sec:main:Stationary mean population activity}.

\section{Estimation of the rate variance from population activities}
\label{Sec:App:Estimation of the rate variance from population activities}

The variance of the population firing rate is not directly accessible from measurements in real biological networks. However, we can estimate the rate variance by the variance of the empirical population activity. We assume that in small time bins $(t,t+\Delta t]$, neurons fire spikes independently with  conditional intensities $\lambda_i(t)$ given the past. Then, from the law of total variance and the conditionally Poisson statistics, we have for the variance of the total number of spikes in a time bin
\begin{align*}
    \var{\Delta Z}&=\lrk{\var{\Delta Z\vert \{\lambda_i\} }}+\var{\lrk{\Delta Z\vert \{\lambda_i\}}}\\
    &=N\lrk{r}\Delta t+N^2\var{r}\Delta t^2,
\end{align*}
where $r(t)=N^{-1}\sum_{i=1}^N\lambda_i(t)$ is the population firing rate and $\lrk{\cdot}$ denotes the trial average.
Hence,
\begin{align*}
    \var{r}&=\frac{\var{\Delta Z}}{N^2\Delta t^2}-\frac{\lrk{r}}{N\Delta t}\\
    &=\var{A_N(t,\Delta t)}-\frac{\lrk{A_N(t,\Delta t)}}{N\Delta t}
\end{align*}
where we used the definition of the empirical population activity, Eq.~\eqref{eq:empir-popact}. For stationary data, the trial average can be replaced by the time average.

\section{Partial derivatives of the function $F$ and approximations thereof}
\label{sec:App:approx_Fh_Fs}
Using the definition of the function $F$ defined in Eq.~\eqref{eq:F_formula}, we find for the partial derivative with respect to $h$ at the fixed point $(\bar h_0, \sigma_0^2)$
\begin{align*}
    F_h&:=\pd{F}{h}(\bar h_0, \sigma_0^2+\sigma_\mu^2)
    = \frac{\phimax \beta\exp\left[-\frac{\beta^2h_0^2}{2\left(1+\beta^2(\sigma_0^2+\sigma_\mu^2)\right)}\right]}{\sqrt{2\pi\left(1+\beta^2(\sigma_0^2+\sigma_\mu^2)\right)}}.
\end{align*}
Note that $\sigma_\mu^2=0$ in the absence of heterogeneity of $\mu_i$.
Similarly, for the partial derivative with respect to the second argument, we find
\begin{align*}
    F_\sigma &:= \pd{F}{\sigma^2}(\bar h_0, \sigma_0^2+\sigma_\mu^2)\nonumber\\
    &= \frac{-\phimax \beta^3h_0\exp\left[-\frac{\beta^2h_0^2}{2(1+\beta^2(\sigma_0^2 + \sigma_\mu^2))}\right]}{2\sqrt{2\pi}\sqrt{1+\beta^2(\sigma_0^2+\sigma_\mu^2)}^3}
\end{align*}
The approximate expressions Eq.~\eqref{eq:approximate_h} and  Eq.~\eqref{eq:sigma0-approx} for $h_0$ and $\sigma_0^2$ at the fixed point 
%, which are valid under the assumption formulated in Sec.~\ref{sec:App:Stationary mean population activity}, 
can be used for further simplifications of the partial derivatives: 
\begin{align*}
    %\label{eq:App:Fh}
    F_h \approx \frac{\phimax\beta Q\left(-\frac{\mu_0}{\phimax w}\right)}{\sqrt{2\pi}\sqrt{1+\beta^2\left(\sigma_0^2 +\sigma_\mu^2\right)}},
\end{align*}
where $\sigma_0^2$ and the function $Q$ are given by Eq.~\eqref{eq:sigma0-approx} and Eq.~\eqref{eq:Q-func}, respectively.
Analogously, we can approximate the partial derivative with respect to $\sigma^2$ at the fixed point:
\begin{align*}
    %F_\sigma &\approx \frac{-\phimax\beta^3hQ\left(\frac{-\mu_0}{\phimax w}\right)}{2\sqrt{2\pi}\sqrt{1+\beta^2(\sigma_0^2+\sigma_\mu^2)}^3}\\
    F_\sigma\approx\frac{-\phimax\beta^2\Phi^{-1}\left(\frac{-\mu_0}{\phimax w}\right)Q\left(\frac{-\mu_0}{\phimax w}\right)}{2\sqrt{2\pi}\left(1+\beta^2(\sigma_0^2+\sigma_\mu^2)\right)}.%\label{eq:App:Fs}
\end{align*}

\section{Linear stability analysis}\label{sec:App:Stabilization of the asynchronous state}

For the linear stability analysis, we consider the linearized system with no noise and constant $\mu(t)=\mu$ which reads 
\begin{subequations}
\label{eq:App:meso_linearized_for_stability}
\begin{align}
    \tau\od{}{t}\delta h(t) &= -\delta h(t) + w\delta r(t-d)\label{eq:App:meso_linearized_for_stability_h}\\
    \tau\od{}{t}\delta\sigma^2(t) &= -2\delta\sigma^2(t) + \frac{w^2(1-p)}{\tau C} \delta r(t-d)\label{eq:App:meso_linearized_for_stability_sigma}\\
    \delta r(t) &= F_h \delta h (t) + F_\sigma \delta \sigma^2(t)\label{eq:App:meso_linearized_for_stability_r},
\end{align}
\end{subequations}
where we neglected the variable $\xi$ as it decays exponentially. To solve the linearized system, we insert the exponential ansatz
\begin{align*}
\begin{aligned}
    \delta h(t)=\hat{h}(\lambda)e^{\lambda t}\\
    \delta \sigma^2(t) = \hat{\sigma}(\lambda)e^{\lambda t},
\end{aligned}
\end{align*}
with a constant parameter $\lambda$. Solving Eq.~\eqref{eq:App:meso_linearized_for_stability_sigma} for $\hat\sigma$ yields
\begin{align}
    \hat{\sigma} = \frac{w^2BF_he^{-\lambda d}}{\lambda \tau +2 - w^2 B F_\sigma e^{-\lambda d}}\hat{h},\quad B:= \frac{1-p}{\tau C}.\label{eq:App:hat_sigma_of_hat_h}
\end{align}
We insert this expression for $\hat{\sigma}$ into the first equation~\eqref{eq:App:meso_linearized_for_stability_h} and obtain an equation for the eigenvalue 
\begin{align}
\label{eq:hopf-condition}
    \lambda \tau = -1 + wF_h e^{-\lambda d} + \frac{\hat{F}_\sigma wF_h e^{-2\lambda d}}{2+\lambda \tau -\hat{F}_\sigma e^{-\lambda d}}
\end{align}
where we used the abbreviation $\hat{F}_\sigma:= w^2BF_\sigma$. This equation depends on the fixed points through the functions $F_h$ and $F_\sigma$. To find an oscillatory instability, we look for a Hopf bifurcation, at which $\lambda = i\omega$ for some real-valued frequency $\omega$. Applying the condition on the complex-valued equation (\ref{eq:hopf-condition}) we obtain two conditions, one for the amplitude 
\begin{equation}
\label{eq:ampl-cond}
    1 = \frac{4+\omega^2\tau^2}{1+\omega^2\tau^2}\frac{w^2F_h^2}{(2-\cos(\omega d)\hat{F}_\sigma)^2+(\omega\tau+\sin(\omega d)\hat{F}_\sigma)^2}
\end{equation}
and one for the phase
\begin{align*}
    \arctan(\omega\tau) &= \arctan\left(\frac{\omega\tau}{2}\right) +\arg(wF_h) -\omega d+2\pi k\nonumber \\ 
    &-\atantwo\left(2-\hat{F}_\sigma\cos(\omega d), \omega \tau + \hat{F}_\sigma\sin(\omega d)\right).
\end{align*}
Both must be satisfied simultaneously.
%\begin{align}
%    &1 = S(\omega)\nonumber \\
%    &:= \frac{4+\omega^2\tau^2}{1+\omega^2\tau^2} \frac{w^2F_h^2}{(2-\cos(\omega d)\hat{F}_\sigma)^2+(\omega\tau+\sin(\omega d)\hat{F}_\sigma)^2}\\
%   & \arctan(\omega\tau) = \arctan\left(\frac{\omega\tau}{2}\right) +\arg(wF_h) -\omega d\nonumber \\ 
%    &-\atantwo\left(2-\hat{F}_\sigma\cos(\omega d), \omega \tau + \hat{F}_\sigma\sin(\omega d)\right) +2\pi k.
%\end{align}
%\begin{align*}
%    1 = \frac{4+\omega^2\tau^2}{1+\omega^2\tau^2}\frac{\cdot w^2F_h^2}{(2-\cos(\omega d)\hat{F}_\sigma)^2+(\omega\tau+\sin(\omega d)\hat{F}_\sigma)^2}\\
%    \arctan(\omega\tau) &= \arctan\left(\frac{\omega\tau}{2}\right) +\arg(wF_h) -\omega d+2\pi k\nonumber \\ 
%    &-\atantwo\left(2-\hat{F}_\sigma\cos(\omega d), \omega \tau + \hat{F}_\sigma\sin(\omega d)\right).
%\end{align*}
Here, $\atantwo$ is the two-argument arctangent and $k\in\N_0$. In the $(d,w)$-parameter space, these conditions correspond to a family of curves, one curve for each $k$. At the $k$-th curve an oscillatory perturbation with frequency $\omega_k$ becomes unstable. For the overall instability boundary it is sufficient that at least one mode becomes unstable. We found empirically that the overall instability boundary is given by the mode $k=0$, whereas higher modes yield regions of instability that are contained already in the instability region given by $k=0$.
 For the solution of both conditions a numerical solution of the fixed point equations is necessary. In \rot{Fig.~\ref{fig:first_order_and_stability_in_main_text}b} we scan through each parameter point and solve the phase condition for $\omega$. In the investigated parameter range we were always able to find a unique solution for the phase condition for each mode. We then determined the boundary in the $(d,w)$-space that separated the regions where the right-hand side of the amplitude condition, Eq.~\eqref{eq:ampl-cond}, is larger and smaller than $1$, respectively. The equations are 
 simpler %more simple 
 for the 1st-order MF \rot{dynamics} where $\hat{F}_\sigma=0$. The simplified amplitude condition reads in this case
\begin{align*}
    1 = \frac{w^2F_h^2}{1+\omega^2\tau^2}\quad\Leftrightarrow\quad
    \omega = \pm \frac{\sqrt{w^2F_h^2-1}}{\tau}.
\end{align*}
This explicit solution for $\omega$ can be used in the phase condition to determine the value of the delay at instability boundary for a given coupling strength $w$:
\begin{align*}
    d=\frac1{\omega} \left(-\arctan(\omega\tau)+\arg(wF_h)+2\pi k\right).
\end{align*}

\section{Susceptibility matrix}
Here, we present explicit expressions for the susceptibility matrix $\tilde{\Mat{\chi}}$ defined in Eq.~\eqref{eq:chi}. Using Cramer's rule we obtain
\begin{subequations}
\label{eq:App:chi_expicit}
\begin{align}
    \tilde{\Mat{\chi}}_{11} &= \frac1{\mathcal{D}}\left(i\omega+\frac{2}{\tau}-F_\sigma\tilde\beta e^{-i\omega d}\right)\\
    \tilde{\Mat{\chi}}_{12} &= \frac1{\mathcal{D}}F_\sigma\tilde\alpha e^{-i\omega d}\\
    \tilde{\Mat{\chi}}_{13} &= \frac1{\mathcal{D}}\tilde\gamma\tilde\alpha e^{-i\omega d} \frac{i\omega + \frac2{\tau}}{i\omega +\frac{1}{\tau}}\\
    \tilde{\Mat{\chi}}_{21} &= \frac{1}{\mathcal{D}}F_h\tilde\beta e^{-i\omega d}\\
    \tilde{\Mat{\chi}}_{22} &= \frac{1}{\mathcal{D}}\left(i\omega+\frac{1}{\tau}-F_h\tilde\alpha e^{-i\omega d}\right)\\
    \tilde{\Mat{\chi}}_{23} &= \frac{1}{\mathcal{D}}\tilde\gamma\tilde\beta e^{-i\omega d}\\
     \tilde{\Mat{\chi}}_{31} &=  \tilde{\Mat{\chi}}_{32} = 0\\
     \tilde{\Mat{\chi}}_{33} &= \frac{1}{\mathcal{D}}\left(i\omega + \frac{2}{\tau}\right)
      -\frac{1}{\mathcal{D}} F_\sigma\tilde\beta e^{-i\omega d}\\
      &- \frac{1}{\mathcal{D}}\frac{i\omega+\frac{2}{\tau}}{i\omega+\frac{1}{\tau}}F_h\tilde\alpha e^{-i\omega d}
\end{align}
\end{subequations}
\commentout{
\begin{multline}
    \tilde \chi(\omega)= \frac1{\mathcal{D}(\omega)}\cdot\\
    \PM{i\omega+2/\tau& F_\sigma\hat\alpha e^{-i\omega d} & \gamma\hat\alpha e^{-i\omega d} \frac{i\omega +2/\tau}{i\omega +\frac{1}{\tau}}\\[2pt]-F_\sigma\hat\beta e^{-i\omega d}&&\\[8pt]
    F_h\hat\beta e^{-i\omega d}&i\omega+\frac{1}{\tau}&\hat\gamma\hat\beta e^{-i\omega d}\\&-F_h\hat\alpha e^{-i\omega d}&\\[8pt]
    &&i\omega + \frac{2}{\tau}\\[2pt]0&0&-F_\sigma\hat\beta e^{-i\omega d}\\[2pt]&&-\frac{i\omega+\frac{2}{\tau}}{i\omega+\frac{1}{\tau}}F_h\hat\alpha e^{-i\omega d}}
\end{multline}
}
with the abbreviations
\commentout{
\begin{multline}
    \mathcal{D}(\omega) = \left(i\omega+\frac1{\tau}\right)\left(i\omega+\frac2{\tau}\right)\\
     -\left(i\omega+\frac{2}{\tau}\right)F_h\frac{w}{\tau}e^{-i\omega d}-\left(i\omega+\frac1{\tau}\right)F_\sigma w^2\frac{1-p}{\tau^2 C}e^{-i\omega d}.
\end{multline}
and
}
\begin{subequations}
\begin{align*}
    \mathcal{D}(\omega) &= \left(i\omega+\frac1{\tau}\right)\left(i\omega+\frac2{\tau}\right)
     -\left(i\omega+\frac{2}{\tau}\right)F_h\tilde\alpha e^{-i\omega d}\nonumber\\
     &\qquad-\left(i\omega+\frac1{\tau}\right)F_\sigma \tilde\beta e^{-i\omega d}\\
    \tilde\alpha &= \frac{w}{\tau},\quad
    \tilde \beta = \frac{w^2}{\tau^2}\left(\frac1{C}-\frac1{N}\right), \quad
    \tilde \gamma = \frac{1}{\sqrt{N}}.
\end{align*}
\end{subequations}

\section{Variance and power spectral density in the stationary state}
\label{sec:App:Variance and power spectral density in the stationary state}
With the susceptibility we can express the power spectral density of the system. From Eq.~\eqref{eq:tildeX}, we obtain
\begin{equation*}
    \left\langle\tilde{\Vect{X}}(\omega)\tilde{\Vect{X}}^*(\omega)\right\rangle = \tilde{\Mat{\chi}} \left\langle \left(\Mat{B}\tilde{\Vect{\zeta}} +\tilde{\Vect{M}}\right) \left(\tilde{\Vect{\zeta}}^*\Mat{B}^* + \tilde{\Vect{M}}^*\right) \right\rangle\tilde{\Mat{\chi}}^*
\end{equation*}
and with the definition of the power spectral density matrix $\Mat{S}(\omega)$, Eq.~\eqref{eq:psd-def},
\begin{equation*}
    \Mat{S}(\omega) =\tilde{\mat{\chi}}(\omega)\lreckig{\Mat{S}_M(\omega)+BB^*}\tilde{\Mat{\chi}}^*(\omega).
\end{equation*}
Here, $\Mat{S}_M$ is the $3\times 3$  spectral density matrix of the external stimulus. In our case, the external stimulus only acts on the $\bar{h}$-variable and thus $(\Mat{S}_M)_{ij} = S_{\mu\mu} \delta_{1i}\delta_{1j}$ has only one non-zero entry with the power spectral density $S_{\mu\mu}(\omega)$ of the Gaussian stimulus $\mu(t)$. Explicitly,
\begin{multline*}
        \Mat{S}_{ij}(\omega)=\frac1{\tau^2}\left[S_{\mu\mu}(\omega) + \frac{w^2r_0}{N}\right]\tilde{\chi}_{i1}(\omega)\tilde{\chi}_{j1}^*(\omega)\\
    +\frac{2G\left(\bar{h}_0, \sigma_{h, 0}^2\right)}{\tau N}\tilde{\chi}_{i3}(\omega)\tilde{\chi}_{j3}^*(\omega).
\end{multline*}
In the main text, we used this expression to compute the power spectrum of the population activity $A(t)$, given by Eq.~\eqref{eq:S_AA}. Taking the limit $p\to 1$ yields an explicit expression for the corresponding power spectrum of the  1st-order MF theory
\begin{equation}
\label{eq:S_AA_1st}
    S_{AA}\approx \frac{r_0}{N} + \frac{S_{\mu\mu} + w^2\frac{r_0}{N}\phi_h^2+2w\frac{r_0}{N}\phi_h(1-\cos(\omega d)w\phi_h)}{(\omega\tau-\sin(\omega d)w\phi_h)^2+(1-\cos(\omega d)w\phi_h)^2}.
\end{equation}
Here, $\phi_h$ is the partial derivative $\partial\phi/\partial h$ at the fixed point.

In principle, the stationary variance of $h(t)$ can be computed from the integral over the power spectral density $S_{11}(\omega)$. However, the integral is difficult to solve for non-zero transmission delay and general spectral statistics $\Mat{S}_M(\omega)$ of the stimulus $\mu(t)$. If we restrict ourselves to the case of zero delay, $d=0$, and a Gaussian white noise stimulus, $\mu(t)=\mu_0+\sqrt{\tau\sigmaext^2}\hat\zeta(t)$, the linearized sytem can be rewritten as
\begin{align}
        \od{}{t}\Vect{X} = \Mat{\Gamma}\Vect{X}+\PM{\sqrt{\frac{\sigmaext^2}{\tau}}\hat{\zeta}(t)\\0\\0}+\PM{\frac{w}{\tau}\sqrt{\frac{r_0}{N}}\eta(t)\\0\\\sqrt{\frac{2G_0}{\tau}}\zeta(t)},\label{eq:App:linear_rewritten}
\end{align}
where $\Mat{\Gamma}=\Mat{T}+\Mat{W}$, or explicitly
\begin{align*}
    \Gamma = \PM{\frac{w}{\tau}F_h-\frac1{\tau}&\frac{w}{\tau}F_\sigma&\frac{w}{\tau}\frac1{\sqrt{N}}\\\frac{w^2(1-p)}{\tau^2 Np}F_h&\frac{w^2(1-p)}{\tau^2 Np}F_\sigma-\frac2{\tau}&\frac{w^2(1-p)}{\tau^2 Np}\frac1{\sqrt{N}}\\0&0&-\frac{1}{\tau}}.
\end{align*}
The two Gaussian white noise processes in the first component of Eq.~\eqref{eq:App:linear_rewritten} can be lumped together into a single Gaussian white noise process:
\begin{align*}
        \od{}{t}\Vect{X} = \Mat{\Gamma}\Vect{X}+\hat{\Mat{B}}\PM{\eta(t)\\\zeta(t)},\quad \hat{\Mat{B}} = \PM{\sqrt{\frac{w^2}{\tau^2}\frac{r_0}{N}+\frac{\sigma_\text{ext}^2}{\tau}}&0\\0&0\\0&\sqrt{\frac{2G_0}{\tau}}}.
\end{align*}
% with 
% \begin{gather*}
% \hat{\Mat{B}} = \PM{\sqrt{\frac{w^2}{\tau^2}\frac{r_0}{N}+\frac{\sigma_\text{ext}^2}{\tau}}&0\\0&0\\0&\sqrt{\frac{2G_0}{\tau}}}.
% \end{gather*}
This is a three-dimensional Ornstein-Uhlenbeck process for which the stationary variance is known to be the solution of the linear system of equations \cite{Ris84} (Lyapunov equation)
\begin{equation}
    \Mat{\Gamma}\Mat{\sigma} + \Mat{\sigma}\Mat{\Gamma}^T = -\hat{\Mat{B}}{\Mat{\hat B}}^T.\label{eq:App:Lyapunov-eq}
\end{equation}
Because the covariance matrix $\Mat{\sigma}$ is symmetric, $\Mat{\sigma}_{ij} = \Mat{\sigma}_{ji}$, we have a system of six linear equations for the variables $\sigma_{11}$, $\sigma_{12}$, $\sigma_{13}$, $\sigma_{22}$, $\sigma_{23}$ and $\sigma_{33}$. For a systematic solution, we notice that the matrix equation~\eqref{eq:App:Lyapunov-eq} reads in the position $(3,3)$
\begin{align}
    2(\Gamma_{31}\sigma_{13}+\Gamma_{32}\sigma_{23}+\Gamma_{33}\sigma_{33}) &= -\frac{2G_0}{\tau},\nonumber    
\end{align}
which reduces to $\sigma_{33} = G_0$
because $\Gamma_{31}=\Gamma_{32}=0$ and $\Gamma_{33}=-1/\tau$. Therefore, we only need to solve a system of 5 linear equations
\begin{multline*}
    \PM{2\Gamma_{11}&2\Gamma_{12}&0&2\Gamma_{13}&0\\
    0&2\Gamma_{21}&2\Gamma_{22}&0&2\Gamma_{23}\\
    \Gamma_{21}&\Gamma_{11}+\Gamma_{22}&\Gamma_{12}&\Gamma_{23}&\Gamma_{13}\\
    0&0&0&\Gamma_{11}+\Gamma_{33}&\Gamma_{12}\\
    0&0&0&\Gamma_{21}&\Gamma_{22}+\Gamma_{33}}\PM{\sigma_{11}\\ \sigma_{12}\\ \sigma_{22}\\ \sigma_{13}\\ \sigma_{23}}\\
    = \PM{-\left(\frac{w^2}{\tau^2}\frac{r_0}{N}+\frac{\sigma_\text{ext}^2}{\tau}\right),&0,&0,&-\Gamma_{13}G_0,&-\Gamma_{23}G_0}^T.
\end{multline*}
We can solve the last two equations separately as they only contain two variables:% and which form the subsystem
% \begin{align}
%     \PM{\Gamma_{11}+\Gamma_{33}&\Gamma_{12}\\ \Gamma_{21}&\Gamma_{22}+\Gamma_{33}}\PM{\sigma_{13}\\ \sigma_{23}} =: \Mat{D}\PM{\sigma_{13}\\ \sigma_{23}} = -G_0\PM{\Gamma_{13}\\ \Gamma_{23}}\\
%     \PM{\sigma_{13}\\ \sigma_{23}} = -\frac{G_0}{\det D} \PM{\Gamma_{13}(\Gamma_{22}+\Gamma_{33})-\Gamma_{23}\Gamma_{12}\\ -\Gamma_{13}\Gamma_{21}+\Gamma_{23}(\Gamma_{11}+\Gamma_{33})}.\label{eq:App:sig13_sig23}
% \end{align}
\begin{align}
%    \PM{\Gamma_{11}+\Gamma_{33}&\Gamma_{12}\\ \Gamma_{21}&\Gamma_{22}+\Gamma_{33}}\PM{\sigma_{13}\\ \sigma_{23}}  = -G_0\PM{\Gamma_{13}\\ \Gamma_{23}}\\
    \PM{\sigma_{13}\\ \sigma_{23}} = -\frac{G_0}{D} \PM{\Gamma_{13}(\Gamma_{22}+\Gamma_{33})-\Gamma_{23}\Gamma_{12}\\ -\Gamma_{13}\Gamma_{21}+\Gamma_{23}(\Gamma_{11}+\Gamma_{33})},\label{eq:App:sig13_sig23}
\end{align}
where $D=(\Gamma_{11}+\Gamma_{33})(\Gamma_{22}+\Gamma_{33})-\Gamma_{21}\Gamma_{12}$. This equation gives us the values of $\sigma_{13}$ and $\sigma_{23}$. 
%Note that these covariances are directly proportional to $G_0$, and thus vanish in the limit $p\to 1$, i.e. in the 1st-order MF theory. 
The remaining three equations of the form
\begin{multline*}
    \PM{2\Gamma_{11}&2\Gamma_{12}&0\\0&2\Gamma_{21}&2\Gamma_{22}\\ \Gamma_{21}&\Gamma_{11}+\Gamma_{22}&\Gamma_{12}}\PM{\sigma_{11}\\ \sigma_{12}\\ \sigma_{22}}\\% =: \Mat{B}\PM{\sigma_{11}\\\sigma_{12}\\ \sigma_{22}}\\
    = \PM{-\frac{w^2}{\tau^2}\frac{r_0}{N}-\frac{\sigma_\text{ext}^2}{\tau}\\0\\0}-\PM{2\Gamma_{13}\sigma_{13}\\2\Gamma_{23}\sigma_{23}\\ \Gamma_{23}\sigma_{13}+\Gamma_{13}\sigma_{23}}
\end{multline*}
have the solution 
\begin{multline}\label{eq:App:sig11_12_22}
%    \PM{\sigma_{11}\\ \sigma_{12}\\ \sigma_{22}} = \\
    \PM{\sigma_{11},& \sigma_{12},& \sigma_{22}}^T = \\
    \frac{-2}{\tau E}\lrrund{\frac{w^2r_0}{\tau N}+\sigmaext^2} \PM{\Gamma_{12}\Gamma_{21}-\Gamma_{22}(\Gamma_{11}+\Gamma_{22})\\ \Gamma_{22}\Gamma_{21}\\ -\Gamma_{21}^2}\\
%    -\frac{2\sigma_\text{ext}^2}{\tau E} \PM{\Gamma_{12}\Gamma_{21}-\Gamma_{22}(\Gamma_{11}+\Gamma_{22})\\ \Gamma_{22}\Gamma_{21}\\ -\Gamma_{21}^2}\\
    -\frac{4\sigma_{13}}{E}\PM{\Gamma_{13}[\Gamma_{21}\Gamma_{12}-\Gamma_{22}(\Gamma_{11}+\Gamma_{22})] + \Gamma_{23}\Gamma_{12}\Gamma_{22}\\\Gamma_{13}\Gamma_{22}\Gamma_{21}-\Gamma_{23}\Gamma_{11}\Gamma_{22}\\-\Gamma_{13}\Gamma_{21}^2+\Gamma_{23}\Gamma_{11}\Gamma_{21}}\\
    -\frac{4\sigma_{23}}{E}\PM{-\Gamma_{23}\Gamma_{12}^2+\Gamma_{13}\Gamma_{12}\Gamma_{22}\\\Gamma_{23}\Gamma_{11}\Gamma_{12}-\Gamma_{13}\Gamma_{11}\Gamma_{22}\\ -\Gamma_{23}[\Gamma_{11}(\Gamma_{11}+\Gamma_{22})-\Gamma_{21}\Gamma_{12}]+\Gamma_{13}\Gamma_{11}\Gamma_{21}}
\end{multline}
with the determinant of the remaining $3\times3$ subsystem
\begin{equation*}
    E = -4(\Gamma_{11}\Gamma_{22}-\Gamma_{12}\Gamma_{21})(\Gamma_{11}+\Gamma_{22})
%    E = -4\det(\hat{\Mat{\Gamma}})\Tr({\hat{\Mat{\Gamma}}}),\quad \hat{\Mat{\Gamma}} = \PM{\Gamma_{11}&\Gamma_{12}\\ \Gamma_{21}&\Gamma_{22}}.
\end{equation*}
The formula for the variances of the mean input potential $\bar h$ and the firing rate $r$ can be more easily interpreted if some further, heuristic simplifications are made. 
%Firstly, the terms for the covariance with the variable $\xi$ are a finite size correction which can be ignored if the network is sufficiently large enough. Secondly, we will focus on the variance of $\bar h$ under the assumption that the term with its variance has the dominant contribution to the linear formula for the variance of the firing rate. 
For the parameters used in our study, we observed that $\Gamma_{11}$ is typically much larger than $\Gamma_{22}$, so that we could safely replace the terms  $\Gamma_{11}+\Gamma_{22}$ in Eq.~\eqref{eq:App:sig11_12_22} by $\Gamma_{11}$. Likewise, we found that the terms proportional to $\sigma_{13}$ and $\sigma_{23}$ in Eq.~\eqref{eq:App:sig11_12_22} are small enough to be ignored. These simplifications yield
\begin{equation*}
    \sigma_{11} \approx \frac12 \frac{1}{(1 - w  F_h)} \left(\frac{w^2r_0}{\tau N}+\sigma_{\text{ext}}^2\right).
\end{equation*}
To simplify even further, we note that in our simulations we have that $\vert w\vert F_h \gg 1$, and hence
\begin{equation}
    \sigma_{11} \approx \left(\frac{-w r_0}{2\tau N} + \frac{\sigmaext^2}{-2w}\right)\frac1{F_h}
    \label{eq:sigma11-simple}
\end{equation}
As mentioned in the main text, the variance of the population firing rate is dominated by the first term in Eq.~\eqref{eq:variance_r_linear_with_finite_size}, so that we obtain from Eq.~\eqref{eq:sigma11-simple}
\begin{equation}
    \var{r} \approx \left(\frac{-w r_0}{2\tau N} + \frac{\sigmaext^2}{-2w}\right)F_h.\label{eq:sigmarr-simple_result}
\end{equation}
The variance in the 1st-order MF theory can be either calculated directly from  Eq.~\eqref{eq:meso_main} or can be understood as a special case of the formulas Eqs.~\eqref{eq:sigma11-simple}, \eqref{eq:sigmarr-simple_result} for $p\to 1$ while keeping $N$ constant. The variances for the mean input and the population firing rate read
\begin{align*}
    \var{\bar{h}} &= \frac{1}{2(1-w\phi_h)}\left(\frac{w^2r_0}{\tau N}+\sigmaext^2\right),\\
    \var{r} &= \frac{1}{2(1-w\phi_h)}\left(\frac{w^2r_0}{\tau N}+\sigmaext^2\right) \phi_h^2,
\end{align*}
where $\phi_h$ is the derivative of the single neuron transfer function at the fixed point. For the the approximation \mbox{$1-w\phi_h\approx -w\phi_h$} we arrive at the same form as Eqs.~\eqref{eq:sigma11-simple}, \eqref{eq:sigmarr-simple_result}.
%Written out and using $\sigma_{12}=\sigma_{21}$ we have a system of linear equations with three variables\begin{align}
%    \begin{pmatrix} 2\Gamma_{11} & 2\Gamma_{12} & 0 \\ \Gamma_{21} & \Gamma_{22}+\Gamma_{11} & \Gamma_{12}\\ 0&2\Gamma_{21}&2\Gamma_{22}\end{pmatrix}\begin{pmatrix} \sigma_{11}\\ \sigma_{12}\\ \sigma_{22}\end{pmatrix} = \begin{pmatrix} -\frac{w^2}{\tau^2} \frac{r_0}{N}\\0\\0\end{pmatrix}.
%\end{align}
%The solution can be obtained with standard methods:
%\begin{align}
%    \begin{pmatrix} \sigma_{11}\\ \sigma_{12}\\ \sigma_{22}\end{pmatrix} = -\frac{2w^2r_0}{D\tau^2N} \begin{pmatrix} (\Gamma_{11}+\Gamma_{22})\Gamma_{22}-\Gamma_{12}\Gamma_{21}\\ -\Gamma_{22}\Gamma_{21}\\ \Gamma_{21}^2\end{pmatrix}
%\end{align}
%where we used $D=4\Tr(\bs{\Gamma})\det(\bs{\Gamma})$ as an abbreviation for the determinant of the matrix above. The variance $\sigma_{rr}$ for the mean firing rate is in the linear approximation
%\begin{align}
%    \sigma_{rr} = F_h^2 \sigma_{11} + F_\sigma^2 \sigma_{22} + 2F_h F_\sigma \sigma_{12}.
%\end{align}

\section{Calculation of the functions $F$ and $G$}
\label{sec:App:Calculation of the function F and G}
For the evaluation of Eq~\eqref{eq:F_formula} and Eq~\eqref{eq:G_formula} for the given non-linearity \eqref{eq:phi-erf}, we use two known formulas~\cite{Owe80} for Gaussian integrals
\begin{align*}
    \int_{-\infty}^{\infty} \Phi(a+bx)g_{0, 1}(x) \mathrm{d}x &= \Phi(\frac{a}{\sqrt{1+b^2}})\\
    \int_{-\infty}^{\infty} \Phi^2(a+bx)g_{0, 1}(x)\mathrm{d}x &=\Phi\left(\frac{a}{\sqrt{1+b^2}}\right)\\
    &-2T\left(\frac{a}{\sqrt{1+b^2}}, \frac{1}{\sqrt{1+2b^2}}\right)
\end{align*}
with $a, b\in\R$ and $g_{0, 1}$ being the standard normal distribution.  Here, 
\begin{equation*}
    T(h,a)=g_{0,1}(h)\int_0^a\frac{g_{0,1}(hx)}{1+x^2}\,dx
\end{equation*}
is the Owen's $T$-function. For our integral \eqref{eq:F_formula} we need to substitute $x:= (h-\bar h)/\sigma$
\begin{align*}
    F(\bar h, \sigma^2) &= \phimax\int_{-\infty}^{\infty}\frac{\Phi\left(\beta(h-\vartheta)\right)}{\sqrt{2\pi\sigma^2}}\exp\left[-\frac{(h-\bar h)^2}{2\sigma^2}\right]\ \mathrm{d}h\nonumber\\
    &=\frac{\phimax}{2\pi}\int_{-\infty}^{\infty}\Phi\left(\beta(\bar h -\vartheta)+\beta\sqrt{\sigma^2}x\right)e^{-x^2/2}\ \mathrm{d}x\nonumber\\
    &=\phimax \Phi\left(\frac{\beta(\bar h-\vartheta)}{\sqrt{1+\beta^2\sigma^2}}\right).
\end{align*}
The second integral \eqref{eq:G_formula} is computed analogously using the same substitution.

\section{There exists exactly one fixed point for inhibitory networks with positive external drive}
\label{sec:unique_FP}
For the calculation of the fixed points of the 2nd-order model~\eqref{eq:meso_main}, we set the left-hand side to zero and solve the following equations for $h_0$:
\begin{align}
    \label{eq:Unique_FP_main}
    r_0 &= F(h_0, \sigma^2(h_0)) = \phimax \Phi\left(\frac{\beta h_0}{\sqrt{1+\alpha(\mu_0-h_0)}}\right) =: F(h_0) ,\\
    r_0 &= \frac{h_0-\mu_0}{w}\label{eq:r0_linear_function}
\end{align}
where $\alpha = -\beta^2w(1-p)/(2\tau C) >0$. The strictly positive non-linear function $F$ in Eq.~\eqref{eq:Unique_FP_main} can only intersect the linear function in Eq.~\eqref{eq:r0_linear_function} at a value for the fixed point $h_0 < \mu_0$. Equation~\eqref{eq:r0_linear_function} represents a strictly monotonically decreasing function of $h_0$ for $w<0$. We will show that for $\mu_0>0$, the function $F(h)$ is strictly monotonically increasing for $h_0<\mu_0$. The consequence is that both graphs will only intersect at exactly one point providing a unique fixed point. The proof for the monotonicity of $F(h_0)$ is straightforward. The derivative is given by the derivative of the error-function:
\begin{equation*}
    \label{eq:F_monoton_increasing}
    \od{F}{h}=  \frac{\phimax\beta\exp\left[-\frac{\beta^2h^2}{1+\alpha(\mu_0-h)}\right]}{\sqrt{2\pi}\sqrt{1+\alpha(\mu_0-h)}}\left(1+\frac12\cdot\frac{\alpha h}{1+\alpha(\mu_0-h)}\right).
\end{equation*}
The first factor is always positive, the remaining factor provides the condition for the monotonic increase:
\begin{equation*}
    \frac2{\alpha}+2\mu_0 > h.
\end{equation*}
For $\mu_0>0$, this condition is fulfilled because
\begin{equation*}
\frac2{\alpha}+2\mu_0> \mu_0 > h.
\end{equation*}
Here, we used $\alpha>0$, $\mu_0>0$ and $\mu_0 > h_0$. With $F(h)>0$ strictly monotonically increasing and the linear function in Eq.~\eqref{eq:Unique_FP_main} monotonically decreasing there must be exactly one fixed point solution $h_0\in]-\infty, \mu_0[$ with a positive firing rate $r_0>0$.

\onecolumngrid
\FloatBarrier
\rot{
\section{Summary tables}
\label{sec:tables}

An overview of the parameters and the population models in different parameter regimes are provided in Table ~\ref{T:summary parameter names} and Table~\ref{tab:MF-models}, respectively.}

\begin{table}[!ht]
\rot{
%\begin{tabularx}{0.45\textwidth} { 
\begin{tabularx}{0.9\textwidth} { 
  | >{\centering\arraybackslash}c 
  | >{\centering\arraybackslash}c 
  | >{\centering\arraybackslash}X | }
 \hline
 \textbf{Parameter} & \textbf{Value / Relations}  & \textbf{Description}\\
 \hline
  $\tau$  & $20$ ms  &  time constant of the low-pass filter dynamics \\
\hline
  $\phimax$  &  $100$ Hz &  maximal firing rate \\
\hline
  $\beta$  & $5$ mV$^{-1}$  &  steepness of the single-neuron transfer function (threshold sharpness)\\
\hline
  $\vartheta$  &  $0$ mV &  inflection point of single-neuron transfer function (soft threshold) \\
\hline
  $d$  &  $0$~ms (unless otherwise noted) &  transmission delay  \\
\hline
 $p$  & see figure captions  &  connection probability \\
\hline
$N$  & see figure captions  &   number of neurons  \\
\hline
$C$  & $C=pN$  &  number of incoming synapses  \\
\hline
$w$  & $w=-1$~mV$\cdot$s (unless noted otherwise)  &   total coupling strength \\
\hline
$J$  &  $J=w/C$, (mVs) &  efficacy of single synapse \\
\hline
$a_{ij}$  &  $\lrrund{a_{ij}}\sim\text{Uniform}\lrrund{\left\{\mathcal{A}\in\{0,1\}^{N\times N}:\sum_{j=1}^N \mathcal{A}_{ij}=C\right\}}$ & adjencency matrix, $a_{ij}=\mathbbm{1}_{\{\text{neuron $j$ connected to $i$}\}}$\\
\hline
$J_{ij}$  & $J_{ij} = Ja_{ij}$ & synaptic efficacy from neuron $j$ to $i$ \\
\hline
$w_{ij}$  & $w_{ij} = NJ_{ij}$, $\lrk{w_{ij}}=w$ & rescaled synaptic efficacy \\
\hline
\end{tabularx}
}
\caption{\label{T:summary parameter names}\rot{Model parameters and their values.
%Some of the parameters can be expressed through other parameters. For flexibility and clarity of our argument, we use all of them alongside. For a better overview for the reader, the parameters are listed in the table above.
}}
\end{table}

\newpage
\onecolumngrid
\rot{
%\section{Performance comparison of mean-field theories}
\newcommand{\lacks}[1]   {\xmark~#1}
}

\begin{table}[!htbp]
\rot{
\begin{tabularx}{\textwidth} { 
  | >{\centering\arraybackslash}X 
  | >{\arraybackslash}X 
  | >{\arraybackslash}X 
  | >{\arraybackslash}X | }
\hline
\textbf{Parameter regime}  & \textbf{MF1, Eq.~\eqref{eq:naive_main}} & \textbf{MF2, Eq.~\eqref{eq:meso_main}} & \textbf{MFsparselim, Eq.~\eqref{eq:meso_sparse_limit}} \\
\hline
$N,C\to\infty$, $p=C/N=O(1)$ (dense) & 
\multicolumn{3}{>{\hsize=\dimexpr3\hsize+4\tabcolsep+2\arrayrulewidth\relax}X|}{
\begin{equation}
\tau\od{}{t}\bar{h} = -\bar{h} + \mu_{\rm ext}(t) + w r(t-d) \ ,\ \ r(t) = \phi\bigl(\bar{h}(t)\bigr) \label{eq:Tab:MF1}
\end{equation}
\centering$\bullet$ exact
} \\
\hline
$N\to\infty$, $C=O(1)$ (sparse)& 

$\to$ Eq.~\eqref{eq:Tab:MF1} \newline\newline
\cmark~stationary population rate (for not too strong coupling)\newline 
\xmark~time-dependent rate response poor & 
\multicolumn{2}{>{\hsize=\dimexpr2\hsize+3\tabcolsep+2\arrayrulewidth\relax}X|}{
\begin{equation}
\begin{aligned}
\tau\od{}{t}\bar{h} &= -\bar{h} + \mu_{\rm ext}(t) + wr(t-d) \\
\tau\od{}{t}\sigma^2 &= -2\sigma^2 + \frac{w^2}{\tau C} r(t-d), \\
r(t) &= F(\bar{h}(t), \sigma^2(t))
\end{aligned}
\label{eq:tab:MFsparselim}
\end{equation}
$\bullet$ exact for annealed network \newline
$\bullet$ for quenched network: \newline 
    \mbox{\quad}\cmark~stationary population rate better than MF1\newline 
    \mbox{\quad}\cmark~dynamical rate response } 
 \\
\hline

$N<\infty$, $p=1$ 
&
\multicolumn{2}{>{\hsize=\dimexpr2\hsize+3\tabcolsep+2\arrayrulewidth\relax}X|}{
\begin{equation}
\begin{aligned}
\tau\od{\bar{h}}{t} &= -\bar{h} + \mu_{\rm ext}(t) + w\lreckig{r(t-d) + \sqrt{\frac{r(t-d)}{N}}\eta(t)}\\
r(t) &=  \phi\bigl(\bar{h}(t)\bigr)
\end{aligned}
\label{eq:tab:MF1full}
\end{equation}
$\bullet$ exact$^*$}
&
$\to$ Eq.~\eqref{eq:tab:MFsparselim} with $C=N$ \newline\newline
\xmark~no finite-size fluctuations\newline
\xmark~stationary and time-dependent rate poor 
 \\
\hline
$N<\infty$, $0<p<1$&
$\to$ Eq.~\eqref{eq:tab:MF1full} \newline\newline
$\bullet$ does capture finite-size fluctuations but with significant errors for annealed  and (much more) for quenched networks & 
\cmark~captures finite-size fluctuations very well for annealed network \newline
\cmark~strong improvement for quenched network compared to MF1 &
$\to$ Eq.~\eqref{eq:tab:MFsparselim} \newline\newline
\xmark~no finite-size fluctuations\newline 
\\
\hline
\end{tabularx}
}
\caption{\label{tab:MF-models} \rot{Performance summary of the mean-field theories for different parameter regimes. $^*$ Exact, apart from the minor diffusion approximation of the shot noise, which is valid for sufficiently large $C$ and which is actually not really necessary: one could go back and replace the term $A(t)\approx r(t)+\sqrt{r(t)/N}\eta(t)$ by a Poissonian shot noise $A(t)=\text{Pois}(Nr(t)dt)/(Ndt)$, which would recover the exact dynamics.}}
\end{table}

\FloatBarrier
\twocolumngrid

%
%\bibliography{my}

\end{document}